\documentclass[11pt]{article}
\usepackage{amsmath,amssymb,graphicx,epsfig,cite,color,hyperref}
\newcommand{\tmop}[1]{\ensuremath{\operatorname{#1}}}

\newcommand{\tmtextrm}[1]{{\rmfamily{#1}}}
\newcommand{\tmtexttt}[1]{{\ttfamily{#1}}}
\newcommand{\be}{\begin{equation}}
\newcommand{\ee}{\end{equation}}
\newcommand{\bea}{\begin{eqnarray}}
\newcommand{\eea}{\end{eqnarray}}
\newcommand{\bi}{\begin{itemize}}
\newcommand{\ei}{\end{itemize}}
\newcommand{\ben}{\begin{enumerate}}
\newcommand{\een}{\end{enumerate}}

\newcommand{\lc}{\left[}
\newcommand{\rc}{\right]}
\newcommand{\lp}{\left(}
\newcommand{\rp}{\right)}

\def\frac#1#2{{{#1}\over {#2}}}
\def\gsim{\mathrel{\rlap{\lower4pt\hbox{\hskip1pt$\sim$}}
    \raise1pt\hbox{$>$}}}         %greater than or approx. symbol
\def\lsim{\mathrel{\rlap{\lower4pt\hbox{\hskip1pt$\sim$}}
    \raise1pt\hbox{$<$}}}         %less than or approx. symbol
\newcommand{\MS}{$\overline{\text{\tmtextrm{MS}}}$}

\newcommand{\draft}[1]{}

\definecolor{grey}{rgb}{0.5,0.5,0.5}

\begin{document}
\begin{flushright}
IFUM-949-FT\\
Nikhef-2010-001\\
ITP-UU-10/03\\
ITFA-2010-01

\end{flushright}
\begin{center}
{\Large \bf Heavy quarks in deep-inelastic scattering
}\\
\vspace{1.6cm}
 Stefano~Forte$^{1,2}$, Eric Laenen$^{3}$, Paolo~Nason$^4$ and 
Juan~Rojo$^2$

\vspace{.8cm}
{\it 
~$^1$ Dipartimento di Fisica, Universit\`a di Milano and\\
~$^2$INFN, Sezione di Milano,\\ Via Celoria 16, I-20133 Milano, Italy\\
~$^3$  ITFA, University of Amsterdam,
  Valckenierstraat 65, 1018 XE Amsterdam, \\
  ITF, Utrecht University, Leuvenlaan 4, 3584 CE Utrecht and\\
  Nikhef Theory Group,  Science Park 105, 1098 XG Amsterdam, The Netherlands\\
~$^4$ INFN, Sezione di Milano-Bicocca,\\Piazza della Scienza 3, I-20126 Milano, Italy\\
}
\vspace{2.4cm}
{\bf \large Abstract:}
\end{center}
We discuss a general framework for the  inclusion  of
heavy quark mass contributions to deep-inelastic structure functions
and their perturbative matching to structure
functions  computed in variable--mass schemes.  
Our approach is based on the so--called
FONLL method, previously introduced and applied to heavy quark
hadroproduction  and photoproduction. We
define our framework, provide expressions up to second order in the
strong coupling, and use them to construct matched expressions for
structure functions up to NNLO.  After checking explicitly the
consistency of our results, we perform a study of the phenomenological
impact of heavy quark terms, and compare 
results obtained at various perturbative orders, 
and with various prescriptions for
the treatment of subleading terms, specifically those related to
threshold behaviour. We also consider the heavy quark structure
function $F_{2\,c}$ and discuss issues related to
the presence of mass singularities in their coefficient functions.

\clearpage

\tableofcontents

\clearpage

\section{Introduction}
\label{sec:intro}

Interest in the inclusion of heavy flavour contributions to deep--inelastic
electroproduction structure functions was recently revived by the
discovery~\cite{Tung:2006tb} that mass-suppressed
terms in global parton fits can affect predictions for the total $W$
and $Z$ production at the LHC by almost 10~\%. Indeed, until very
recently, many global studies of parton distributions (PDFs, henceforth)
were performed assuming  heavy quarks to decouple from
 the structure functions for $Q^2<m_h^2$, but to be massless when
$Q^2>m_h^2$, despite the availability of more refined approaches where
contributions with full mass dependence are consistently taken
into account.

A technique for the inclusion of heavy mass-suppressed
contributions
to structure functions was developed long 
ago~\cite{Collins:1998rz,Aivazis:1993pi},
based upon a %  previous (P.N.Dec10)
renormalization scheme with
explicit heavy quark decoupling~\cite{Collins:1978wz}.
Several variants of this  method  (usually called ACOT) 
were subsequently proposed, such as 
S-ACOT~\cite{Kramer:2000hn} and
ACOT-$\chi$~\cite{Tung:2001mv,Kretzer:2003it}.
However,  the ACOT method was 
first used for an actual global parton fit only recently, in
Refs.~\cite{Kretzer:2003it,Tung:2006tb}.
An alternative method (sometimes called TR) has also been 
advocated~\cite{Thorne:1997ga,Thorne:1997uu,Thorne:2006qt}, and used for parton
fits~\cite{Martin:1998sq,Martin:2001es,Martin:2002aw}. 
Recently, however, the methods used by
the CTEQ~\cite{Tung:2006tb} and MSTW~\cite{Martin:2009iq} groups for
their current parton fits, based respectively on the
ACOT~\cite{Collins:1998rz,Aivazis:1993pi} and
TR~\cite{Thorne:1997ga,Thorne:1997uu,Thorne:2006qt} procedures, have adopted at
least in part a common framework: they have been compared recently in
Refs.~\cite{Thorne:2008xf,Olness:2008px}, thereby elucidating
differences and common aspects.

A somewhat different technique for the inclusion of heavy quark
effects, the so called FONLL method, 
was introduced in Ref.~\cite{Cacciari:1998it} in the context of
hadroproduction of  heavy
quarks. The FONLL method only relies on
standard QCD factorization and calculations with massive quarks in the
decoupling scheme of Ref.~\cite{Collins:1978wz} and with massless quarks in the
 \MS\ scheme. The name FONLL is motivated by the fact
that the method was originally %  (P.N.Dec10)
used to combine a fixed (second) order calculation
with a next-to-leading log one; however, the method is entirely
general, and it can be used to combine consistently a fixed order with
a resummed
calculation to any order of either.

It is the purpose of this paper to present the application of the
FONLL scheme to deep--inelastic structure functions. We shall see
that, thanks to its simplicity, the method actually provides a
framework for understanding differences between other existing
approaches, and for a study of the effect of different choices in the
inclusion of subleading terms. We shall % then (P.N.Dec10)
use this scheme to
perform detailed studies of the phenomenological impact of the
treatment of heavy quarks up to NNLO, and of the ambiguities involved
in the procedure.

In section~2 we will describe in detail the theory of the FONLL
method for structure functions.
In section~3 we will provide
explicit expressions up to NNLO, and discuss issues related to
perturbative ordering.  In section~4 we shall
compare the structure functions
determined with different
heavy quark matching procedures  at various orders, and with various
treatments of threshold subleading terms. Finally, in section~5 we
shall discuss issues related to the
presence of mass singularities in the definition of the heavy quark
structure functions $F_{2\,c}$ and $F_{2\,b}$. For ease of reference
and for definiteness, the complete explicit FONLL expressions for the structure
function   $F_{2}$ are collected in an Appendix.

\section{The FONLL method in deep-inelastic scattering}
\label{sec:FONLL}

The basic problem in the treatment of heavy quarks in a QCD process
stems from the fact that QCD calculations are usually performed in a
decoupling scheme~\cite{Collins:1978wz}, rather than in the
\MS\ scheme. 
Indeed, in the standard \MS\
scheme, heavy quarks contributions are present at all scales: for
instance, the $\beta$ function depends on $n_f$ with $n_f=6$ at all
scales. In a decoupling scheme,
instead, in the computation of a process characterized by the hard
scale $Q^2$ all  quarks with mass $m_q^2>Q^2$ (``heavy'', henceforth) 
are not treated
in the \MS\ scheme. Rather,  `heavy flavour graphs' are subtracted at
zero momentum.\footnote{Here `heavy flavour graphs' are
  defined~\cite{Collins:1978wz} 
(recursively) as either graphs which
contain a heavy flavour line, or counterterms to heavy flavour
graphs.}
The important consequence of this definition is that 
heavy quarks decouple
for scales much lower than the heavy quark mass.
This implies that the
GLAP evolution equations, and the
running of $\alpha_s$  are identical
to those which one would get in the
\MS\ scheme, but with $n_f$ equal to the number of light flavours,
$n_f=n_l$ .
If for $Q^2<m_h^2$
one neglects all terms suppressed by  powers of $Q^2/m_h^2$,
 one obtains a so--called
zero--mass 
scheme, where all quarks are treated as
massless, but heavy quarks are absent at $Q^2\le m^2_h$.
This scheme can be combined with the usual \MS{} scheme for $n_l+1$
flavours, including the heavy quark, when $Q^2\ge m^2_h$; if  terms
suppressed by  powers of $m_h^2/Q^2$ are neglected throughout this
yields the so called
zero--mass variable--flavour number scheme (ZMVFN from now on) of
Ref.~\cite{Aivazis:1993pi}.\footnote{Note that here and henceforth we
 call ``variable--flavour number'' scheme (VFN, and thus also ZMVFN) 
 a scheme where
  all large $Q^2$ logs are resummed, so in particular, when $Q^2>>m^2_h$ all
  $\ln\frac{Q^2}{m_h^2}$ terms are resummed. In contrast, in some
  previous references, VFN or ZMVFN is used to denote schemes in which
  $\ln\frac{Q^2}{m_h^2}$ are not resummed to all orders: in particular
  in Ref.~\cite{Buza:1996wv} VFN denotes the scheme
 in which logs are not resummed, while that in which logs are resummed
 is called ``PDF''. In Ref.~\cite{Alekhin:2009ni} ZMVFN is used
 despite the fact that $\ln\frac{Q^2}{m_h^2}$ are never resummed to
 all orders in this reference~\cite{blumpriv}.}

The ZMVFN scheme
is not accurate near the threshold region, where
$\frac{m_h^2}{Q^2}\sim 1$. The problem in this region 
is easily remedied by simply
using   the decoupling scheme with $n_f=n_l$,
but retaining explicitly the full
dependence on the heavy quark mass in the computation of hard
cross sections. This way of computing however looses accuracy in
comparison to the previous ZMVFN when
$Q^2\gg m_h^2$ so that $L\equiv\ln Q^2/m_h^2 \gg 1$, because, 
in the decoupling scheme with $n_f=n_l$, these large
logs  are only included to
fixed order in $\alpha_s$ while in the ZMVFN scheme they are resummed
to all orders. 

Heavy quark schemes are all based on the idea of matching these two
different ways of calculating, each of which is more accurate in some
kinematic region. The basic idea in the ACOT~\cite{Aivazis:1993pi}
scheme  is to retain explicitly
the mass dependence in Wilson coefficients of the ZMVFN scheme
calculation (based on the massive quark factorization
theorem~\cite{Collins:1998rz}), while the basic idea in the 
TR~\cite{Thorne:1997ga,Thorne:1997uu} scheme is to require continuity
of physical observables in the threshold region, where one switches
between the decoupling scheme and the massless calculation with
$n_{f}=n_l+1$ flavours. In both cases, % P.N. 5-1-2010 ~\cite{Thorne:1997ga}
the matching
conditions ensue from the requirement that computations of the same
observable within different renormalization schemes give the same
answer within the respective accuracy.

The FONLL scheme is instead simply based on the idea of combining the
decoupling scheme computation with the ZMVFN computation, and
subtracting double counting terms between the two order by order in an
expansion in powers of $\alpha_s$ and
$L$. Effectively, this means that for
$Q^2>m^2_h$  one
performs the calculation in the massless
scheme, but then one replaces
all terms whose mass-dependence is known with their exact massive
expression. This can then be done to any desired order.

\subsection{FONLL expressions for DIS structure functions}
\label{sec:fonlldissf}

Let us now see explicitly how the FONLL scheme works in the case of
a generic deep-inelastic structure function $F(x,Q^2)$: in
Sects.~\ref{sec:implement}-\ref{sec:pheno} 
 we
shall specifically discuss both $F_2$ and $F_L$, but for the time
being the distinction is irrelevant, so  we will refer
to a generic structure function $F$. 
For the rest of the paper we will assume $n_l$ light flavours, with
 a single heavy
flavour of
mass $m_h=m$. In Sec.~\ref{sec:pheno}, when discussing
phenomenological implications, we will study specifically the case of
charm.

The
expression of $F(x,Q^2)$ in the fully massless scheme, which is accurate
when $W \gg 4 m^2$, % P.P.Dec10
with $W \equiv \sqrt{Q^2 (1 - x) / x}$, is factorized in terms of
PDFs\footnote{Note that we define PDFs $f(x)$ in such a way that 
  $xf(x)$ is a
  momentum distribution; a different definition such that $f(x)$ is
  the momentum distribution itself, which differs from our own by a
  factor of $x$, has also been sometimes used (see
  e.g. Ref.~\cite{Laenen:1992zk}).}
\begin{equation}
  F^{(n_l+1)} (x, Q^2) = x \int_x^1 \frac{dy}{y} \sum_{i = q, \bar{q}, h,
  \bar{h}, g} C_i^{(n_l+1)} \left( \frac{x}{y}, \alpha_s^{(n_l+1)} (Q^2) \right)
  f_i^{(n_l+1)} (y, Q^2), \label{eq:Fnf}
\end{equation}
where $q$ is any light quark, $h$ is the heavy quark, and $n_l+1$ is the total
number of flavours. In this expressions, the PDFs include the
heavy flavour as a light parton, and % comma, P.N. Dec10
the strong coupling $\alpha_s^{(n)}$ and
the PDFs $f_i^{(n)}$ satisfy standard GLAP
equations with $n_f=n_l+1$. Henceforth, we shall refer to Eq.~(\ref{eq:Fnf}) as the determination
of $F(x,Q^2)$ in the massless scheme.

The expression of $F(x,Q^2)$ in the decoupling scheme with
$n_f=n_l$, which is accurate when $W \approx 4 m^2$, is instead given by
\begin{equation}
  F^{(n_l)} (x, Q^2) = x \int_x^1 \frac{dy}{y} \sum_{i = q, \bar{q}, g}
  C_i^{(n_l)} \left( \frac{x}{y}, \frac{Q^2}{m^2}, \alpha_s^{(n_l)} (Q^2) \right)
  f_i^{(n_l)} (y, Q^2). \label{eq:Fnl}
\end{equation}
The coefficient functions are computed fully retaining the mass
dependence,
while now
$\alpha_s^{(n_l)}$ and $f_i^{(n_l)} (y, Q^2)$ obey standard $\overline{\tmop{MS}}$
evolution equations with $n_l$ flavours. When  $W^2
\ll 4 m^2$  the heavy quark mass dependence drops out of the coefficient
functions $C_i^{(n_l)}$, which then reduce to the
standard massless \MS\ 
coefficient
functions with $n_l$ flavours. Henceforth, we shall refer to Eq.~(\ref{eq:Fnl}) as the determination
of $F(x,Q^2)$ in the massive scheme.

In order to carry out
the FONLL procedure, we need to express the decoupling scheme cross
section, Eq.~(\ref{eq:Fnl}) in terms of $\alpha_s^{(n_l+1)}$ and $f_i^{(n_l+1)}$ for
$i\ne h,\bar{h}$. %P.N. Dec10
The coupling constant and PDFs are
related in the two schemes by equations of the form
\begin{eqnarray}
  \alpha_s^{(n_l+1)} (Q^2) & = & \alpha_s^{(n_l)} (Q^2) + \sum^{\infty}_{i = 2} c_i (L)
  \times \left( \alpha_s^{(n_l)} (m^2) \right)^i \ ,  \label{eq:matchingalpha}\\
  f_i^{(n_l+1)} (x, Q^2) & = & \int_x^1 \frac{dy}{y} \sum_{j = q, \bar{q}, g} K_{ij} \left( \frac{x}{y}, L,
  \alpha^{(n_l)}_s (Q^2) \right) f_j^{(n_l)} (y, Q^2) \ ,  
  \label{eq:matchingpdf}
\end{eqnarray}
where 
\begin{equation}
L \equiv \log Q^2 / m^2 .
\label{ldef}
\end{equation}
The coefficients $c_i (L)$ are polynomials in $L$, and
 the functions $K_{i j}$ can be expressed as an expansion in
powers of
$\alpha_s$, with
coefficients that are polynomials in $L$. 

The first $2n_l+1$ equations in Eq.~(\ref{eq:matchingpdf}) relate the light
quark and gluon PDFs in the two schemes, and can be inverted to
express the massive--scheme PDFs  in terms of the massless-scheme
ones. The last two equations in Eq.~(\ref{eq:matchingpdf}) express heavy
quark PDFs in terms of the light flavour ones, under the assumption
that the heavy flavour PDF is generated
perturbatively. A possible  intrinsic heavy flavour
contribution~\cite{Brodsky:1981se,Pumplin:2007wg}
could be added as a separate contribution on the right-hand side of
 Eq.~(\ref{eq:matchingpdf}) for $i=h,\>\bar h$.

Inverting Eqs.~(\ref{eq:matchingalpha}-\ref{eq:matchingpdf}) and
substituting in
Eq.~(\ref{eq:Fnl}), one can obtain an expression of $F^{(n_l)} (x,
Q^2)$ 
in terms of $\alpha_s^{(n_l+1)}$ and $f^{(n_l+1)}$:
\begin{equation}
  F^{(n_l)} (x, Q^2) = x \int_x^1 \frac{dy}{y} \sum_{i = q, \bar{q}, g}
  B_i \left( \frac{x}{y}, \frac{Q^2}{m^2}, \alpha_s^{(n_l+1)} (Q^2) \right)
  f_i^{(n_l+1)} (y, Q^2), \label{eq:Fnlbar}
\end{equation}
where the coefficient functions $B_i$ are such that
substituting the matching relations
Eqs.~(\ref{eq:matchingalpha}-\ref{eq:matchingpdf})
in  Eq.(\ref{eq:Fnlbar}) one gets back the original expression
Eq.~(\ref{eq:Fnl}).  We can thus use for $F^{(n_l)}$ the
expression given in Eq.~(\ref{eq:Fnlbar}), and avoid any further
reference to $\alpha_s^{(n_l)}$ and $f_i^{(n_l)}$: from now on, we
shall use Eq.~(\ref{eq:Fnlbar}) as the determination of $F$ in the
massive scheme.

In order to match  the two expressions
for $F$ in the massless scheme,
Eq.~(\ref{eq:Fnf}), and in the massive scheme, Eq.~(\ref{eq:Fnlbar}),
we now work out their perturbative expansion.
Using GLAP evolution in the absence of intrinsic heavy quark
contributions, the heavy quark PDFs $f_{h }^{(n)}$, $f_{\bar h
}^{(n)}$  at the scale
$Q^2$ which appear in the massless--scheme expression
Eq.~(\ref{eq:Fnf})  can be  determined in terms of
the light-quark PDFs, 
$f^{(n_l)_{}}_i$ with $i \neq h,\,   \bar{h}$ at the scale
$m$,  convoluted with coefficient functions  which  can be expressed
as a power series in $\alpha_s^{(n_l)} (m)$,
with coefficients that are polynomials in $L$, or, alternatively,
in terms of
the light-quark parton distributions $f_i^{(n_l+1)}$ at the scale
$Q^2$ convoluted with coefficient functions  expressed
as a power series in $\alpha_s^{(n_l+1)} (Q^2)$, %P.N. Dec10
again with (different)
coefficients that are polynomials in $L$. 
%
% the first half of the sentence is not strictly necessary - one could
% say directly that everything is in terms of alpha(Q) and f_i(Q) -
% but I find the extra step useful because in an actual code it is the
% expression in terms of f(m) that one would have most naturally. SF
% But I won't object if it is deleted.
% 

Thus, the
massless-scheme expression
 Eq.~(\ref{eq:Fnf}) may be written entirely in terms of light-quark
PDFs:
\begin{equation}
 F^{(n_l+1)} (x, Q^2) = x \int_x^1 \frac{dy}{y} \sum_{i = q,
   \bar{q}, g} 
  A_i^{(n_l+1)} \left( \frac{x}{y}, L,\alpha_s^{(n_l+1)} (Q^2) \right)
  f_i^{(n_l+1)} (y, Q^2), \label{eq:Fnflh}
\end{equation}
where the $A_i^{(n_l+1)}$ coefficient functions are given by a
perturbative expansion of the form
\begin{equation}
A_i^{(n_l+1)} \left( z, L,\alpha_s^{(n_l+1)} (Q^2) \right)=
\sum^N_{p = 0} \left( \alpha_s^{(n_l+1)} (Q^2) \right)^p  
  \sum_{k = 0}^{\infty} A_i^{p, k} (z) \left( \alpha_s^{(n_l+1)} (Q^2) L \right)^k,
  \label{eq:logexpnf}
\end{equation}
where at leading order $N=0$, and  at N$^k$LO $N=k$.

On the other hand, the massive-scheme expression Eq.~(\ref{eq:Fnlbar})
is also written in terms of the light quark PDFs, with coefficient
functions $B_i$ which admit a fixed order expansion of the form
\begin{equation}
B_i \left(z, \frac{Q^2}{m^2}, \alpha_s^{(n_l+1)} (Q^2) \right)
= \sum^P_{p = 0} \left(\frac{\alpha_s^{(n)} (Q^2)}{2\pi} \right)^p B_i^p
  \left( z, \frac{Q^2}{m^2} \right), \label{eq:logexpnl}
\end{equation}
where $P$ is the order of the expansion needed to reach the desired accuracy.
It follows that the sum $B^{(0),\,p}_i $ of
all contributions to the massive-scheme expression 
Eq.~(\ref{eq:logexpnl}) which do not vanish when
  $Q^2 \gg m^2$ must also be present in the massless-scheme result,
i.e. they must correspond to the logarithmic and constant terms in
Eq.~(\ref{eq:logexpnf}).  Namely,  
to $p$--th order in $\alpha^{(n_l+1)}_s(Q^2)$
\begin{equation}
B^{(0),\,p}_i \left( x, \frac{Q^2}{m^2} \right) \equiv \sum_{k = 0}^p A^{p - k, k}_i (x)
  L^k,\label{eq:masszlim}
\end{equation}
where we denote by $B_i^{(0),\,p}$ the massless limit of $B_i^{p}$, in the
sense that
%P.N.Dec6
\begin{equation}
  \lim_{m \rightarrow 0} \left[ B^p_i \left( x, \frac{Q^2}{m^2} \right) -
B^{(0),\,p}_i \left( x, \frac{Q^2}{m^2} \right)
\right] = 0. \label{eq:asympmatch}
\end{equation}
In other words, $B_i^{(0),\,p}$ is obtained from $B_i^{p}$ by retaining all
logarithmic and constant terms, and dropping all terms suppressed by powers
of $m/Q$.
%P.N.Dec6

The FONLL method can be simply stated as follows: in the
massless-scheme expression Eq.~(\ref{eq:Fnflh}) one replaces all
contributions to the expansion~(\ref{eq:logexpnf}) of the 
coefficient functions $A_i^{(n_l+1)} \left( z,L, \alpha_s^{(n_l+1)}
(Q^2)\right)$ which appear in $B^{(0),\,p}_i \left( x, \frac{Q^2}{m^2}
\right)$, Eq.~(\ref{eq:masszlim}), with their fully massive expression
$B^p_i \left( x, \frac{Q^2}{m^2} \right)$ from Eq.~(\ref{eq:Fnlbar}). 
In this way, all the mass suppressed
effects  that
are not present in Eq.~(\ref{eq:Fnf}) but are known from Eq.~(\ref{eq:Fnl})
are included. 

In order to do this in a systematic way, we define thus the massless
limit of the massive-scheme expression Eq.(~\ref{eq:Fnlbar}), namely
\begin{equation}
  F^{(n_l,\,0)} (x, Q^2) = x \int_x^1 \frac{dy}{y} \sum_{i = q, \bar{q}, g}
  B^{(0)}_i \left( \frac{x}{y}, \frac{Q^2}{m^2}, \alpha_s^{(n_l+1)} (Q^2) \right)
  f_i^{(n_l+1)} (y, Q^2), \label{eq:Fnlbarzero}
\end{equation}
where
\begin{equation}
B^{(0)}_i \left( z, \frac{Q^2}{m^2}, \alpha_s^{(n_l+1)} (Q^2) \right)
= \sum^P_{p = 0} \left( \alpha_s^{(n_l+1)} (Q^2)
  \right)^m B^{(0),\,p}_i \left( z, \frac{Q^2}{m^2} \right),
\label{eq:massivezero}
\end{equation}
with  $B^{(0),\,p}_i$ given by Eq.~(\ref{eq:masszlim}), and the sum
over $p$ consistently performed including all terms 
up to the order $P$ in $\alpha_s^{(n_l+1)} (Q^2)$ to which
the massive-scheme expression has been determined.

The FONLL approximation for $F_2$ is then given by
\begin{eqnarray}
&&  F^{\tmop{FONLL}} (x, Q^2) =  F^{(d)} (x, Q^2)  + F^{(n_l)} (x,
  Q^2), \label{eq:FONLL}
\\
 &&\quad F^{(d)} (x, Q^2)\equiv
\left[ F^{(n_l+1)} (x, Q^2) - {F}^{(n_l,\,0)} (x,
  Q^2) \right]
\label{eq:fdiff}
\end{eqnarray}
where   $F^{(n_l)}$
is the massive-scheme expression Eq.~(\ref{eq:Fnlbar}), and the
``difference'' contribution Eq.~(\ref{eq:fdiff}) is constructed out of the
massless-scheme expression $F^{(n_l+1)}$ Eq.~(\ref{eq:Fnflh}),
and 
the massless limit  $F^{(n_l,\,0)}$ 
Eq.~(\ref{eq:Fnlbarzero}) of the massive-scheme
expression.

Because of Eq.~(\ref{eq:asympmatch}), when $Q^2 \gg m^2$ the FONLL
expression reduces to the massless-scheme one  $F^{(n_l+1)}$. When instead
$Q^2 \approx m^2$ the FONLL expressions differ from the massive-scheme
one  $F^{(n_l)}$ through the ``difference'' term $F^{(d)}$
Eq.~(\ref{eq:fdiff}), which is however subleading in $\alpha_s(Q^2)$.
Note that in the FONLL scheme
continuity does not play a role in the matching conditions: at NLO
continuity ensues accidentally from the matching conditions, but at
higher orders subleading discontinuities may arise.

When $Q^2<m^2$, but $W^2>4 m^2$ i.e. above
the threshold for heavy quark production there are various options. A
simple possibility is to just use
 the massive scheme result
Eq.~(\ref{eq:Fnlbar}). It is easy to see that this amounts to
replacing  the ``difference'' term in
Eq.~(\ref{eq:FONLL}) by
\begin{equation}
  F^{(d')} (x, Q^2) \equiv \left[ F^{\rm ZMVFN} (x, Q^2) -
    \Theta(Q^2-m^2)
{F}^{(n_l,\,0)} (x,
  Q^2) \right]  \label{eq:FONLLp}
\end{equation}
where $F^{\rm ZMVFN} (x, Q^2)$ denotes the massless calculation with
$n_f=n_l+1$ when $Q^2>m^2$, and with $n_f=n_l$ when $Q^2<m^2$: this
is identical to  Eq.~(\ref{eq:FONLL}) for $Q^2>m^2$, and it reduces to the
massive calculation for $Q^2<m^2$  because the term in square brackets
vanishes there. Of course, when $Q^2\ll m^2$ and $W^2\ll 4 m^2$ 
the massive calculation
 in turn coincides with  the massless calculation with $n_f=n_l$, so
 one recovers the same result as the ZMVFN in this region. However, if
 one only wishes to use the result at scales $Q^2$ which are never
 much lower than $m^2$, a simpler option may be to just use
 Eq.~(\ref{eq:FONLL}) everywhere. Then, when $Q^2< m^2$ but $W^2>4 m^2$
 the term in square brackets will contain some small 
 contributions in the $n_f=n_l+1$ scheme evolved backwards below the
 heavy quark threshold.

The FONLL formula
Eq.~(\ref{eq:FONLL}) may look similar to a prescription
suggested in Ref.~\cite{Buza:1996wv}, and then further discussed in
Ref.~\cite{Buza:1997nv} (in particular Eq.~(5) of the latter), 
sometimes referred to as the BMSN
prescription, which is also based on the idea of combining
computations performed in schemes which differ in the number of active
flavours. However, in the BMSN method the issue of using PDF defined
in a single factorization scheme in all terms is not addressed (unlike
in FONLL, where it  is
accomplished expressing everything, Eq.~(\ref{eq:Fnflh}), 
in terms of $\alpha_s^{(n_l+1)}$ and
  $f^{(n_l+1)}$). This leads to inconsistent results beyond
  ${\mathcal O}(\alpha_s)$, as stated in Ref.~\cite{Buza:1997nv}, where it is
  argued that the inconsistency is however numerically small in
  practice. Also, contributions proportional to the light or heavy
  quark electric charge are not easily separated in the BSMN method,
  again leading to (possibly small~\cite{Buza:1997nv}) inconsistencies
  beyond ${\mathcal O}(\alpha_s)$, unlike in FONLL where this separation can be
  treated in a fully consistent way as we will do explicitly to
  ${\mathcal O}(\alpha^2_s)$ in Sect.~\ref{sec:fonllm} below.
As for the comparison of FONLL to the ACOT and TR schemes, we will
come back to it in the end of Sect.~\ref{sec:fonllsf}, after fully
specifying the FONLL scheme up to ${\mathcal O}(\alpha^2_s)$.

Finally, it should be observed that near threshold the $F^{(d)}$ term
Eq.~(\ref{eq:fdiff}), though subleading, could in
practice be non--negligible, 
and it does not provide any information
because it contains higher--order logarithmic contributions in a
region in which  $L$ is not large.  In practice, it may thus 
be convenient to suppress this term through a suitable 
kinematic factor when $Q$ is near $m_h$, 
which is allowed without modifying the accuracy of
the calculation because this term is subleading, as we shall discuss
in the next section.

\subsection{Mismatch in accuracy}\label{sec:mismatch}

In order for Eq. (\ref{eq:FONLL}) and (\ref{eq:fdiff})
to work properly, it is necessary to determine
$F^{(n_l+1)}$  with an accuracy which is at least as high as that used in the
computation of $F^{(n_l)}$.
Only in this case, in fact, in the difference expression Eq. (\ref{eq:fdiff})
we are subtracting terms that are actually present in $F^{(n_l+1)}$.

However, it is 
possible to generalize Eq.~(\ref{eq:fdiff}) in such a way that
it can still be used even when $F^{(n_l)}$ is known with higher accuracy
than $F^{(n_l+1)}$. For this purpose, it is sufficient to retain
in $F^{(n_l,0)}$ in (\ref{eq:fdiff}) only those terms that are also present in
$F^{(n_l+1)}$. In this case, it is no longer true that when $Q^2\gg m^2$
the FONLL expression reduces to the massless scheme one $F^{(n_l+1)}$
up to mass suppressed terms only. In fact, $F^{(n_l,0)}$ in (\ref{eq:fdiff})
and $F^{(n_l)}$ in (\ref{eq:FONLL}) no longer cancel in this limit, since some
terms have been excluded from  $F^{(n_l,0)}$. It is however still true
that the FONLL reduces to the massless scheme one $F^{(n_l+1)}$ up
to mass suppressed terms and terms of higher order in ${\cal O}(\alpha_s)$.
Further on, we will illustrate an application of this
approach to the charm structure function, where $F_c^{(n_l+1)}$ is evaluated
at NLO, and $F_c^{(n_l,0)}$ is evaluated at order ${\cal O}(\alpha_s^2)$.

\subsection{The heavy quark threshold}
\label{sec:threshold}

As already stated, when $Q \approx m$ the $F^{(d)}$ term becomes totally
unreliable, since it contains higher--order contributions enhanced by
powers of $L$, in a region in which $L$ is not large.
 It may then be advisable to suppress $F^{(d)}$
 Eq.~(\ref{eq:fdiff}) in the threshold region. Because $F^{(d)}$ is
formally
 subleading, this of course does not
 change the nominal accuracy of the calculation,  but it may in
 practice improve the perturbative stability and smoothness of the result.

This suppression can be obtained in various ways. Two classes of
possibilities which have been considered in previous studies consist
of introducing a threshold factor or a rescaling variable.
In the former case, one replaces $F_2^{(d)}$ 
 Eq.~(\ref{eq:fdiff}) with 
\begin{equation}
  F^{(d,\,th)} (x, Q^2) = f_{\text{\tmtextrm{thr}}} (x, Q^2) F^{(d)} (x
, Q^2),
\end{equation}
where $f_{\tmop{thr}}$ is such that $ F^{(d,\,th)} (x, Q^2)$ only
differs from $F^{(d)} (x
, Q^2)$ by terms that are power--suppressed for large $Q^2$, namely
\begin{equation}
  f_{\text{\tmtextrm{thr}}} (x, Q^2) = 1 + \mathcal{O} \left( 
 \frac{m^2}{Q^2}
   \right)\label{eq:thrfacts}
\end{equation}
but it enforces vanishing of $F^{(d,\,th)} (x, Q^2)$ below the
threshold at $Q^2=m^2$. A suitable choice is
\begin{equation}
  f_{\tmop{thr}} (x, Q^2) =  \Theta (Q^2 - m^2) \left(1 - \frac{Q^2}{m^2}\right)^2,
\label{eq:threshold}
\end{equation}
where the factor in brackets ensure that $f_{\tmop{thr}} (x, Q^2)$ and
its first derivative with respect to $Q^2$ are continuous at
$Q^2=m^2$.

When using a rescaling variable, one instead replaces  in all
convolutions which enter the expression for $F^{(d)}$ the variable $x$
with  a new rescaling variable $\chi
  (x, Q^2)$, so that the factorized expression of all structure
functions 
is now given by
\begin{equation}
\label{eq:chiscaling}
F^{(\chi)}(x,Q^2)= x\int_{\chi(x, Q^2)}^1\frac{dy}{y} C\left(\frac{\chi(x,
  Q^2)}{y},\alpha(Q^2)\right) f(y,Q^2),
\end{equation}
where the rescaling variable $\chi
  (x, Q^2)$  is chosen in such a way that again
$ F^{(d,\,\chi)} (x, Q^2)$ only
differs from $F^{(d)} (x
, Q^2)$ by power suppressed terms, namely

\begin{equation}
\quad \chi (x, Q^2) = x + \mathcal{O} \left( \frac{m^2}{Q^2} \right),
\label{eq:rescfac}
\end{equation}
but it is such that, viewed as a function of $x$, $F^{(d,\,th)} (x, Q^2)$
only has support above threshold. The threshold can be set at the
physical production value $W^2=m^2$ by choosing
\begin{equation}
\chi = x \left( 1 + \frac{4 m^2}{Q^2} \right).
\label{eq:chidef} 
\end{equation}
This choice was adopted in Ref.~\cite{Tung:2001mv,Kretzer:2003it}, and
the ACOT method supplemented by it is called the ACOT-$\chi$
prescription; it is used among others in the recent CTEQ  parton
fits~\cite{Tung:2006tb}.\footnote{
  Note  that a slightly
  different definition of rescaling is also possible, whereby one
  simply lets \begin{equation}\label{eq:altchi}F^{(d,\,\chi)}=F^{(d)}(\chi(x,Q^2),Q^2).\end{equation} This
    definition, which differs by the above one by a factor
    $\frac{\chi}{x}$, has been for instance discussed in
    Ref.~\cite{Martin:1998sq}; however, the definition
    Eq.~(\ref{eq:chiscaling}) is used by CTEQ~\cite{Nadolsky}.}

A generalization of the $\chi$ variable was introduced in 
Ref.~\cite{Nadolsky:2009ge}, by defining a
one-parameter family of variables which interpolate between $x$ and
$\chi$, but all of which set the threshold at the physical value. In
fact, it was argued in Ref.~\cite{Thorne:2008xf} that a full treatment
of heavy quark threshold such as ACOT or the FONLL method discussed
here could be well--approximated by simply using the ZMVFN, but with
$x$ replaced by a suitable fine--tuned $\chi$--like variable. Be that
as it may, we note that within the FONLL (or ACOT) method there is no
conceptual advantage in introducing a suppression at the physical
threshold $W^2$ over a suppression at $Q^2=m^2$, given that for most
of the interesting kinematic region the physical threshold correspond
to values of $Q^2$ which are lower than $m^2$, and where thus the massive
expression is being used anyway.

In Ref.~\cite{Cacciari:1998it},
which dealt with hadroproduction of heavy quarks, yet another form of
a suppression factor was proposed, based on the transverse momentum of
the heavy quark pair. We remark here that there is no straightforward
generalization of the $\chi$-scaling prescription to the hadroproduction
case.

All these threshold modifications of higher order terms, being subleading,
can neither be justified nor be excluded a priori. Their purpose is to
mimic in a phenomenological way the effect
of finite mass effects not included in the calculation. They can thus
be validated to a certain extent by comparing results obtained using
these prescriptions with exact results when the latter are known,
though of course this does not prove that their effect will be the
same or similar when the exact result is not known. 
 We will perform
this validation in section~\ref{sec:compth} below.

\section{Implementation of the FONLL method}
\label{sec:implement}

We shall now work out explicitly the FONLL prescription
Eq.~(\ref{eq:FONLL}) up to $\mathcal{O}(\alpha_s^2)$.  
 We will discuss separately the contributions to
the structure function in
which the virtual photon couples to the heavy quark only (``heavy''
structure function $F_{h}$) and the rest.
The basic ingredient in the FONLL construction is the scheme change of
Eqs.~(\ref{eq:matchingalpha}-\ref{eq:matchingpdf}). We will first work
this out up to order $\alpha_s^2$, then use it to construct up to the
same order the various contributions to the FONLL expression of
Eq.~(\ref{eq:FONLL}). We will then combine these contributions into
 expressions for the structure functions. Throughout %P.N.Dec10
 this 
section we will discuss a generic structure function
$F(x,Q^2)$, and all formulae will be valid for both $F_2$
and $F_L$ unless otherwise stated.

It is important at this point to recall that the meaning
of ``NNLO'' in this
 context is
partly a matter of convention: one may use an absolute definition,
where LO is ${\cal O}(1)$,
NLO the  ${\cal O}(\alpha_s)$ and so on, or a relative definition, in
which
the LO is defined as the first non--vanishing order, or a combination
of the above according to the observable which is being considered.
Various options which are relevant for PDF determination up to NNLO
will be discussed in Sec.~\ref{sec:fonllsf}. Finally, we will provide explicit
numerical checks of the consistency of our procedure in the various cases.

\subsection{Scheme change}
\label{sec:scheme}

First of all let us discuss the details of the change between the
scheme with $n_l$ flavours (massive scheme) and that with $n_f=n_l+1$
flavours (massless scheme).
For future convenience, we define
\begin{eqnarray}
  a_s (Q^2) \equiv\frac{\alpha_s (Q^2)}{2 \pi};\quad
  \beta_0 \equiv2 \pi b_0,\label{twopifacs} 
\end{eqnarray}
with
\begin{equation}
  b_0 = \frac{33 - 2 n_f}{12 \pi} \label{betazero} \ .
\end{equation}
The 
running of the coupling with $n_f$ flavours to lowest nontrivial order
is given by
\begin{equation}
\alpha_s (Q^2) = \alpha_s(m^2) - b_0 \log
  \frac{Q^2}{m^2} \alpha_s^2 + \mathcal{O}
  (\alpha_s^3). \label{eq:alphaevol}
\end{equation}
% Moved below eq. 28, P.N.5-1-2010 
%Since all coefficient functions at  
%NNLO are at most of order $\mathcal{O}(\alpha_s^2)$,
%we will not need terms of order ${\cal O} (\alpha_s^3)$ in the matching condition for $\alpha_s$.

\subsubsection{Matching}
\label{sec:match}

At a given value of $Q^2$
the relation between the coupling constant in the massless and
massive schemes, 
Eq.~(\ref{eq:matchingalpha}), reads
\begin{equation}
  a_s^{(n_l+1)} (Q^2) = a_s^{(n_l)} (Q^2) + \frac{2 T_R}{3} a_s^2 L + \mathcal{O}
  (\alpha_s^3) , \label{eq:alphanl}
\end{equation}
which immediately implies that the two schemes 
coincide when the scale is equal to the heavy quark mass:
\begin{equation}
a_s^{(n_l+1)} (m)= a_s^{(n_l)} (m)+ \mathcal{O}
  (\alpha_s^3). \label{eq:alphanlmq}
\end{equation}
Since all coefficient functions at  
NNLO are at most of order $\mathcal{O}(\alpha_s^2)$,
we will not need terms of order ${\cal O} (\alpha_s^3)$
in the matching condition for $\alpha_s$.

The relation between parton distributions, Eq.~(\ref{eq:matchingpdf}), %P.N.Dec10
for light partons
in the two schemes at $Q^2=m^2$
is also trivial  up to order $\alpha_s$,
\begin{equation}
  f_i^{(n_l+1)} (x, m^2) = f_i^{(n_l)} (x, m^2) +{\mathcal O}(\alpha_s^2), \label{eq:qmatchmz}
\end{equation}
 but it becomes nontrivial
starting at  ${\cal O}(\alpha_s^2)$, because at this order 
contributions from heavy quark loops arise, which can thus contribute
to the renormalization of light flavours and gluons in the massless
scheme where  $n_f=n_l+1$. We have thus
\begin{equation}
  f_i^{(n_l+1)} (x, m^2) = f_i^{(n_l)} (x, m^2) + a_s^2 \int \frac{d z}{z} \sum_j
  K_{i j} (z) f_j^{(n_l)} \left( \frac{x}{z}, m^2 \right) +\mathcal{O}
  (\alpha_s^3), \label{eq:qmatchm}
\end{equation}
where the indices $i,\>j$ take the values $q,\,\bar q,\,g$, and the functions
$K_{qq}$, $K_{gq}$ and  $K_{gg}$ were computed in
Ref.~\cite{Buza:1996wv}, where they are respectively given in
Eqs.~(B4, B5, B7). 
On the other hand, in the absence of an intrinsic heavy quark
contribution, the matching conditions Eq.~(\ref{eq:matchingpdf}),
in the case of the heavy quark distribution are
\begin{equation}
  f_h^{(n_l+1)} (x, m^2) = f_{\bar h}^{(n_l+1)} (x, m^2)=
 {\mathcal O}(\alpha_s^2). \label{eq:qmatchmh}
\end{equation}

As already mentioned, the %P.N.Dec10
coefficient functions $C_i$ start at order
one only in the case of the light quark contribution to $F_2$ (in the
absence of intrinsic heavy quark contributions). Therefore,
up to order $\alpha_s^2$, the only nontrivial contribution to matching
Eq.~(\ref{eq:qmatchm}) is due to light quarks: gluon contributions to
structure functions start at ${\cal O}(\alpha_s)$, thus the contribution
from  $K_{gq}$ and  $K_{gg}$ to structure functions only start at 
${\cal O}(\alpha_s^3)$. Therefore, in this paper we will only use
Eq.~(\ref{eq:qmatchm}) for light quarks, in the form
\begin{equation}
  f_q^{(n_l)} (x, m^2) = f_q^{(n_l+1)} (x, m^2) - a_s^2 \int \frac{d z}{z} K_{q q}
  ( z) f_q^{(n_l+1)} \left( \frac{x}{z}, m^2 \right)+\mathcal{O}
  (\alpha_s^3),\label{eq:qmatchmq}
\end{equation}
with\footnote{Note that the expression given here differs by a factor
  4 from that obtained setting $\mu=m$ in 
Eq.~(B4) of Ref.~\cite{Buza:1996wv} because we expand in
powers of $\frac{\alpha_s}{2\pi}$ and not $\frac{\alpha_s}{4\pi}$ as
in that reference; note also that we have absorbed all distributions
into the $+$ prescription.} 
\begin{equation}
K_{q q} = C_F T_R \left[ \frac{1 + z^2}{1 - z} \left( \frac{1}{6} \log^2 z
   + \frac{5}{9} \log z + \frac{28}{27} \right) + (1 - z) \left( \frac{2}{3}
   \log z + \frac{13}{9} \right) \right]_+ .
\label{eq:kappaqq}
\end{equation}

\subsubsection{Scale dependence}
\label{sec:scdep}

While the matching condition for
parton distributions at the scale $Q^2=m^2$, Eqs.~(\ref{eq:qmatchm}), is
sufficient in order to fix the relation between PDFs in the massive
and massless schemes, in order to implement the FONLL approximation we
need to express the massive-scheme PDFs in terms of the
massless-scheme ones at a generic scale $Q^2$, as required in
Eq.~(\ref{eq:Fnlbar}). The relation  at any scale can be obtained from
the matching condition Eq.~(\ref{eq:qmatchm}) by solving the evolution
equation for both $f_i^{(n_l+1)}$ and $f_i^{(n_l)}$ in the
respective schemes. Because only the light quark coefficient function
starts at ${\mathcal O}(\alpha_s^0)$, up to order $\alpha_s^2$
matching conditions at 
${\mathcal O}(\alpha_s)$ will be sufficient for gluons, while 
${\mathcal O}(\alpha_s^2)$ will be required for quarks.

Evolving up to $Q^2$ Eq.~(\ref{eq:qmatchmz}) to order $\alpha_s$ gives 
\begin{equation}
  f_i^{(n_l+1)} (x, Q^2) = f_i^{(n_l)} (x, Q^2) +
a_s L \int_x^1 \frac{d z}{z} 
  \sum_j \left[ P^{(n_l+1),\,0}_{i j} (z) - P^{(n_l),\,0}_{i j} (z) \right] f_j^{(n_l)}
  \left( \frac{x}{z}, Q^2 \right) + \mathcal{O} (\alpha_s^2) ,
\label{eq:qmatchq}
\end{equation}
where $P^{0}_{i j} (z)$ are leading--order Altarelli-Parisi
splitting functions in the two schemes, and the sum runs over all
PDFs. Noting that
\begin{equation}
  P^{(n_l+1),\,0}_{i j} (z) - P^{(n_l),\,0}_{i j} (z) =  - \delta_{i g} \delta_{j
  g}  \frac{2 T_R}{3} \delta (1 - z) . \label{eq:p0nldiff} 
\end{equation}
it immediately follows that the relation between gluon distributions
at scale $Q^2$ is
\begin{equation}
 f_g^{(n_l+1)} (x, Q^2) = f_g^{(n_l)} (x, Q^2) - a_s L \frac{2
  T_R}{3} f_g^{(n_l+1)} (x, Q^2) + \mathcal{O} (\alpha_s^2)  \label{eq:ntolg}
\end{equation}

To order $\alpha_s^2$ Eq.~(\ref{eq:qmatchq}) is modified because the
initial condition at $Q^2=m^2$ becomes Eq.~(\ref{eq:qmatchm}), and
also because the evolution mismatch receives three further
contributions: from the change in running coupling
Eq.~(\ref{eq:alphanl}), from the second iteration of the leading--order
splitting functions $P^{(0)}_{ij}$, and from the next--to--leading
order splitting functions $P^{(1)}_{ij}$. Combining these
contributions, the matching condition Eq.~(\ref{eq:qmatchm}) becomes
\begin{eqnarray}
  f_q^{(n_l)} (x, Q^2) &=& f_q^{(n_l+1)} (x, Q^2) - a_s^2 \int \frac{d z}{z} K_{q q}
  ( z) f_q^{(n_l+1)} \left( \frac{x}{z}, m^2 \right)\nonumber\\
& - & a_s^2 L \int_x^1 \frac{d z}{z}  
 \frac{2 T_R}{3}  \frac{L}{2}\sum_{j \in q \bar{q} g}
  P^{(n_l),\,0}_{q j} (z) \times f_j^{(n_l+1)} \left( \frac{x}{z}, Q^2
  \right) \label{eq:ntolq} \\ 
& + &  \frac{(a_s L)^2}{2}  \frac{2 T_R}{3} \int_x^1 \frac{d z}{z}
  P^{(n_l),\,0}_{q g} (z) f_g^{(n_l+1)} \left( \frac{x}{z}, Q^2 \right) \nonumber\\
  & - & a_s^2 L \int_x^1 \frac{d z}{z}  \sum_{j \in q \bar{q} g} \left[
  P^{(n_l+1),\,1}_{q j} (z) - P^{(n_l),\,1}_{q j} (z)\right] \times f_j^{(n_l)} \left( \frac{x}{z}, Q^2
  \right) +\mathcal{O}
  (\alpha_s^3) , \nonumber
\end{eqnarray}
where each line in Eq.~(\ref{eq:ntolq}) corresponds respectively
to each of
the contributions listed above. Note that, because of Eq.~(\ref{eq:p0nldiff}),
the ${\mathcal O}(\alpha_s)$ correction vanishes for light quarks.
Note also that the $f_g$ contribution cancels between the second and third line in
Eq.~(\ref{eq:ntolq}).
Using the fact that~\cite{cfp}
\begin{equation}
  P^{(n_l+1),\,1}_{q j} (z) - P^{(n_l),\,1}_{q j} (z) =- \delta_{j q} \Delta_{q q}
  (z),
\label{eq:nlodiff}
\end{equation}
with
\begin{equation}
 \Delta_{q q} (z) = C_F T_R \left[ \frac{1 + z^2}{1 - z} \left(
  \frac{2}{3} \log z + \frac{10}{9} \right) + \frac{4}{3} (1 - z)
  \right]_+ 
\label{eq:deltaexp}
\end{equation}
we reproduce the result of Ref.~\cite{Buza:1996wv}:
\begin{eqnarray}
&& f_q^{(n_l)} (x, Q^2) = f_q^{(n_l+1)} (x, Q^2) 
   -  a_s^2 \int \frac{d z}{z} K_{q q} (z,L) f_j^{(n_l+1)} \left( \frac{x}{z},
  Q^2 \right) \nonumber \\ &&
\quad K_{q q} (z,L) =K_{q q} (z) 
   +   \frac{L^2}{2}  \frac{2 T_R}{3}
  P^{(n_l),\,0}_{q q} (z) 
   -  L \Delta_{q q} (z),\qquad
\label{eq:qltonfin}
\end{eqnarray}
with $K_{q q} (z)$ and $\Delta_{q q} (z)$ given respectively by
Eqs.~(\ref{eq:kappaqq}-\ref{eq:deltaexp}).

\subsection{The FONLL approximation up to ${\cal O }(\alpha_s^2)$}
\label{sec:fonllm}

It is now convenient to separate off explicitly 
the light and heavy contributions to
the generic structure function $F$ and 
its corresponding coefficient functions $C_i$:
\begin{eqnarray}
  F(x, Q^2) & =&  F_{l} (x, Q^2) + F_{h}
  (x, Q^2);\nonumber
\\ C_i (x, \alpha_s(Q^2))  &=& C_{i,\, l} (x,
  \alpha_s(Q^2))+C_{i,\, h} 
  (x, \alpha_s(Q^2)) 
\label{eq:hlsep}
\end{eqnarray}
where we define $F_{h}$ and $C_{i,\, h}$ as the contributions to
$F$ and $C_i$ respectively which are
obtained when only the electric charge of the heavy quark is nonzero. Note
that therefore, up to ${\mathcal O} (\alpha_s^2)$,  in the coefficient
functions the label $l,\,h$  denotes the
quark to which the virtual photon couples, whereas the label $i$
denotes the parton that enters the hard scattering
process.\footnote{Note that up to ${\mathcal O} (\alpha_s^2)$ the
  coefficient functions are the sum of contributions which are
  quadratic in the quark electric charge; however  
at order ${\mathcal O} (\alpha_s^3)$, interference terms proportional
to the product of different electric charges do arise 
(see Ref.~\cite{Vermaseren:2005qc}).} %P.N.Dec10
The separation of Eq.~(\ref{eq:hlsep}) will be carried through also for the
corresponding FONLL expressions (Eqs.~(\ref{eq:FONLL}, \ref{eq:fdiff})), i.e. we will write
\begin{eqnarray}
&&  F_{l}^{\tmop{FONLL}} (x, Q^2) =   F_{l}^{(n_l)} (x,Q^2) + F_{l}^{(d)} (x, Q^2) , \label{eq:FONLLl}
\\
&&  F_{h}^{\tmop{FONLL}} (x, Q^2) =   F_{h}^{(n_l)} (x,Q^2) + F_{h}^{(d)} (x, Q^2). \label{eq:FONLLh}
\end{eqnarray}
In the
following Sects.~\ref{sec:alpha}-\ref{sec:alphasq} we will provide
expressions for the various contributions to Eq.~(\ref{eq:hlsep}) up
to $\mathcal{O}
  (\alpha_s^2)$, with more explicit formulae collected in 
Appendix~\ref{sec:appendix}.

For the rest of this section we shall assume that the generic structure function
$F(x,Q^2)$ refers to the structure function $F_2(x,Q^2)$ in
electromagnetic deep-inelastic scattering. The specific differences for the
cases of both $F_L(x,Q^2)$ and weak-mediated DIS structure functions
will be discussed at the end of Sec.~\ref{sec:alphasq}.

\subsubsection{Order $\alpha_s$}
\label{sec:alpha}

At order $\alpha_s$ the
matching condition for parton distributions Eq.~(\ref{eq:qmatchm}) is the
trivial relation $f_q^{(n_l)} (x, m^2) = f_q^{(n_l+1)} (x, m^2)$.
This means that at this order 
all light quark and the gluon PDFs  should be
matched by assuming them to be continuous 
at $Q=m$, while the (non-intrinsic) heavy quark PDFs should be
simply taken to vanish at $Q=m$. 
The matching condition at a generic scale $Q^2$, used to determine the
coefficient functions $B_i$ of Eq.~(\ref{eq:Fnlbar}) in terms of the
original massive--scheme ones $C_i^{(n_l)}$ starts at
order $\alpha_s$, Eq.~(\ref{eq:qmatchq}). However,
Eqs.~(\ref{eq:ntolg}, \ref{eq:ntolq})
show that the correction is
nonzero only for the gluon PDF. 
%Because only the quark P.N.Dec10
Since the gluon 
coefficient function starts at ${\mathcal O}(\alpha_s)$,
it follows
that the coefficient functions $B_i$ start differing from the
$C_i^{(n_l)}$ only at ${\mathcal O}(\alpha_s^2)$: 
\begin{equation}
B_i \left(z, \frac{Q^2}{m^2}, \alpha_s^{(n_l+1)}(Q^2)\right) =C^{(n_l)}_i \left(z, \frac{Q^2}{m^2}, \alpha_s^{(n_l+1)}(Q^2)\right)+ {\mathcal O}(\alpha_s^2).
 \label{eq:btocrel}
\end{equation}

%%%%%%%%%%%%%%%%%%%%%
\begin{center}
\begin{figure}
\begin{center}
\epsfig{width=0.49\textwidth,figure=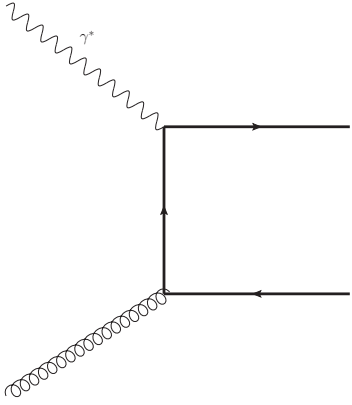,clip}
\vspace{-4cm}
\end{center}
\caption{\small 
The ${\mathcal O}(\alpha_s)$ diagram for heavy quark production.}
\label{fig:feynlo}
\end{figure}
\end{center}
%%%%%%%%%%%%%%%%%%%%%%

At this order all ``light'' coefficient functions are
the same in the massless and massive scheme:
\begin{equation}
C^{(n_l)}_{i,\, l}\lp x,\frac{Q^2}{m^2}, \alpha_s(Q^2)\rp
=C^{(n_l+1)}_{i,\, l}\lp x, \alpha_s(Q^2)\rp + \mathcal{O}
  \lp \alpha_s^2\rp \ ,
\label{eq:cqfonllnl}
\end{equation}
which immediately implies that the FONLL expression of $F_{l} (x,
Q^2)$ trivially reduces to the massless scheme one:
\begin{equation}
F_{l}^{\tmop{FONLL}} (x,Q^2)=
F_l^{(d)}(x,Q^2)+F_{l}^{(n_l)}(x,Q^2)= %P.N.Dec10
F_{l}^{(n_l+1)}(x,Q^2) +\mathcal{O}
  (\alpha_s^2).\label{eq:fonllnll}
\end{equation}
where we have
\begin{eqnarray}
F_{l}^{(d)}(x,Q^2) &=& x\sum_{i=h,\bar{h}}\int_x^1 C_{i,l}^{(n_l+1)}\left(\frac{x}{y},\alpha_s(Q^2)\right) f_i^{(n_l+1)}(y,Q^2), \\
F_{l}^{(n_l)}(x,Q^2)&=&x\sum_{i\ne h,\bar{h}}
\int_x^1 C_{i,l}^{n_l+1}\left(\frac{x}{y},\alpha_s(Q^2)\right) f_i^{(n_l+1)}(y,Q^2).
\end{eqnarray}
The ``heavy'' coefficient functions $C_{i,h}$,
for $i$ corresponding to any light quark are also trivially
the same at order $\mathcal{O}(\alpha_s)$ in the massless and massive scheme, 
since they both vanish at that order, but the gluon one is different. We have
\begin{equation}
 C^{(n_l)}_{g,\, h} \left( z, \frac{Q^2}{m^2}, \alpha_s(Q^2) \right) =
  a_s(Q^2) 2 e_h^2 C_g^{(n_l),1} \left( z, \frac{Q^2}{m^2} \right) + 
\mathcal{O}\lp \alpha_s^2\rp, \label{eq:C0nl}
\end{equation}
where
\begin{eqnarray}
  C_{g}^{(n_l),1} \left( z, \frac{Q^2}{m^2} \right) & = & \theta \left( W^2 - 4
  m^2 \right) \times T_R [ (z^2 + (1 - z)^2 + 4 \epsilon z (1 - 3 z) - 8
  \epsilon^2 z^2) \log \frac{1 + v}{1 - v} \nonumber\\
  &  & + (8 z (1 - z) - 1 - 4 \epsilon z (1 - z)) v \left. \right], \label{eq:C0nlc}
\end{eqnarray}
with
\begin{equation}
 \epsilon \equiv m^2 / Q^2,\quad v
  \equiv \sqrt{1 - 4 m^2 / W^2} .\label{eq:kindefns}
\end{equation}
The factor of $2$ associated with the squared charges in
Eq.~(\ref{eq:C0nl})
accounts for the contribution of both the quark and the
antiquark. This Feynman
diagram for this contribution is shown in Fig.~\ref{fig:feynlo}.

The massless limit Eqs.~(\ref{eq:masszlim}-\ref{eq:asympmatch}) of $C^{(n_l)}_{g,\, h}$
Eqs.~(\ref{eq:C0nl}-\ref{eq:C0nlc}) is 
\begin{equation}
  {B}^{(0),\,1}_{g,\, h} \left( z, \frac{Q^2}{m^2} \right) =  2 e_h^2 
  {C}_{g}^{(n_l,0), 1} \left( z, \frac{Q^2}{m^2} \right), \label{eq:mznlo}
\end{equation}
where, according to Eq.~(\ref{eq:masszlim}) the index $1$ denotes the
first order in $\alpha_s$, and
\begin{equation}
   {C}_{g}^{(n_l,0), 1}\left( z, \frac{Q^2}{m^2} \right) =  T_R \left[
  (z^2 + (1 - z)^2) \log \frac{Q^2 (1 - z)}{m^2 z} + (8 z (1 - z) - 1) \right]
  , \label{eq:mznloexp}
\end{equation}
which in the limit $Q^2=m^2$ reproduces the massless scheme coefficient function:
\begin{equation}
 C_{g, h}^{(n_l+1)} \left( z, \alpha_s(Q^2) \right) = a_s 2
   e_h^2  {c}_{g}^{(n_l,0), 1} \left( z, 1 \right) \ . \label{eq:masslessg}
\end{equation}

The FONLL expression for
 the ``heavy'' component is given by
\begin{equation}
  F_{h}^{\tmop{FONLL}} (x, Q^2) =F_{h}^{(n_l)} (x,
  Q^2)+F^{(d)}_{h} (x, Q^2). 
 \label{eq:FONLLnlo}
\end{equation}
The two contributions on the right-hand side of
Eq.~(\ref{eq:FONLLnlo}) are respectively the massive-scheme
heavy-quark structure function, given by 
\begin{equation}
  F_{h}^{(n_l)} (x, Q^2)  =  x \int_x^1 \frac{dy}{y} C^{(n_l)}_{g, h} \left(
  \frac{x}{y}, \frac{Q^2}{m^2},\alpha_s(Q^2) \right) f_g^{(n_l+1)} (y,Q^2) ,
\label{eq:hqnlo}
\end{equation}
with $C_{g, h}^{(n_l)} \left(
  \frac{x}{y}, \frac{Q^2}{m^2}, \alpha_s \right)$ given by
  Eqs.~(\ref{eq:C0nl},\ref{eq:C0nlc}); and the  
``difference'' contribution
\begin{eqnarray}
  F^{(d)}_{h} (x, Q^2) & = & x \int_x^1 \frac{dy}{y} \left[ C_{q,\,h}^{(n_l+1)} \left(
  \frac{x}{y}, \alpha_s(Q^2) \right) \lc f^{(n_l+1)}_h (y, Q^2) + f^{(n_l+1)}_{\bar{h}} (y, Q^2)\rc  +
  \right. \phantom{aaaaa} \nonumber\\
  &  & \left. \left( C_{g,\, h}^{(n_l+1)} \left(
  \frac{x}{y}, \alpha_s(Q^2) \right) -   {B}^{(0)}_{g,\, h} \left( \frac{x}{y},
  \frac{Q^2}{m^2},\alpha_s(Q^2)  \right) \right) f^{(n_l+1)}_g (y, Q^2) \right],  \label{eq:fdnlo}
\end{eqnarray}
where $C_{g,\, h}^{(n_l+1)}$  and ${B}^{(0)}_{g,\, h}$ are given by
Eqs.~(\ref{eq:masslessg}),(\ref{eq:mznlo}), respectively, and
$C_{q,\,h}^{(n_l+1)}$ is the 
${\mathcal O}(\alpha_s)$ quark coefficient function
for production of a (massless) quark of charge $e_h$. 

It is easy to see explicitly that, in the region where
$L$ is not large,
 the ``difference''  term is of order
$\mathcal{O} (\alpha_s^2)$, so that to $\mathcal{O} (\alpha_s)$ the
FONLL expression coincides with the massive--scheme one. Indeed,
Eq.~(\ref{eq:qmatchmh}) implies that the heavy quark distribution is
$\mathcal{O} (\alpha_s)$. Use of the leading--order QCD evolution equations
immediately leads to
\begin{equation}
  f_h (y, Q^2)=f_{\bar{h}} (y, Q^2) = \frac{\alpha_s(Q^2)}{2 \pi} L 
\int \frac{d z}{z} T_f (z^2 + (1 -
  z)^2) g \lp \frac{x}{y}, Q^2\rp  + \mathcal{O} (\alpha_s^2).
\end{equation}
Noting that $
C_{q,\,h} \left(
  x, \alpha_s(Q^2) \right)=\delta(1-x)+ {\mathcal O} (\alpha_s)$ it
  immediately follows that
the two terms in Eq.~(\ref{eq:fdnlo}) cancel up to terms of order
$\mathcal{O} (\alpha_s^2)$.

%%%%%%%%%%%%%%%%%%%%%%%%%%%%%%%%%%%%%%%%%%%%%%%%
\subsubsection{Order $\alpha^2_s$}
\label{sec:alphasq}

We will now work out the FONLL expressions for structure
functions at ${\mathcal O}(\alpha_s^2)$.
We expand structure functions and coefficient functions in powers of
$a_s^{(n_l+1)}(Q^2)$, Eq.~(\ref{twopifacs}), as follows:
\begin{equation}
C_i= \sum^P_{p = 0} \left(\frac{\alpha_s^{(n_l+1)} (Q^2)}{2\pi} \right)^p C_i^{p};\quad
F^{\tmop{FONLL}}
 = \sum^P_{p = 0} \left(\frac{\alpha_s^{(n_l+1)} (Q^2)}{2\pi} \right)^p F^{\tmop{FONLL},\,p},
\label{eq:expall}
\end{equation}
and similarly for other quantities.
In order to deal
with manageable expressions we will collect in
Appendix~\ref{sec:appendix} the expressions for
the ${\mathcal O}(\alpha_s^2)$ coefficients $C_i^2$,
  $F^{\tmop{FONLL},\,2}$, and so on, while the contributions up to
  ${\mathcal O}(\alpha_s)$ were given 
explicitly in Sec.~\ref{sec:alpha}.

%%%%%%%%%%%%%%%%%%%%%%%%%%%%%%%%%
\begin{center}
\begin{figure}
\begin{center}
\epsfig{width=0.49\textwidth,figure=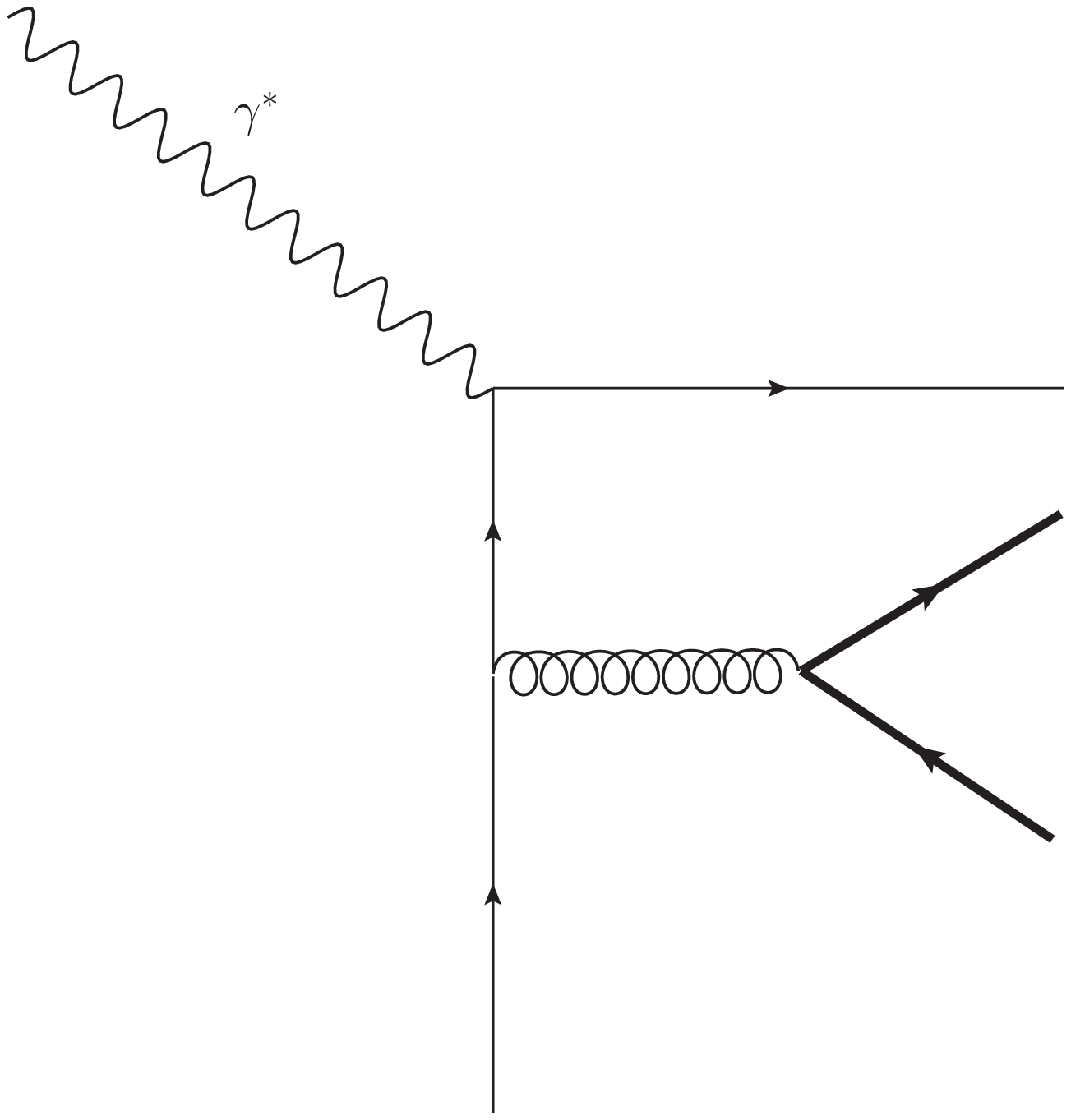,clip}
\epsfig{width=0.49\textwidth,figure=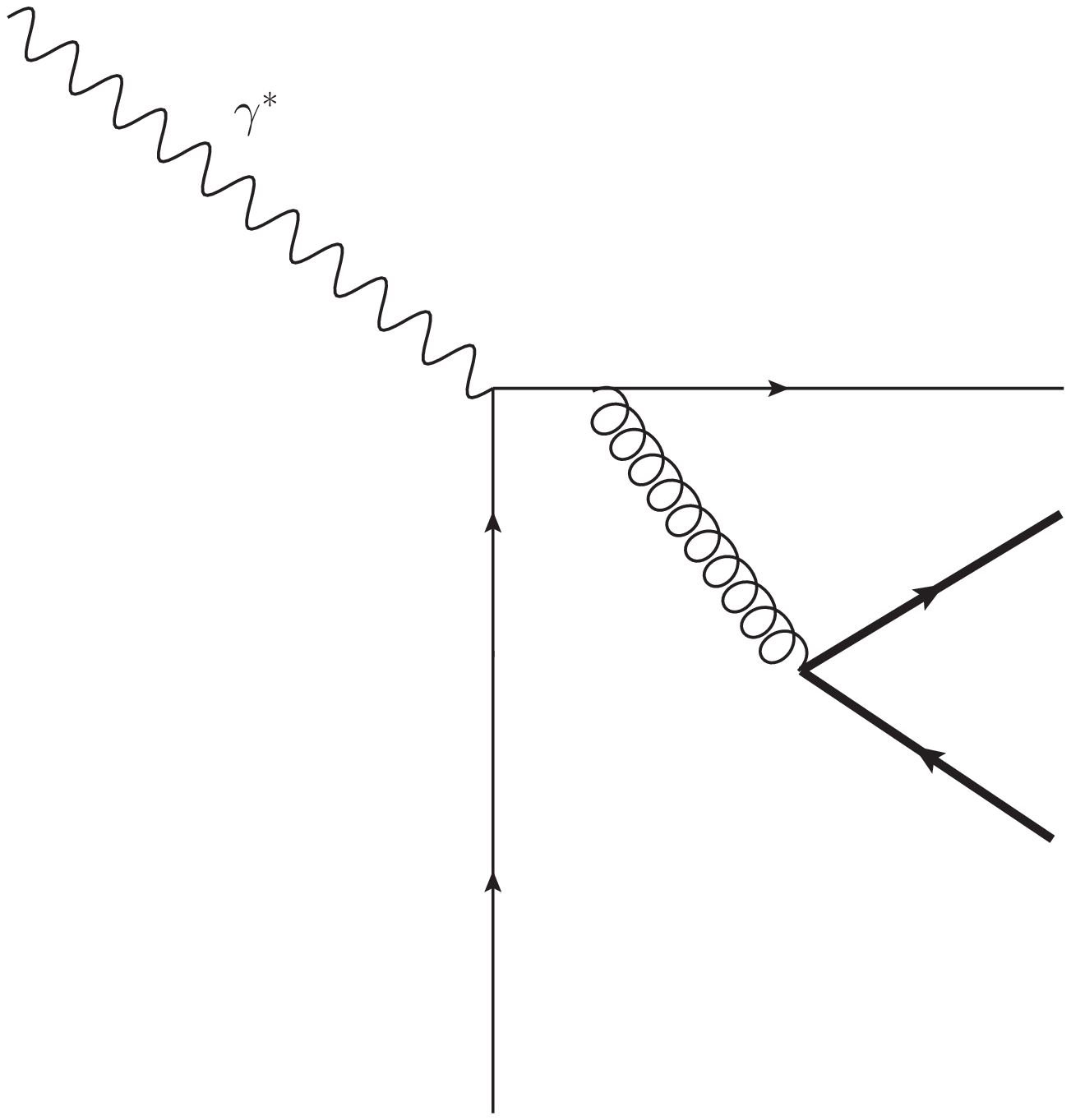,clip}
\vspace{-4cm}
\end{center}
\caption{\small 
The ${\mathcal O}(\alpha_s^2)$ diagrams for heavy quark production
which  contribute to the ``light'' coefficient function $F_{l}$
Eq.~(\ref{eq:hlsep}).}
\label{fig:feyn-nlo-ir}
\end{figure}
\end{center}
%%%%%%%%%%%%%%%%%%%%%%%%%%%%%%%%%%

At the NNLO order, both ``light'' and ``heavy'' coefficient functions
differ in the massive and massless scheme. As far as ``light'' coefficient
functions are concerned, the gluon one is the same 
\begin{equation}
C^{(n_l),\,2}_{g,\, l}\left(x,\frac{Q^2}{m^2}
\right)=C^{(n_l+1),\,2}_{g,\, l}(x),\label{eq:cqfonllnnl}
\end{equation}
but the quark one $C^{(n_l),\,2}_{i\ne g,\, l}$ is different, because of 
the contribution
coming from the emission of a gluon which in turns radiates a
$h$--$\bar h$ pair, shown in Fig.~\ref{fig:feyn-nlo-ir}.

Further contributions may arise when 
re--expressing the strong coupling and parton distribution in
terms of their massless--scheme counterparts for quarks 
we find that the ${\mathcal
  O}(\alpha_s^2)$ contribution $B^2_{i,\,l}$ for $i\ne g$
(Eq.~(\ref{eq:logexpnl}))
is given in terms of the coefficients (Eq.~(\ref{eq:expall}))
of the expansion of the 
massive--scheme coefficient function by
\begin{equation}
B^2_{i,\,l} \left(z,\frac{Q^2}{m^2}\right) =  C^{(n_l),\,2}_{i,\,l} \left(z,\frac{Q^2}{m^2}\right)
  -  \left[ e_i^2 K_{q q} (z, L) + \frac{2 T_R}{3} L C^{(n_l),\,1}_{i} (z) \right],
\label{eq:expcq}
\end{equation}
where $C^{(n_l),\,2}_{i,\,l}$ is the coefficient function for the 
$i$--th light quark or antiquark,
$K_{q q} (z, L)$ is given in Eq.~(\ref{eq:qltonfin}) and
$C^{(n_l),\,1}_{i}$ is the standard~\cite{cfp} (massless) NLO contribution to
the coefficient function for the $i$--th quark or antiquark.
The term proportional to $C^{(n_l),\,1}_{i}$ arises when expressing
$\alpha_s^{(n_l)}$ in terms of $\alpha_s^{(n_l+1)}$ (using Eq.~(\ref{eq:alphanl}))
in the  $\mathcal{O}(\alpha_s)$ correction to $C^{(n_l)}_{i} (z)$.
The
massless limit of $B^2_{i,\,l} $ Eq.~(\ref{eq:expcq}), for $i\ne g$
is given by
\begin{equation}
B^{(0),\,2}_{i,\,l} \left(z,\frac{Q^2}{m^2}\right) =  C^{(n_l,0),\,2}_{i,\,l} \left(z,\frac{Q^2}{m^2}\right)
  -  \left[ e_i^2 K_{q q} (z, L) + \frac{2 T_R}{3} L C^{(n_l),\,1}_{i} (z) \right],
\label{eq:expcqzm}
\end{equation}
where $C^{(n_l,0),\,2}_{i,\,l}$ denotes the massless limit of the
coefficient function in the massive scheme.

In the gluon sector,  even though the coefficient function
$C^{(n_l),\,2}_{g,\,l}$ coincides with its massless counterpart,
according to
Eq.~(\ref{eq:cqfonllnnl}),  
both the expression in
terms of their massless--scheme counterparts 
of the strong coupling Eq.~(\ref{eq:alphanl}) and  of the gluon
distribution Eq.~(\ref{eq:ntolg}) contribute to the determination of
$B^2_g$ (Eq.~(\ref{eq:logexpnl})), as discussed in
Sec.~\ref{sec:scdep}. 
The former contribution is of course
the same as in Eq.~(\ref{eq:expcq}), and the latter turns out to
cancel it exactly, so that in the end we get
\begin{equation}
B^2_{g,\,l} \left(z,\frac{Q^2}{m^2}\right) =  C^{(n_l),\,2}_{g,\,l} \left(z\right)
,\label{eq:expcg}
\end{equation}
namely, the coefficient function is the same as in the massless scheme
after all.
%%%%%%%%%%%%%%%%%%%%%%%%%%%%%%%%%
\begin{center}
\begin{figure}
\begin{center}
\epsfig{width=0.49\textwidth,figure=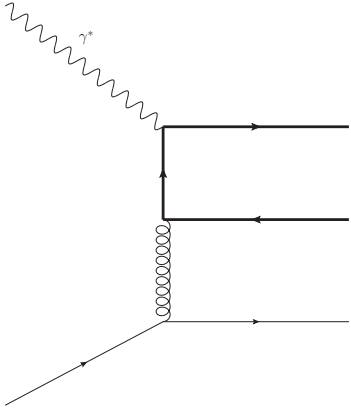,clip}
\vspace{-4cm}
\end{center}
\caption{\small 
The light-quark initiated ${\mathcal O}(\alpha_s^2)$ diagram 
for heavy quark production
which contributes to the ``heavy'' coefficient function $F_{h}$
Eq.~(\ref{eq:hlsep}). 
%Right:
%the analog Feynman diagram which belongs to the 
%``light'' coefficient functions 
}
\label{fig:feyn-nlo-qinit}
\end{figure}
\end{center}
%%%%%%%%%%%%%%%%%%%%%%%%%%%%%%%%%

\begin{center}
\begin{figure}
\begin{center}
\epsfig{width=0.49\textwidth,figure=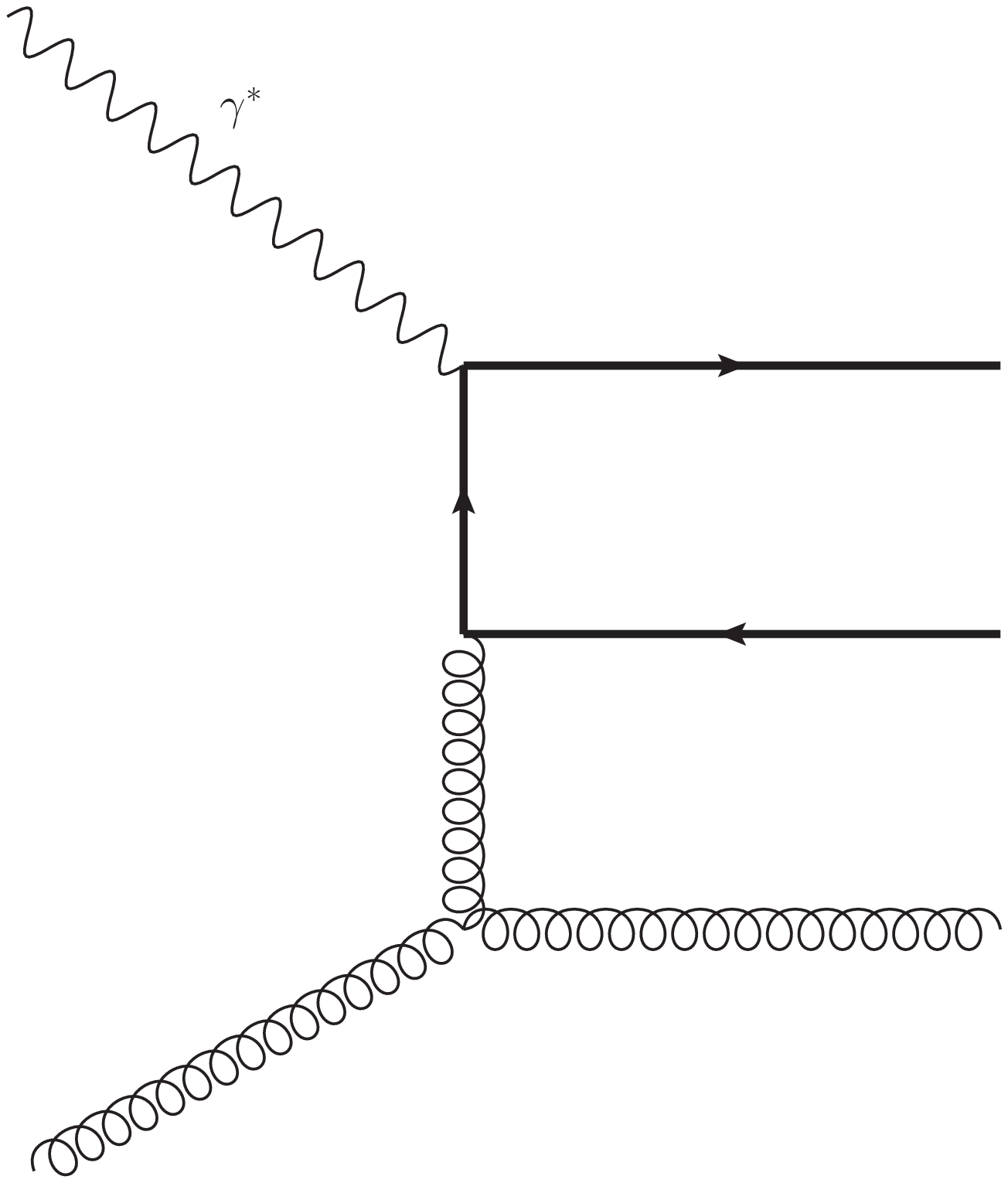,clip}
\epsfig{width=0.49\textwidth,figure=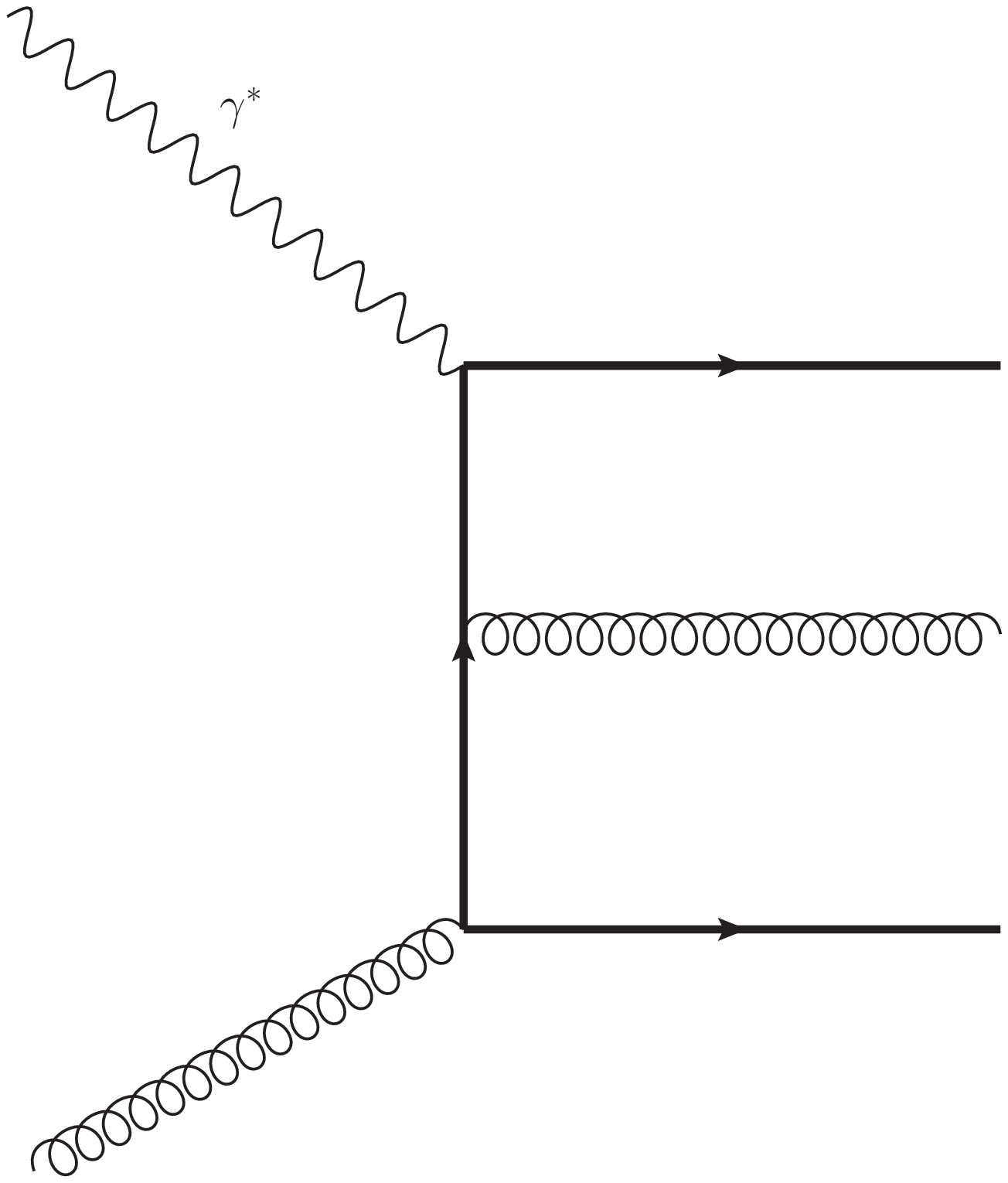,clip}
\vspace{-4cm}
\end{center}
\caption{\small 
The gluon initiated ${\mathcal O}(\alpha_s^2)$ real diagrams for heavy quark production
which contribute to the ``heavy'' coefficient function $F_{h}$
Eq.~(\ref{eq:hlsep}).}
\label{fig:feyn-nlo-ginit}
\end{figure}
\end{center}

Turning now to ``heavy'' coefficient functions, 
both the quark and gluon ones,  $C^{(n_l),\,2}_{i,\, h}$, with $i=q,\,g$
are nontrivial in the massive scheme, because heavy quarks contribute
both to quark--initiated (Fig.~\ref{fig:feyn-nlo-qinit}) and gluon--initiated processes
(Fig.~\ref{fig:feyn-nlo-ginit}).
When re-expressing in terms of the massless-scheme strong coupling and
PDFs, we now find
\begin{equation}
B^{2}_{i,\, h}\left(z, \frac{Q^2}{m^2}\right) =C^{(n_l),\,2}_{i,\, h}\left(z, \frac{Q^2}{m^2}\right).
\label{eq:btocrelnnlo}
\end{equation}
Indeed, for an incoming light quark the ``heavy'' contribution only
starts at ${\mathcal O}(\alpha_s^2)$, so up to ${\mathcal
  O}(\alpha_s^2)$ it is unaffected by changes in the coupling and PDFs,
while for an incoming gluon the argument leading to
Eq.~(\ref{eq:expcg}) also applies, and leads to the same conclusion.
%%%%%%%%%%%%%%%%%%%%%%%%%%%%%%%%%
%%%%%%%%%%%%%%%%%%%%%%%%%%%%%%%%%

Collecting this information, we can construct
the FONLL expression for the various contributions to the  structure function.
For the ``light'' component
$F_{l}(x,Q^2)$, we find that the  ${\mathcal O} (\alpha_s^2)$ 
contribution to the FONLL expression is given by
\begin{equation}
  F_{l}^{\tmop{FONLL},\,2} (x, Q^2) = 
F^{(d),\,2}_{l} (x, Q^2) + F_{l}^{(n_l),\,2} (x, Q^2),
 \label{eq:FONLLnnlo}
\end{equation}
Here the contribution to the massive-scheme structure function is
\begin{equation}
  F_{l}^{(n_l),\, 2} (x, Q^2)  =  x \sum_{i\ne h,\bar{h}}\int_x^1 \frac{dy}{y}
  B^2_{i, l} \left(
  \frac{x}{y}, \frac{Q^2}{m^2} \right) f_i^{(n_l+1)} (y,Q^2) ,
\label{eq:hqnnlod}
\end{equation}
where the sum runs over light quark flavours and antiflavours and
$B^2_{i, l}$ and  $f_i^{(n_l+1)}$ are respectively the massive
coefficient function Eq.~(\ref{eq:expcq}) and  PDF for the $i$--th
light flavour. The ${\mathcal O}(\alpha^2)$ term of the ``difference''
contribution is then given by
\begin{eqnarray}
  F^{d,\, 2}_{l} (x, Q^2) & = & x \sum_{i\ne h,\bar{h},g}\int_x^1  \frac{dy}{y}
\left[ C_{i,\, l}^{(n_l+1),\, 2} \left(
  \frac{x}{y}\right) -   {B}^{(0),\,2}_{i,\, l} \left( \frac{x}{y},
  \frac{Q^2}{m^2} \right)\right]  f^{(n_l+1)}_i (y, Q^2) 
\nonumber \\
&&+x \sum_{i=h,\bar{h}}\int \frac{dy}{y}
 C_{i,\, l}^{(n_l+1),\, 2} \left(
  \frac{x}{y}\right) f^{(n_l+1)}_i (y, Q^2) \ ,
\label{eq:fdnnlo}
\end{eqnarray}
where ${B}^{(0),\,2}_{i,\, h}$ are given by Eqs.~(\ref{eq:expcqzm}, \ref{eq:expcg}), and
the NNLO massless light quark and gluon coefficient
functions $C_{i,\,
  h}^{(n_l+1),\, 2}$, 
first computed in Refs.~\cite{vanNeerven:1991nn,Zijlstra:1991qc}
 are respectively given  in Eqs.~(B.2-B.4) and
Eq.~(B.6) 
of Ref.~\cite{Zijlstra:1992qd}.\footnote{Note that in Ref.~\cite{Zijlstra:1992qd} coefficient
  functions are given as a series in powers of
  $\frac{\alpha_s}{4\pi}$.}

\begin{center}
\begin{figure}
\begin{center}
\epsfig{width=0.49\textwidth,figure=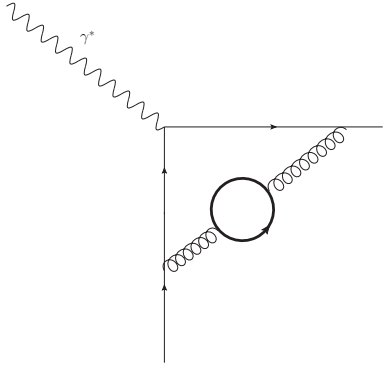,clip}
\vspace{-4cm}
\end{center}
\caption{\small 
The light-quark initiated ${\mathcal O}(\alpha_s^2)$ diagram
in which heavy quarks contribute to the gluon
self-energy.}
\label{fig:feyn-nlo-self}
\end{figure}
\end{center}

Now that the structure function is computed to ${\mathcal
  O}(\alpha_s^2)$,
when $L$ is not  large, the difference contribution $F^d$ must start at the
order $\alpha_s^3$. The
second term in Eq.~(\ref{eq:fdnnlo}) manifestly starts at this order, 
since it is proportional to the product of the heavy flavour 
parton density, which
starts at order $\alpha_s$, and the coefficient function for
 an incoming heavy quark
to produce a light quark that couples to the photon, which is of
 order $\alpha_s^2$.
In order for the first term in Eq.~(\ref{eq:fdnnlo}) not to be of   ${\mathcal
  O}(\alpha_s^2)$, the difference of coefficients in the square bracket
must vanish identically. Notice therefore that in actual fact the
coefficient ${B}^{(0),\,2}_{i,\, l} \left( \frac{x}{y},
  \frac{Q^2}{m^2} \right)$ does not depend on $Q^2/m^2$.
In order to verify this cancellation,
 we rewrite the first term on the
r.h.s of Eq.~(\ref{eq:expcqzm}) as
\begin{equation}
C^{(n_l,0),2}_{i,l}\lp z,\frac{Q^2}{m^2}\rp 
=\bar{C}^{(n_l),2}_{i,l}(z)+e_i^2 C^{\rm NS,2} \lp z,\frac{Q^2}{m^2}\rp  \ ,
\label{eq:splitCmassive}
\end{equation}
where $\bar{C}^{(n_l),2}_{i,l}(z)$ stands for the \MS{} coefficient function of flavour $i$
with $n_l$ light flavours, and 
\begin{equation}
C^{\rm NS,2}\lp z,\frac{Q^2}{m^2}\rp\equiv\frac{1}{4}\left[L^{\rm NS,(2)}_{q}
\lp z,\frac{m^2}{Q^2}\rp+
L^{\rm NS,S+V,(2)}_{q}\lp z,\frac{m^2}{Q^2}\rp\right],
\label{eq:mustvanishb}
\end{equation}
where $L^{\rm NS}_{q}$ is
given in equations (D.8) and (D.10) of Ref.~\cite{Buza:1995ie} and the factor
of $1/4$ accounts 
for the fact that $\alpha_s/(4\pi)$ rather than $\alpha_s/(2\pi)$ is used
in that reference. In other words, Eq.~(\ref{eq:splitCmassive}) splits the coefficient functions
into terms not involving heavy flavours at all, and terms involving a
heavy flavour,\footnote{By ``terms not involving heavy flavours'' we also mean here terms where
the heavy flavour contribution totally decouples in the decoupling scheme, which is typically the
case of heavy flavour loop corrections to on--shell gluon propagators.}
 like the diagrams shown in
 Figs.~\ref{fig:feyn-nlo-ir},\ref{fig:feyn-nlo-self}. 
%%%%%%%%%%%%%%%%%%%%%%%%%%%%%%%%%
%%%%%%%%%%%%%%%%%%%%%%%%%%%%%%%%%

Since the ${C}^{(n_l+1),2}_{i,l}$ coefficients are also \MS{} coefficients (and since they
depend linearly upon $n_l$), we have
\begin{equation}\label{eq:mustvanisha}
C^{(n_l+1),2}_{i,l}(z)-\bar{C}^{(n_l),2}_{i,l}(z)=\frac{\partial}{\partial n_l}C^{(n_l+1),2}_{i,l}(z),
\end{equation}
so that, in order for the first term of Eq.~(\ref{eq:fdnnlo}) to be
of order $\alpha_s^3$ we must have
\begin{equation}\label{eq:mustvanish}
\frac{\partial}{\partial n_l}C^{(n_l+1),2}_{i,l}(z)-e_i^2 C^{\rm NS,2}(z,m^2/Q^2)
+\left[ e_i^2 K_{q q} (z, L) + \frac{2 T_R}{3} L C^{(n_l),\,1}_{i} (z) \right]=0.
\end{equation}
Using Eqs.~(\ref{eq:qltonfin}), (D.8) and (D.10) of
Ref.~\cite{Buza:1995ie}, and 
Eq.~(4) of Ref.~\cite{vanNeerven:1991nn} for the coefficient of $n_l$ in $C^{(n_l+1),2}_{i,l}$,
we can easily verify Eq.~(\ref{eq:mustvanish}).\footnote{We remark here that the first moments
of $C^{(n_l+1),2}_{i,l}$, $\bar{C}^{(n_l),2}_{i,l}$, $C^{\rm NS,2}(z,m^2/Q^2)$, $K_{q q} (z, L)$
and $C^{(n_l),\,1}_{i}$ all vanish individually. This fact, which
can be explicitly verified, is a consequence of the Adler sum rule.}

The  structure of the ${\mathcal O} (\alpha_s^2)$ 
contribution to the FONLL expression for the ``heavy''
component of $F$ is similar to that of its  
${\mathcal O} (\alpha_s)$ counterpart Eq.~(\ref{eq:FONLLnlo}), namely
\begin{equation}
  F_{h}^{\tmop{FONLL},\,2} (x, Q^2) = F^{(d),\,2}_{h} (x, Q^2) + F_{h}^{(n_l),\,2} (x, Q^2),
 \label{eq:FONLLnnloh}
\end{equation}
except
that now both light quarks and gluons contribute to the
massive--scheme term:
\begin{equation}
  F_{h}^{(n_l),\,2} (x, Q^2)  =  x \int_x^1 \frac{dy}{y} 
\sum_{i \ne h,\bar{h}}
  B^{(n_l),\,2}_{i, h} \left(
  \frac{x}{y}, \frac{Q^2}{m^2} \right) f_i^{(n_l+1)} (y,Q^2) ,
\label{eq:hqnnlo}
\end{equation}
with the sum running over all light quark and antiquarks and gluons,
and $B^{(n_l),\,2}_{g, h}$ given by Eq.~(\ref{eq:btocrelnnlo}). The real diagrams
corresponding to the gluon initiated contributions to
Eq.~(\ref{eq:hqnnlo}) are 
shown
in Fig.~\ref{fig:feyn-nlo-ginit}, while the corresponding light-quark
initiated contribution is shown in Fig.~\ref{fig:feyn-nlo-qinit}.
 
The difference contribution is now given by
\begin{eqnarray}
&&  F^{(d),\,2}_{h} (x, Q^2)  =  x \sum_{i = h, \bar{h}} \int_x^1 \frac{dy}{y} 
  C^{(n_l+1),2}_{i, h} \left(
  \frac{x}{y}\right) f_i^{(n_l+1)} (y,Q^2) 
\nonumber\\&&\qquad
   +    \sum_{i \ne h, \bar{h}, g} \int_x^1 \frac{dy}{y} 
\left[ C^{(n_l+1),2}_{i, h}\left(
  \frac{x}{y}\right) - {B}^{(0),\,2}_{i,\, h} \left(
  \frac{x}{y}, \frac{Q^2}{m^2} \right)\right]
 f_i^{(n_l+1)} (y,Q^2).
  \label{eq:fdnnloh}
\end{eqnarray}
Here $C^{(n_l+1),2}_{i, h}$ are the massless-scheme  ${\mathcal O}(\alpha_s^2)$
contributions
to the standard coefficient functions for production of
a quark of electric charge $e_h$, from
Refs.~\cite{vanNeerven:1991nn,Zijlstra:1991qc,Zijlstra:1992qd} and
already used in Eq.~(\ref{eq:fdnnlo}), with the heavy quark treated as
another massless flavour, while ${B}^{(0),\,2}_{i,\, h}$ are the
massless limits of the massive coefficient functions
Eq.~(\ref{eq:btocrelnnlo}), given in Eqs.~(D4,D6) of
Ref.~\cite{Buza:1995ie} for gluons and quarks respectively.

As mentioned in the beginning of Sec.~\ref{sec:fonllm}, all results
so far apply to  the structure function $F_2(x,Q^2)$. However,
they also hold for  the longitudinal 
structure function $F_L (x, Q^2)$, whose
coefficients are given in the references cited above along with those
for $F_2$. The only difference is that, because the structure
function $F_L (x, Q^2)$ starts at ${\mathcal O}(\alpha_s)$ also in the
light quark sector, for this structure function one finds
the simplified expression
\begin{equation}
B^2_{i,\,l;\,L} \left(z,\frac{Q^2}{m^2}\right) =  C^{(n_l),\,2}_{i,\,l;\,L}
\left(z,\frac{Q^2}{m^2}\right)-  \frac{2 T_R}{3} L C^{(n_l),\,1}_{i} (z) 
\label{eq:expcqfl}
\end{equation}
instead of the more complicated relation Eq.~(\ref{eq:expcq}).

While we have provided up to now explicit results for electromagnetic
DIS structure functions, it is straightforward to generalize
the above discussion to the case of weak-mediated neutral current DIS:
indeed, the only modifications consists of the replacement of
the electromagnetic charges by the corresponding weak charges.

The generalization to charged--current (CC) DIS is slightly more complicated.
Firstly, full  ${\mathcal O}(\alpha_s^2)$  CC coefficient
coefficient functions are not available: only their asymptotic
$Q^2\gg m^2$ limits~\cite{Buza:1997mg} and their 
threshold behaviour~\cite{Corcella:2003ib} are currently known, so
that the FONLL can only be implemented to this order.
For CC scattering initiated by a heavy
quark, like strangeness production though $W-$boson exchange, the FONLL
formulae of Sec.~\ref{sec:alpha} straightforwardly 
generalize, taking into account
only the different charges and kinematics. On the other hand, for
CC scattering with a light quark in the initial state, such as
charm production in neutrino scattering (commonly known as  
``dimuon production'')
the heavy quark PDF does not enter, so the FONLL results reduces
to the massive calculation (with scheme ambiguities only entering at
${\mathcal O}(\alpha_s^2)$).

\subsection{Structure functions: schemes and perturbative ordering}
\label{sec:fonllsf}

Various possibilities for the definition of perturbative ordering are
possible and have been advocated, especially in the context of parton
fits including DIS data. This is due to the fact that LO parton
distributions can be used for the computation of any hard process
at the first nonvanishing oder: it is then
natural to
adopt a ``relative'' definition of perturbative ordering, where LO is
the lowest nontrivial order at which the process starts
occurring. This is to be contrasted with an ``absolute'' definition of
perturbative ordering, such that LO refers to ${\mathcal
  O}(\alpha^0_s)$ (parton model), NLO to ${\mathcal
  O}(\alpha_s)$,
and so forth. The issue is nontrivial because in a global parton fit
one may want to combine data for the total structure function $F$,
which starts at ${\mathcal
  O}(\alpha^0_s)$, and those for the heavy structure function $F_h$, which
starts at ${\mathcal
  O}(\alpha_s)$.
 %%%%%%%%%%%%%%%%%%%%%%%%%
\begin{center}
\begin{figure}
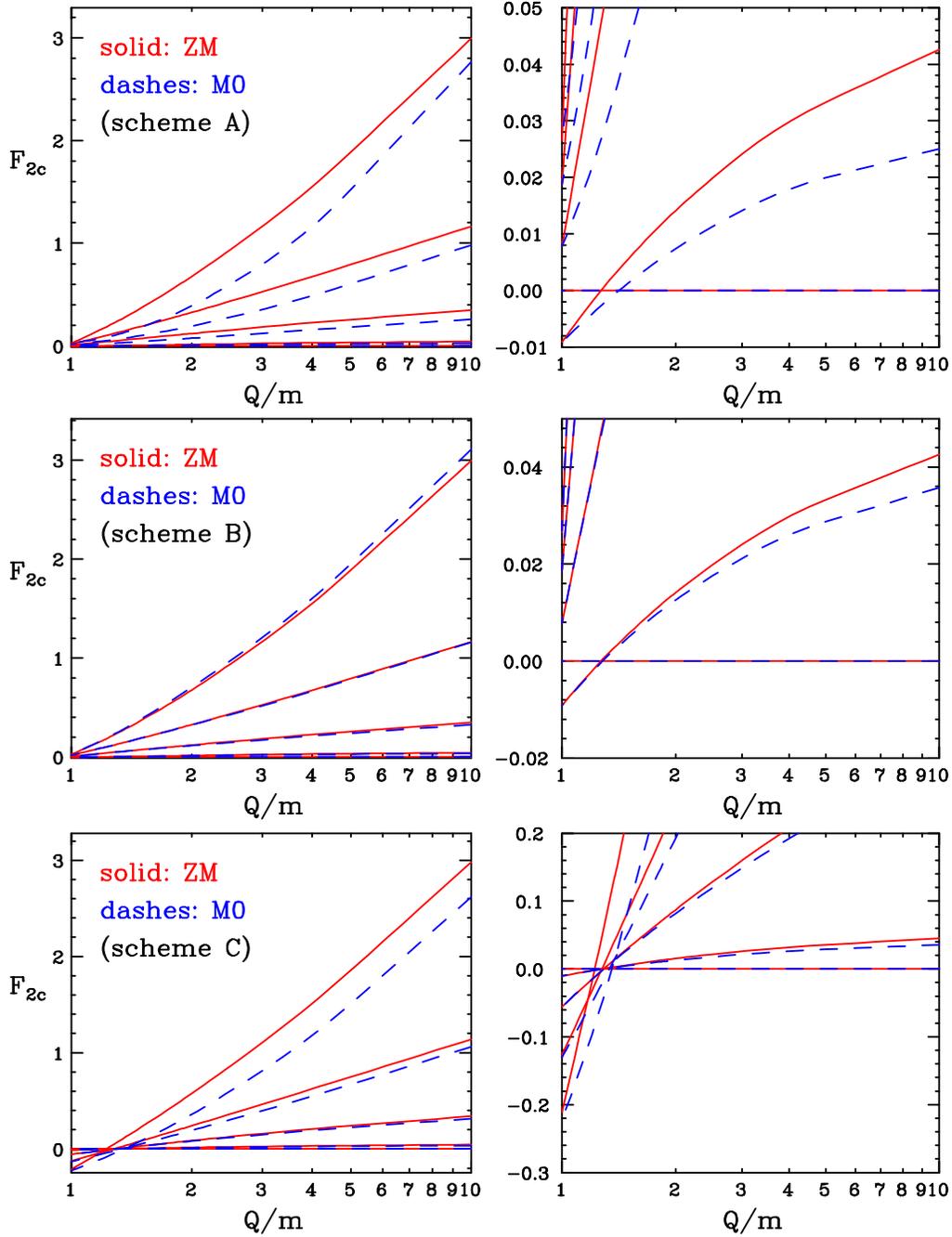

\begin{center}
\epsfig{width=0.9\textwidth,figure=F2-21-035.ps,clip}
\epsfig{file=F2-22-035.ps,width=0.9\textwidth}\\
\epsfig{file=F2-32-035.ps,width=0.9\textwidth}
\end{center}
\caption{\small\label{fig:F2-035}
 The $F_{2\,c}$ structure function computed in
the 
massless scheme Eq.~\ref{eq:Fnf} (ZM) and in the massless limit of
the massive scheme Eq.~(\ref{eq:Fnlbarzero}) (M$0$),
in the three perturbative ordering schemes A-C of Sec.~\ref{sec:fonllsf}.
The structure function is plotted as
a function of $Q/m$, for $x=10^{-\frac{5}{4}(j-1)}$,
with integer $1\le j\le 5$ (from top to bottom at large $Q/m$). The
plots on the right show a magnification of the threshold region.}
\end{figure}
\end{center}

One option is then to simply adopt an absolute perturbative ordering, 
whereby at NLO all quantities are computed at
${\mathcal O}(\alpha_s^2)$. This choice has been adopted for instance
in the CTEQ parton fits\cite{Tung:2006tb}. We will refer to it as
scheme A henceforth.
However, as discussed in Sec.~\ref{sec:mismatch} we can also
consistently
use different accuracies in the computation of different
contributions. At NLO we can then compute the total structure function at
$O(\alpha)$, but 
use the ${\mathcal
  O}(\alpha_s^2)$ results of Sec.~\ref{sec:alphasq} for the
determination of $F_h$ (both with NLO PDFs). In such case, it
is not necessary to compute any of the massless-scheme coefficient
functions $C^{(n_l+1)}_i$ beyond NLO (i.e. beyond ${\mathcal
  O}(\alpha_s)$) because this does not improve the accuracy of the NLO
calculation in the massless sector, but all massive coefficient
functions $B^{2}_i$ and their massless limits $B^{(0),\,2}_i$ should
be computed to ${\mathcal
  O}(\alpha_s^2)$ in order to have this accuracy in the massive
sector. In order to ensure consistency of the subtraction it is
sufficient to include only logarithmic terms in
$B^{(0),\,2}_i$, because the corresponding
non-logarithmic terms are not included in the massless coefficient
functions $C^{(n_l+1)}_i$. We will refer to this as scheme B.

This scheme B is reminiscent of the method proposed in
Refs~.\cite{Thorne:1997ga,Thorne:1997uu,Thorne:2006qt}, and
used e.g. in MSTW PDF fits~\cite{Martin:2009iq}: there too,
the ${\cal O}(\alpha_s^2)$
expression of the massive cross section is used in conjunction with
NLO parton densities and coefficient functions, although within a totally
different approach.

We can then also pursue a full ${\mathcal
  O}(\alpha_s^2)$  computation, by simply performing the absolute-orderd
computation of scheme A to one extra order, with NNLO PDFs. 
We will refer to this as
scheme C. However, we cannot pursue scheme B to one extra order
because ${\mathcal O}(\alpha_s^3)$ massive coefficient functions are
not known yet, though  their large $Q^2$ limit was recently
determined in Ref.~\cite{Bierenbaum:2009mv}. In NNLO MSTW
fits~\cite{Martin:2009iq} they have been modelled based on their known
large-- and small $x$ behaviour.

%%%%%%%%%%%%%%%%%%%%%%%%

%%%%%%%%%%%%%%%%%%%%
\begin{center}
\begin{figure}
\begin{center}
\epsfig{file=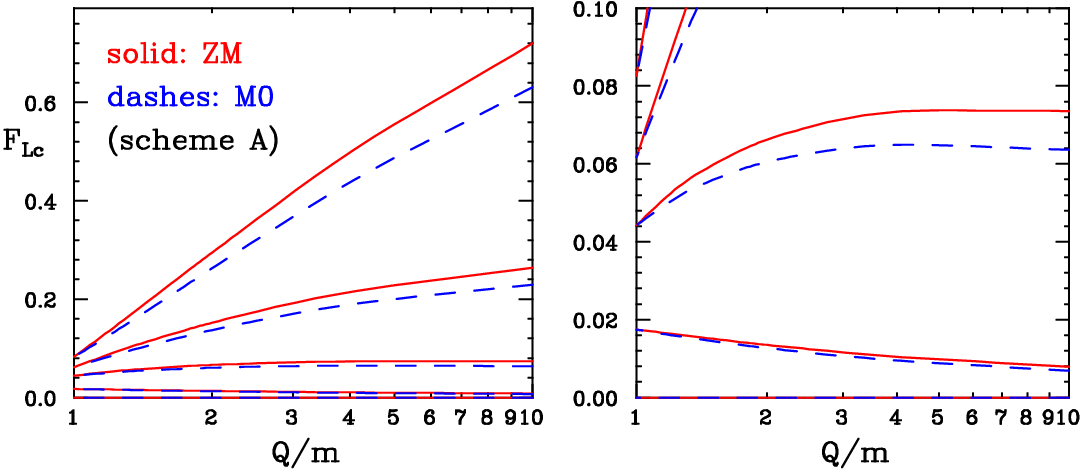,width=0.9\textwidth}\\
\epsfig{file=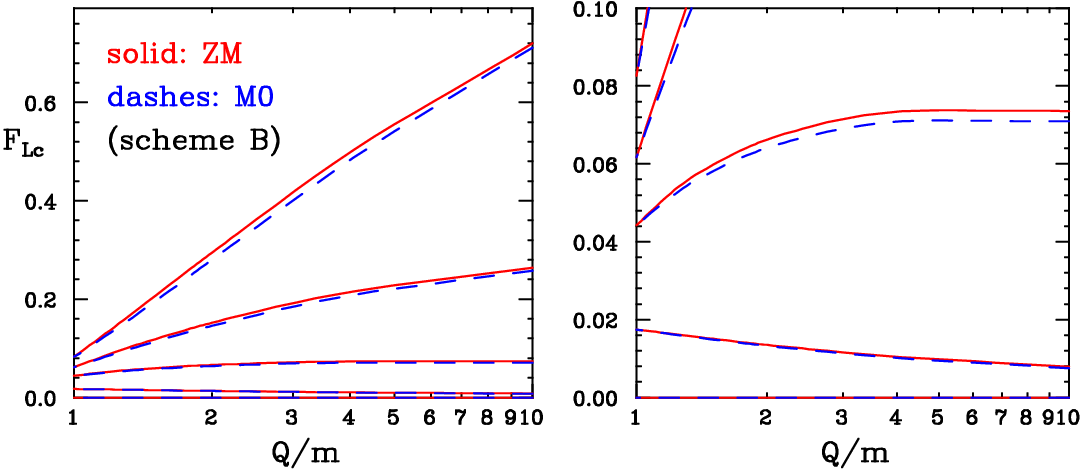,width=0.9\textwidth}\\
\epsfig{file=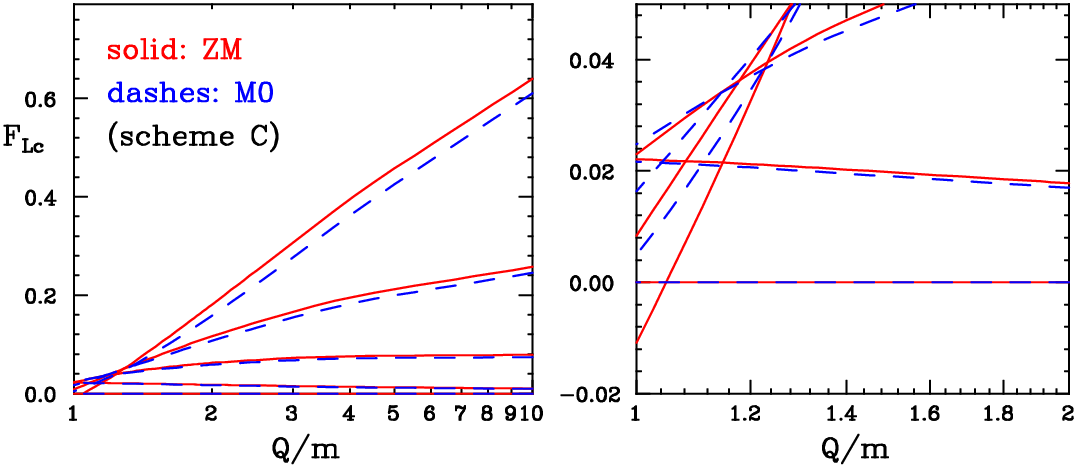,width=0.9\textwidth}
\end{center}
\caption{\small\label{fig:FL-035} Same as Fig.~\ref{fig:F2-035},
but for the structure function $F_{L\,c}(x,Q^2)$.}
\end{figure}
\end{center}
In summary, we will consider three options for perturbative ordering, 
whose consistency will be
checked explicitly in the following Sec.~\ref{sec:fonlloverlap},
and whose phenomenological implications will be discussed in 
Sec.~\ref{sec:pheno}, namely:
\begin{itemize}
\item Scheme A: the ${\mathcal
  O}(\alpha_s)$  FONLL expressions of Sec.~\ref{sec:alpha} with 
NLO parton densities.
\item Scheme B: the ${\mathcal
  O}(\alpha_s^2)$  FONLL expressions of Sec.~\ref{sec:alphasq}, but
using NLO parton densities and ${\mathcal
  O}(\alpha_s)$ massless coefficient functions, and retaining the ${\mathcal
  O}(\alpha_s^2)$ contributions to the
massive coefficient functions. In this case, the massive expression exceeds
in accuracy the massless one when $L$ is not large. Thus, according
to section \ref{sec:mismatch}, in the massless limit expression of
Eq. (\ref{eq:fdiff}), the non-logaritmic  ${\cal O}(\alpha_s^2)$ term in $F^{(n_l,0)}$
should not be included.
\item Scheme C: the ${\mathcal
  O}(\alpha_s^2)$ FONLL expressions of Sec.~\ref{sec:alphasq} with 
NNLO parton densities.
\end{itemize}

%\subsection{Comparison of FONLL with other approaches}
We address now the question of the relation of the FONLL scheme
with other approaches. As already said, our B scheme is reminiscent
of the TR scheme, although it is certainly not identical to it.
The complexity of the TR implementation has prevented us from
obtaining a clearer view of the differences.

\begin{center}
\begin{figure}
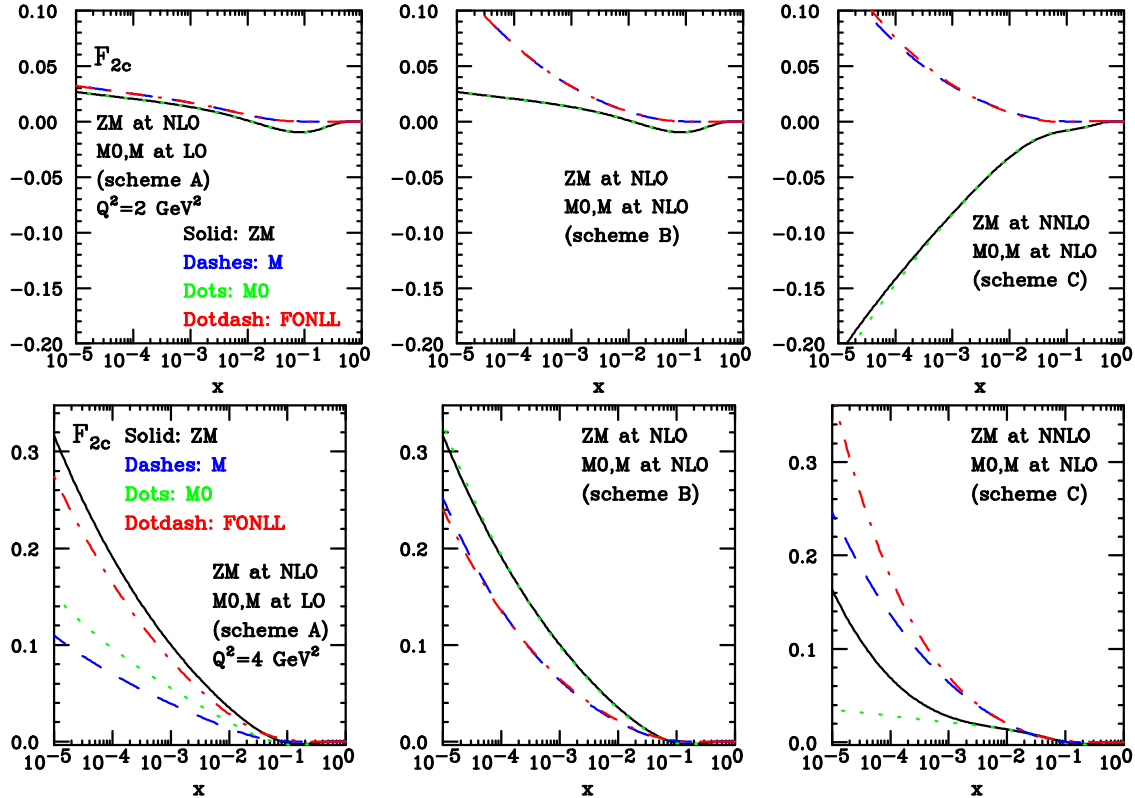

\epsfig{file=F2-0002.ps,width=\textwidth}
\epsfig{file=F2-0004.ps,width=\textwidth}
\caption{\small
\label{fig:F2-0020} The $F_{2\,c}$ structure function computed in
the massive scheme Eq.~\ref{eq:Fnl} (M),
massless scheme Eq.~\ref{eq:Fnf} (ZM), massless limit of
the massive scheme Eq.~(\ref{eq:Fnlbarzero}) (M$0$)
and FONLL scheme Eq.~(\ref{eq:FONLL}). The schemes adopted in the
three columns   correspond to
the different combinations of perturbative orders discussed in
Sec.~\ref{sec:fonllsf}. Results at the scales
$Q^2=m_c^2=2$~GeV$^2$ and $Q^2=4$~GeV$^2$ are given in the two rows.}
\end{figure}
\end{center}
Our A scheme is easily shown to be equivalent to the SACOT variant
of the ACOT scheme. If a $\chi$-scaling prescription is also applied
(with the same definition used by the ACOT group), the A scheme
becomes equivalent to the SACOT-$\chi$ one. There is no analog
of our B scheme within ACOT.
A full discussion of ACOT at NNLO as not yet appeared in the literature.
We believe, however, that
our C scheme should be equivalent to SACOT at the NNLO level,
or at least we do not see any reason why it should differ from it.
A benchmarking of the FONLL, ACOT and TR prescription, which in
particular confirms that FONLL and SACOT-$\chi$ agree to NLO, has been
recently performed in Ref.~\cite{leshouches10}

Recently, the BMSN prescription which we briefly discussed in
Sect.~\ref{sec:fonlldissf} after Eq.~(\ref{eq:FONLLp}), has also
been used for PDF determination in
Ref.~\cite{Alekhin:2009ni}. On top of the differences already
mentioned there between the FONLL and BMSN approaches, in
Ref.~\cite{Alekhin:2009ni} the resummation of powers of $L$
Eq.~(\ref{ldef}) which we discussed at length in
Sect.~\ref{sec:fonlldissf} is not performed~\cite{blumpriv}: this may be adequate in
PDF determinations which do not use data at large $Q^2$, but it
requires a subsequent transformation to a scheme where such a
resummation is included for the sake of phenomenology at large $Q^2$,
as indeed done in Ref.~\cite{Alekhin:2009ni}.
\begin{center}
\begin{figure}
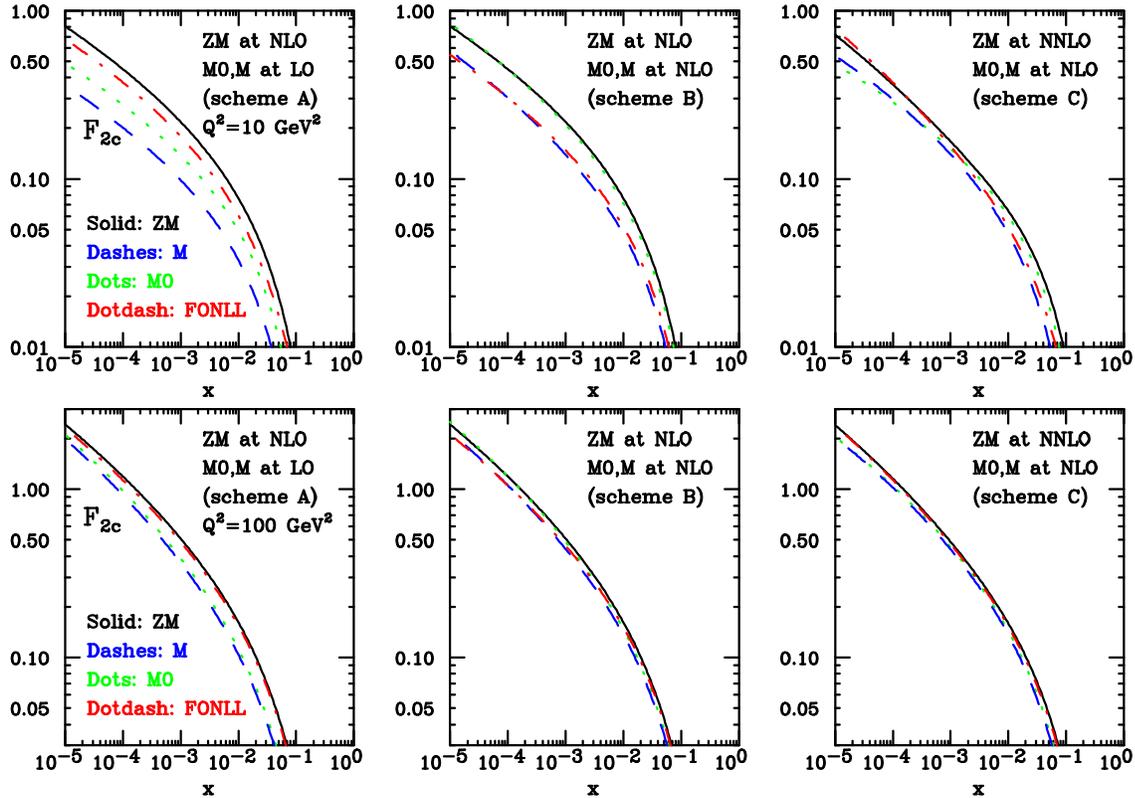

\epsfig{file=F2-0010.ps,width=\textwidth}
\epsfig{file=F2-0100.ps,width=\textwidth}
\caption{\small
\label{fig:F2-0100} Same as Fig.~\ref{fig:F2-0020}, but at the  higher
scales $Q^2=10$ and $Q^2=100$~GeV$^2$. Note that a log scale is now
adopted in the $y$--axis.}
\end{figure}
\end{center}

\subsection{Consistency of the FONLL procedure}
\label{sec:fonlloverlap}

The consistency of the FONLL method is guaranteed by the subtraction
of the overlapping terms from the massless result, as
implemented in Eq.~(\ref{eq:FONLL}). Ideally, one should be able to
verify that, in the  $m \to 0$ limit and with 
light flavour PDFs fixed at the scale $Q^2$, 
the massless scheme (ZM, henceforth) result Eq.~(\ref{eq:Fnf})   
and the zero--mass limit
of the massive (M0, henceforth) result Eq.~(\ref{eq:Fnlbarzero}) 
should coincide up to higher orders in
$\alpha_s$. This is in practice quite difficult in general~\cite{Chuvakin:1999nx}; 
however, several
features that must hold in
the relation between the ZM and M0 result near threshold are rather
easier to verify: in this section we will perform some of these checks,
thereby verifying the consistency of the FONLL procedure.
The discussion which follows applies  to a generic deep-inelastic
``heavy''structure function $F_h(x,Q^2)$ Eq.~(\ref{eq:hlsep}):
we will show explicit
results  for both the structure functions $F_{2\,h}(x,Q^2)$ and
$F_{L\,h}(x,Q^2)$.

Recall how the perturbative expansion and the logarithms of $Q^2/m^2$
are organized in the ZM and M0 result. 
The ``heavy'' 
structure function $F_{h}(x,Q^2)$ in the massless
scheme Eqs.~(\ref{eq:Fnf},\ref{eq:hlsep}) is given by
\bea
 &&F^{(n_l+1)}_{h} (x, Q^2) = x \int_x^1 \frac{dy}{y}\left[C_{h,h}^{(n_l+1)} \left( \frac{x}{y}, \alpha_s(Q^2) \right) 
  \lp h (y, Q^2) + \bar{h} (y, Q^2)\rp \right. \\&&\quad\left.+  C_{q,h}^{(n_l+1)} \left( \frac{x}{y}, \alpha_s (Q^2) \right)
  \sum_{q} \lp q (y, Q^2) + \bar{q} (y, Q^2)\rp  + C_{g,h}^{(n_l+1)} \left( \frac{x}{y}, \alpha_s(Q^2) \right) 
  g(y, Q^2) \right], \nonumber
\eea
where the choice of scheme for $\alpha_s$ and PDFs is immaterial.
\begin{center}
\begin{figure}
\epsfig{file=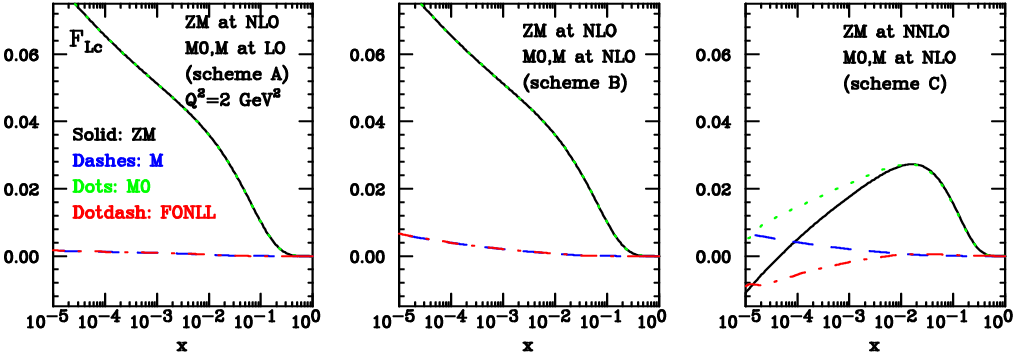,width=\textwidth}
\epsfig{file=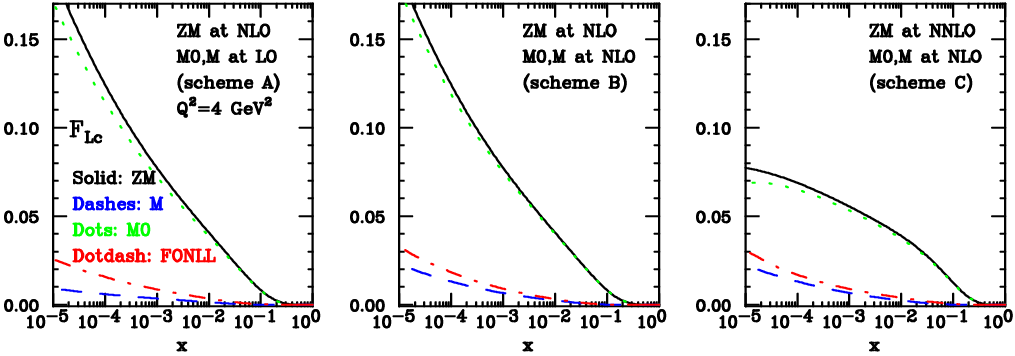,width=\textwidth}
\caption{\small 
\label{fig:FL-0020} Same as Fig.~\ref{fig:F2-0020}, but for the
structure function $F_{L\,c}$.}
\end{figure}
\end{center}

%We write schematically
%\begin{equation}
%{\rm ZM}=h\otimes C_h + g\times C_g + q\otimes C_q,
%\end{equation}
%with $h$, $g$ and $q$ representing the heavy, gluon and light parton densities,%and $C_{h/g/q}$ the relative coefficient functions for $F_{h}$.

For the sake  of a  comparison between the ZM  and M0
results,
the heavy quark PDFs $h,\bar{h}$ can be viewed as a function 
of $L=\log\lp Q^2/m^2\rp$, with the light parton densities $g(y,Q^2)$ and $q(y,Q^2)$ treated
as constants. At NLO we get
\begin{equation}
h(y,Q^2)\equiv \sum_{i=1,\infty}h^{(0)}_i(y)\alpha_s^i L^i+\sum_{i=2,\infty} h^{(1)}_i(y) 
\alpha_s^i L^{i-1}.
\end{equation}
Decomposing the coefficient functions as
\begin{equation}
\quad\quad C_{h,h}^{(n_l+1)}=C_{h,h}^{(n_l+1),0}+\alpha_s C_{h,h}^{(n_l+1),1},\quad\quad
C_{g,h}^{(n_l+1)}=\alpha_s C_{g,h}^{(n_l+1),1},\quad\quad C_{q,h}^{(n_l+1)}=0\;.
\end{equation}
the ZM structure function is, up to NLO
%\begin{equation}
%ZM_{\rm NLO}=h^{(0)}_1 C_h^{(0)} \alpha_s L
%+h^{(0)}_1   C_h^{(1)}\alpha_s^2 L +h^{(1)}_2 
% C_h^{(0)}\alpha_s^2 L  +h^{(1)}_2   C_h^{(1)}\alpha_s^3 L + 
%g C_g^{(1)}\alpha_s +{\cal O}(L^2).
%\end{equation}
\bea
\label{eq:fmassless-expL}
&& F^{(n_l+1)}_{h} (x, Q^2) = x \int_x^1 \frac{dy}{y}\left[  
C_{h,h}^{(n_l+1),0} \left( \frac{x}{y} \right)\lp \alpha_s(Q^2) h_1^{(0)}(y)
+ \alpha_s^2(Q^2) h_2^{(1)}(y)
\rp L \right.\\&&\;\left.
+ C_{h,h}^{(n_l+1),1} \left( \frac{x}{y} \right) \lp 
h_1^{(0)}(y)\alpha_s^2(Q^2) +
h_2^{(1)}(y)\alpha_s^3(Q^2) \rp L 
+ C_{g,h}^{(n_l+1),1} \left( \frac{x}{y} \right) \alpha_s(Q^2) g(y,Q^2)
+\mathcal{O}\lp L^2\rp \right] .\nonumber
\eea

On the other hand, the ZM result  can be written as a power series in
$L$ by expanding the  coefficient function as
\be
C^{(n_l,0)}\lp x,L,\alpha_s(Q^2)\rp \equiv
\sum_{i=1}^{N}\sum_{j=0}^{i} C^{(n_l,0),i,j}(x)\alpha_s(Q^2)^iL^j .
\ee
%\begin{equation}
%M0=g\times \left(C_{1,0}^{(m0)}\alpha_s+ C_{1,1}^{(m0)}\alpha_s L \right)
%\end{equation}

In scheme A of Sec.~\ref{sec:fonllsf}, which corresponds to using
NLO with fixed--order results consistently included up to ${\mathcal O} (\alpha_s)$ result, one gets
\bea
\label{eq:fmassive0-expL}
&& F^{(n_l,0)}_{h} (x, Q^2) = x \int_x^1 \frac{dy}{y}\left[   
C_{g,h}^{(n_l,0),1,0} \left( \frac{x}{y} \right) \alpha_s(Q^2)g(y,Q^2)\right.
\nonumber \\&&\left.\qquad
+ C_{g,h}^{(n_l,0),1,1} \left( \frac{x}{y} \right) \alpha_s(Q^2)g(y,Q^2)L\right] .
\eea
Therefore, because of Eq.~(\ref{eq:masslessg})
the ZM  result Eq.~(\ref{eq:fmassless-expL}) and the
M0 result Eq.~(\ref{eq:fmassive0-expL})
must coincide when $L=0$, while their 
 slope in $L$ near $L=0$ should 
differ 
by terms of order $\alpha_s^2$, which are present in the ZM
Eq.~(\ref{eq:fmassless-expL}) but not
in the M0 result Eq.~(\ref{eq:fmassive0-expL}).
This is borne out by the plots  Fig.~\ref{fig:F2-035}
for $F_{2\,c}(x,Q^2)$ and
Fig.~\ref{fig:FL-035}
for $F_{L\,c}(x,Q^2)$, in which  the structure function is plotted as
a function of $Q/m$ for several values of $x$ (with $m=m_c=\sqrt{2}$ GeV).
We have also checked that
the difference in slope is indeed 
of order $\alpha_s^2$, by repeating the plot with different values 
of $\alpha_s^2$.

In scheme B, which corresponds to NLO but with   ${\mathcal O}
(\alpha_s^2)$ contributions included in massive coefficient functions,
the M0 result becomes
%
% There are also quark terms, maybe they do not have logs?
%
%\begin{equation}
%M0=g\times \left(C_{1,0}^{(m0)}\alpha_s+ C_{1,1}^{(m0)}\alpha_s L
%+C_{2,1}^{(m0)}\alpha_s^2 L+C_{2,2}^{(m0)}\alpha_s^2 L^2 \right),
%\end{equation}
\bea
&& F^{(n_l,0)}_{h} (x, Q^2) = x \int_x^1 \frac{dy}{y}  \left[
C_{g,h}^{(n_l,0),1,0} \left( \frac{x}{y} \right) \alpha_s(Q^2)g(y,Q^2)
\nonumber \right.\\&&\quad
+ C_{g,h}^{(n_l,0),1,1} \left( \frac{x}{y} \right) \alpha_s(Q^2) L g(y,Q^2)
+ C_{g,h}^{(n_l,0),2,1} \left( \frac{x}{y} \right) \alpha_s(Q^2)^2 Lg(y,Q^2)
\nonumber \\&&\quad \label{eq:fmassive0-schemeB-expL}
\left.+ C_{q,h}^{(n_l,0),2,1} \left( \frac{x}{y} \right) \alpha_s(Q^2)^2L\sum_{q}\lp q(y,Q^2)+\bar{q}(y,Q^2) \rp
 + \mathcal{O}\lp L^2\rp\right] \ ,
\eea
which now  also receives a contribution from the light--quark
initiated terms of Fig.~\ref{fig:feyn-nlo-ir}, which start at $\mathcal{O}\lp \alpha_s^2\rp$.
Note that, as already mentioned in Sec.~\ref{sec:fonllsf}, in scheme B 
non-logarithmic $\mathcal{O}\lp \alpha_s^2\rp$ are not included in the
M0 result, because they must match the ZM result, which in this
scheme is determined at NLO, so that beyond $\mathcal{O}\lp
\alpha_s\rp$ it
only contains logarithmically enhanced terms.
Comparing the M0 result Eq.~(\ref{eq:fmassive0-schemeB-expL}) to the
ZM one of Eq.~(\ref{eq:fmassless-expL}) one sees that in scheme B
not only the value  at threshold but also the slope 
should agree, up to a small correction $\mathcal{O}\lp \alpha^3_s\rp$
due to contribution proportional to $h_2^{(1)}$ in Eq.~(\ref{eq:fmassless-expL}). This is borne out by
the plots in Figs.~\ref{fig:F2-035}--\ref{fig:FL-035}.

The same reasoning can be pursued to NNLO. In this case, we only have
results in scheme C, in which fixed--order results are included up to
${\mathcal O} (\alpha_s^2)$. In this case the ZM and M0 results
differ at threshold both in value and slope
through  terms of order
$\alpha_s^3$ arising both from the PDF and the coefficient function.  
This mismatch is apparent in
Figs.~\ref{fig:F2-035}--\ref{fig:FL-035}, and turns out to 
be quantitatively
different in $F_{2\,c}$ and in $F_{L,\,c}$. Indeed, for $F_{2\,c}$
(Fig.~\ref{fig:F2-035}) 
the difference in slope  is very visible, but
the difference
in value at $L=0$ is very small and 
 difficult to appreciate. On the other hand
for $F_{L\,c}$ (Fig.~\ref{fig:FL-035}), 
higher order corrections 
are larger and  the differences
also in values at threshold are more visible.
\begin{center}
\begin{figure}
\epsfig{file=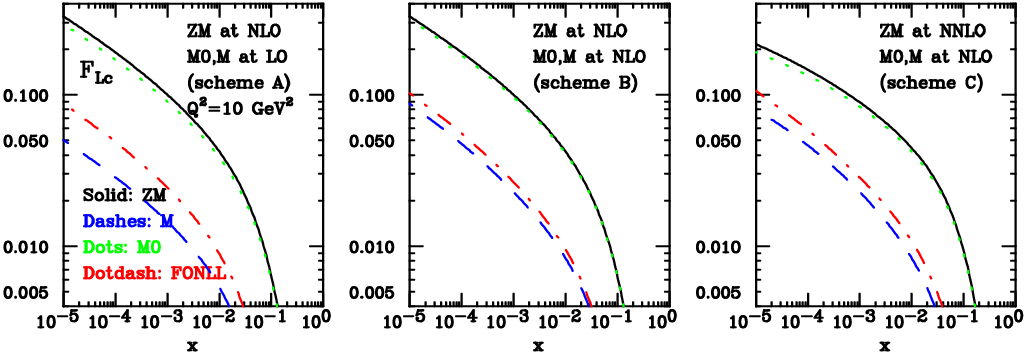,width=\textwidth}
\epsfig{file=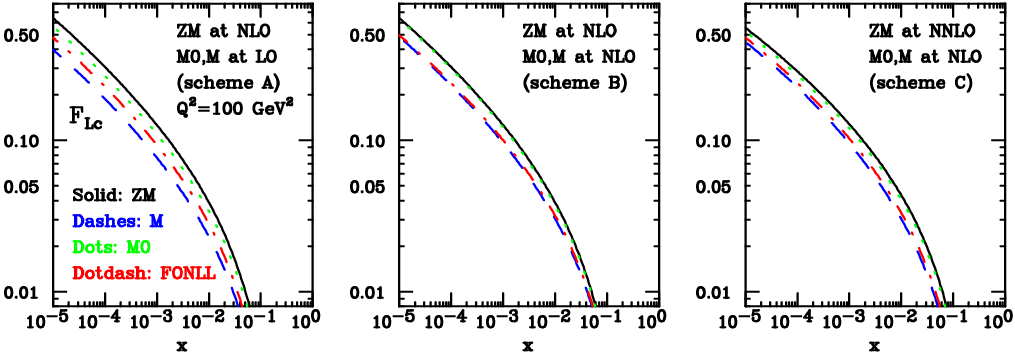,width=\textwidth}
\caption{\small 
\label{fig:FL-0100} Same as Fig.~\ref{fig:F2-0100}, but for the
structure function $F_{L\,c}$.}
\end{figure}
\end{center}

\section{Phenomenological impact of the FONLL method}
\label{sec:pheno}

We will now examine the phenomenological impact of the FONLL method
on structure functions. We will consider  the case, both for $F_2$ and
$F_L$, of the total
structure function, as well as  the
``heavy'' contribution $F_h$ to structure functions in the particular case
of charm, namely $F_{2\,c}$  and  $F_{L\,c}$. The purpose of our
study is to assess the impact of heavy quark terms at various
perturbative orders,  to compare and validate different prescriptions
for the treatment of subleading terms close to threshold, and finally
to provide a general assessment of the impact of heavy quarks on
structure functions. A study of the impact of heavy quark terms on a
determination of PDFs should be part of a global PDF fit and it is thus
beyond the scope of our work. 

In order to ease comparison with other
studies, we will use the PDFs of the standard reference Les
Houches~\cite{Giele:2002hx} set at the scale $Q_0^2=2$~GeV$^2$, 
evolved using the {\tt QCDNUM}~\cite{qcdnum}
package, and tabulated in
Refs.~\cite{Giele:2002hx,Dittmar:2005ed} both in the massive and
massless scheme (with, in the latter case, PDFs matched
according to the results given in Sec.~\ref{sec:match}). Massive
${\mathcal O}(\alpha_s^2)$ contributions to
coefficient functions have been determined using the numerical
implementation  of the results of
Ref.~\cite{Laenen:1992zk}, and their massless limit using the
numerical implementation of the results of
Ref.~\cite{Buza:1995ie}. All the analytic expressions which we have
used and the computer codes used to implement them are illustrated in
the Appendix~\ref{sec:appendix}. As in Ref.~\cite{Giele:2002hx}, we take
the charm mass at
$m^2=2$~GeV$^2$ and $\alpha_s(m^2_c)=0.35$, which corresponds to
values in the ballpark $\alpha_s(M^2_z)\sim 0.120$ at NLO and
$\alpha_s(M_z)\sim 0.110$ at NNLO  according to the scheme used (see  
Refs.~\cite{Giele:2002hx,Dittmar:2005ed} for the exact values). 
We will furthermore 
neglect the $b$ threshold, by letting $m_b\to\infty$.

\subsection{Comparison between FONLL, massive and massless results}
\label{sec:compfmm}

At first, we compare structure functions 
in the massive scheme Eq.~(\ref{eq:Fnl}) (M, henceforth),
in the massless scheme Eq.~(\ref{eq:Fnf}) (ZM, henceforth), in the
massless limit of
the massive scheme Eq.~(\ref{eq:Fnlbarzero}) (M$0$, henceforth),
and using the FONLL  prescription Eq.~(\ref{eq:FONLL}).
We begin by showing  in the first three plots of Fig.~\ref{fig:F2-0020}
the structure function $F_{2\,c}$, computed using the three different
combinations of perturbative orders discussed in
Sec.~\ref{sec:fonllsf} and referred to as schemes A--C, 
at the charm mass scale $Q^2=m_c^2=2$~GeV$^2$.
We see that in all cases the ZM and M$0$ results either coincide,
or else are very close, as we have already seen in Sec.~\ref{sec:fonlloverlap}.  It follows that the ``difference'' contribution
Eq.~(\ref{eq:fdiff}) is
vanishingly small, and thus FONLL result coincides with the massive
calculation.

However, the
same quantities at  $Q^2=4$~GeV$^2$ (also displayed in  Fig.~\ref{fig:F2-0020})
show remarkable differences.
In the scheme~A, the ZM and M$0$ results differ considerably,
so FONLL no longer simply reduces  to the massive
result, and it is in fact rather closer to the ZM result.  
This is not the case in scheme B, in which
the ZM and M$0$ results again nearly
coincide. The  case of scheme C is similar to that of scheme B for
$x>10^{-3}$, but  for smaller values of $x$ the ZM and M$0$ result again differ
considerably.  In Fig.~\ref{fig:F2-0100} we see that this pattern
persists at higher energies, although with a reduced impact of mass
effects. This means that inclusion  of the ${\mathcal
  O}(\alpha_s^2)$ 
contribution to the
massive terms in scheme  B significantly 
reduces the impact of the unreliable ${\mathcal
  O}(\alpha_s^2)$  terms in the difference contribution which are
present in scheme A, and whose effect persists even at larger scales. 
However, in scheme C further ${\mathcal
  O}(\alpha_s^3)$  terms are introduced, which may become large at
small $x$ due to large small $x$ logs~\cite{Catani:1990eg}.

From the plot on the first column of Fig.~\ref{fig:F2-0100}, we 
also see that the difference between the M$0$ and M
curves  (i.e. the impact of mass effects)
in the first order heavy flavour coefficient is quite large
even at a scale of $10\;{\rm GeV}^2$.  This difference seems to be
reduced in the third column, which suggests a cancellation of mass
effects between the first and second order in $\alpha_s$. This
cancellation should be considered accidental, and it cannot be taken as
evidence that mass effects are negligible at this scale.
\begin{center}
\begin{figure}
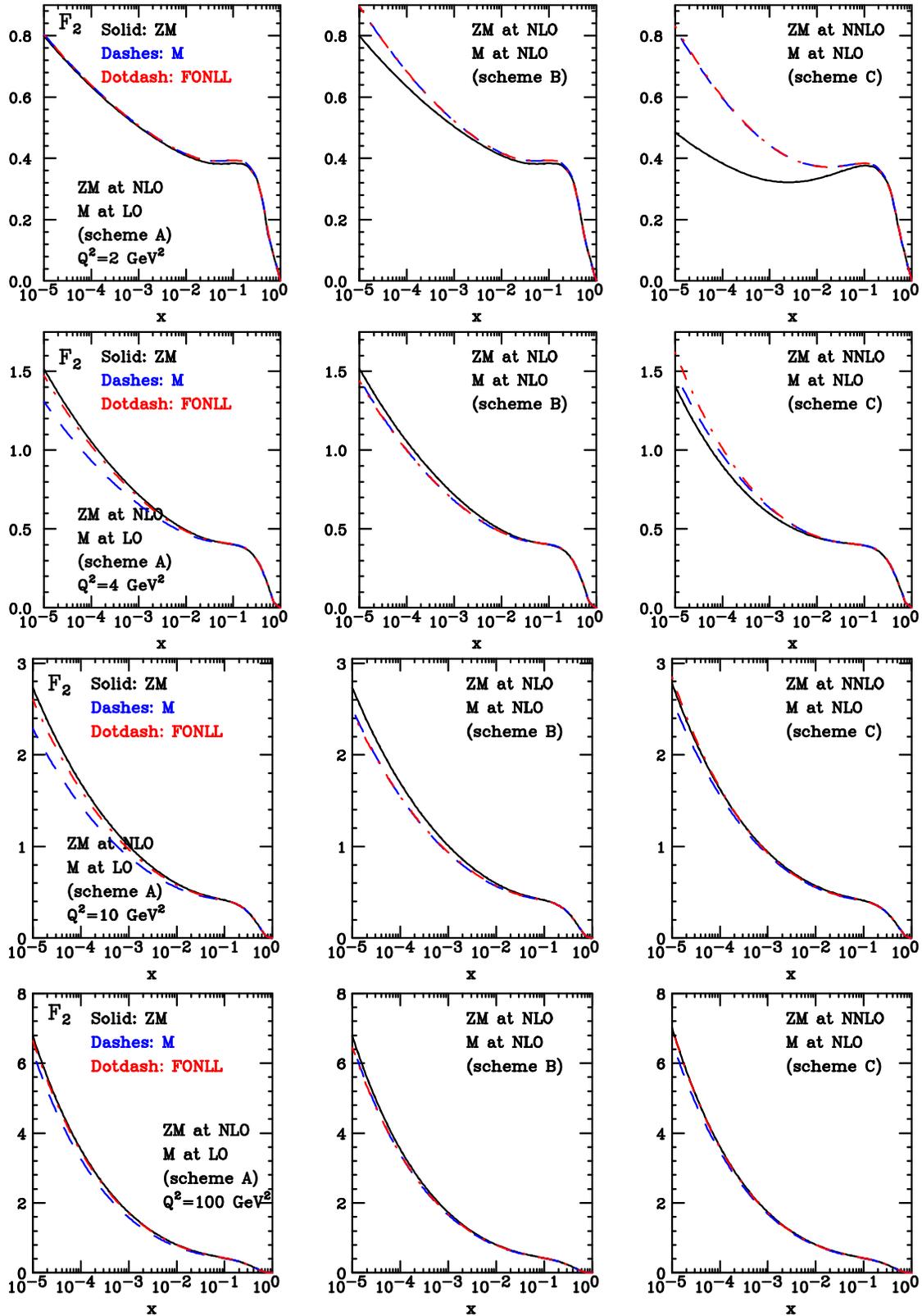

\epsfig{width=0.99\textwidth,figure=f2tot-fonll-100.ps}\\
\epsfig{width=0.99\textwidth,figure=f2tot-fonll-200.ps}\\
\epsfig{width=0.99\textwidth,figure=f2tot-fonll-500.ps}\\
\epsfig{width=0.99\textwidth,figure=f2tot-fonll-5000.ps}
\caption{\small \label{fig:f2tot} Same as
  Figs.~\ref{fig:F2-0020}--\ref{fig:F2-0100} but for the
total structure function $F_2$.}
\end{figure}
\end{center}

We now turn to the structure function $F_{L\,c}$. Also in this case, 
right at threshold (Fig.~\ref{fig:FL-0020})  the ZM and
M$0$ results compensate exactly in the A and B schemes, though now in
scheme C  they start  to deviate at $x\lsim 10^{-2}$, and only coincide for
larger $x$ values. As the scale  is raised to $Q^2=4$~GeV$^2$, the ZM
and M$0$ results start to differ, though by a much smaller amount than
in the case of $F_{2\,c}$.
In this case, however,
the ZM and M$0$ results at and close to threshold 
are much larger in magnitude than the M
result. This means that for $F_{L,c}$ the massless approximation is
completely inadequate close to threshold. This is clarly seen in
Fig.~\ref{fig:FL-0020}, where the massless result at threshold 
in  all cases A-C is
totally different from the massive one, to which instead FONLL
essentially reduces in cases A-B, because of the complete cancellation
between the M and M0 terms. In case C, however, the M and M0 terms do
not cancel completely at threshold (recall Fig.~\ref{fig:FL-035},
especially the lower right plot) 
and thus, because of the the large relative size of these (subleading
and unphysical) terms  
in comparison to the massive result, the FONLL result  ends up having
also an unphysical behaviour at threshold. This means that in this
case it is crucial to implement a suppression of these subleading
terms as discussed in Sect.~\ref{sec:threshold}, and that this
suppression must be very strong at threshold (as it is using the
threshold factor Eq.~(\ref{eq:threshold})): only in such case
the difference  contribution Eq.~(\ref{eq:fdiff}) is removed 
 and FONLL reproduces the M result again.
Obviously, this means that the impact of the threshold prescription
for $F_L$ in scheme C is very strong.  
Once again,
at $Q^2=10\;{\rm GeV}^2$ (Fig.~\ref{fig:FL-0100}) the M and
M$0$ results differ considerably. Even at $Q^2=100\;{\rm GeV}^2$
mass effects are not at
all negligible for $F_{L\,c}$.
%-
\begin{center}
\begin{figure}
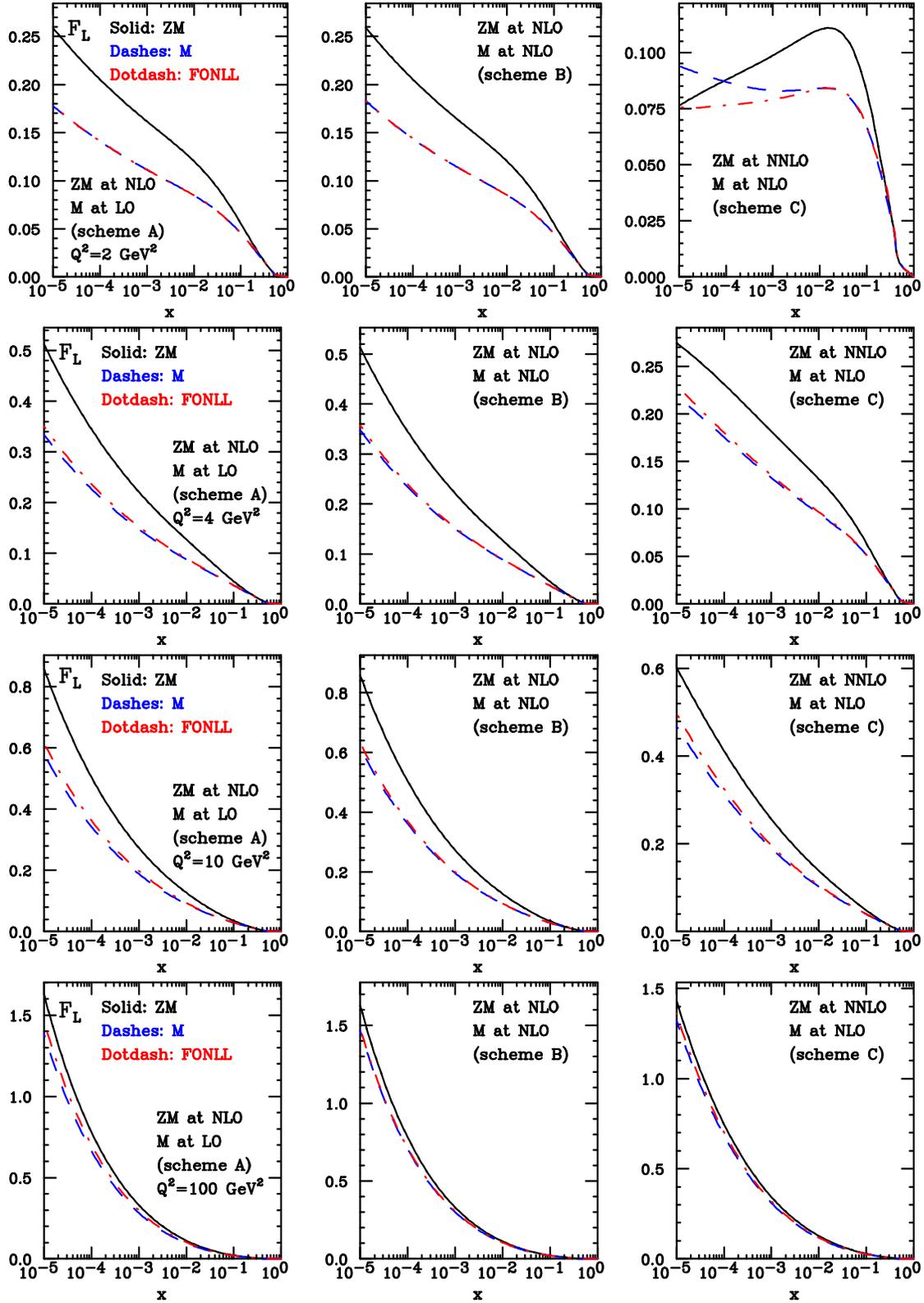

\epsfig{width=0.99\textwidth,figure=FLtot-fonll-1100.ps}\\
\epsfig{width=0.99\textwidth,figure=FLtot-fonll-1200.ps}\\
\epsfig{width=0.99\textwidth,figure=FLtot-fonll-1500.ps}\\
\epsfig{width=0.99\textwidth,figure=FLtot-fonll-6000.ps}
\caption{\small \label{fig:fltot}
Same as
  Figs.~\ref{fig:FL-0020}--\ref{fig:FL-0100} but for the
total structure function $F_L$.}
\end{figure}
\end{center}
%-

In Figs.~\ref{fig:f2tot}-\ref{fig:fltot} we turn to the to the total
  structure functions, i.e. the sum of the light and heavy
  contributions Eq.~(\ref{eq:hlsep}). 
As discussed in Sec.~\ref{sec:fonllm}, the FONLL prescription has no
effect on $F_l$ to $O(\alpha_s)$, and thus is scheme A. Its effect to
$O(\alpha_s^2)$, which comes from  the Feynman diagrams of
Figs.~\ref{fig:feyn-nlo-ir}-\ref{fig:feyn-nlo-self}  is very small:
the contribution from these diagrams is small in the first place, and
its contribution to the FONLL prescription is further suppressed by
the cancellation
Eqs.~(\ref{eq:mustvanisha}-\ref{eq:mustvanish}). Therefore,  the
dominant effect of the FONLL prescription is on $F_h$, and thus
diluted in the total structure functions. Nevertheless, the main qualitative
  features remain the same. Namely, in scheme A
a significant difference between the
  massive and FONLL results arises just above the threshold, and
  persists up  to scales as large as 10--20~GeV$^2$, with the FONLL
  result rather closer to the massless one for $F_2$ and to the
  massive one for $F_L$. This difference
  is driven by subleading (unreliable) threshold contributions, and thus  
sensitive to the threshold treatment discussed in
Sec.~\ref{sec:threshold}, to be
studied phenomenologically in Sec.~\ref{sec:compth}. Its effects are
greatly reduced in schemes B-C (but with some small $x$ instability in
scheme C) and in general for the structure function $F_L$. 

\subsection{Comparison and impact of threshold prescriptions}
\label{sec:compth}

As explained in Sec.~\ref{sec:threshold} the purpose of threshold
prescriptions is to modify subleading contributions to the FONLL
expression which are certainly not accurate but may be non-negligible
close to threshold. As such, from a theoretical point of view 
all threshold prescriptions (including
the choice of not implementing any threshold prescription) are equally
justified, and varying the threshold prescription merely provides an
estimate of unknown subleading terms. However, when the higher order
exact result is known we may validate the threshold prescription by
comparing the subleading terms which are affected by the threshold
prescription with their exact form. Here we perform such a validation;
then we assess the impact of different choices for the threshold
prescription on the final FONLL result.

The terms which are affected by the threshold prescription are the
difference between the ZM and M0 expressions: they are nonzero because
the ZM contains contributions to all orders, while the M0 is
determined at fixed perturbative order. In the simplest (trivial) case
in which the structure function is determined at ${\mathcal
  O}(\alpha_s^0)$, the massive contribution to the FONLL expression
vanishes: the FONLL expression then reduces to the massless scheme one
which we know to be inaccurate close to threshold. A threshold
prescription applied at this order 
might then reduce the difference between the massless scheme
expression and one in which mass effects are properly treated. In
fact, in Ref.~\cite{Thorne:2008xf} the use of a massless scheme
supplemented by the $\chi$--scaling threshold prescription was
advocated as a possible approximation to schemes were mass effects are
fully included such as the FONLL scheme discussed  here.
\begin{center}
\begin{figure}
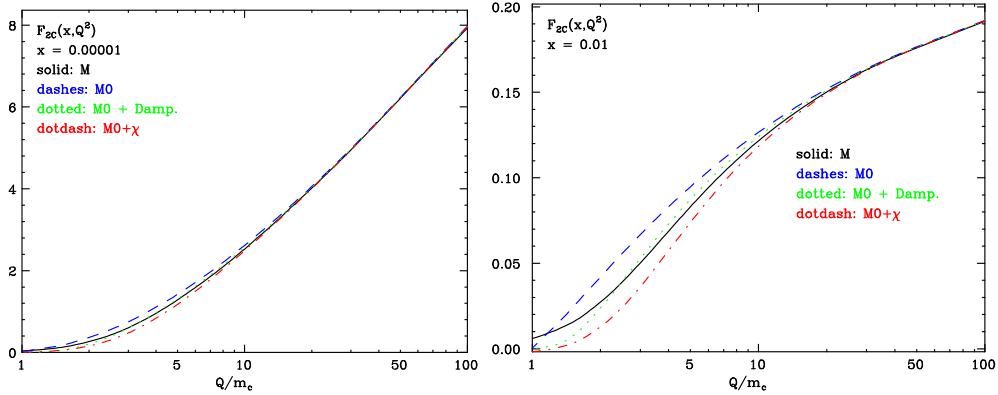

\begin{center}
\epsfig{width=0.42\textwidth,figure=F2-cthres-1000010.ps}
\epsfig{width=0.45\textwidth,figure=F2-cthres-1010000.ps}\\
\caption{\small \label{fig:thresholdlosx}
The  ${\mathcal O}(\alpha_s)$ contribution to the charm structure function, 
$F_{2\,c}(x,Q^2)$ with
  $m_c^2=2$~GeV$^2$  plotted as a function of
$Q^2$ at fixed $x$ at small $x$. The four curves correspond to the 
the full massive--scheme (M) result Eq.~(\ref{eq:Fnlbar}), its
$m\to0$ limit (M0) Eq.(~\ref{eq:Fnlbarzero}), and the results obtained
applying to  the M0 a damping factor Eq.~(\ref{eq:thrfacts}) or $\chi$--scaling
Eq.~(\ref{eq:chiscaling}).}
\end{center}
\end{figure}
\end{center}

\begin{center}
\begin{figure}
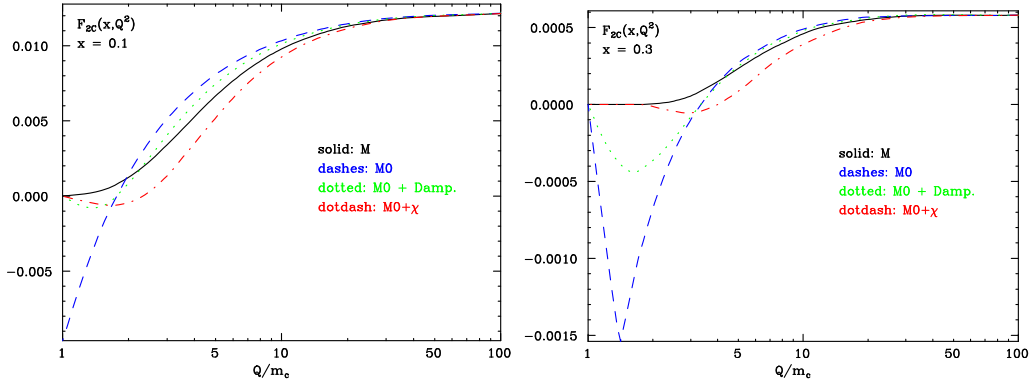

\begin{center}
\epsfig{width=0.45\textwidth,figure=F2-cthres-1100000.ps}
\epsfig{width=0.45\textwidth,figure=F2-cthres-1300000.ps}\\
\caption{\small \label{fig:thresholdlo}
Same as Fig.~\ref{fig:thresholdlosx} but at large $x$.}
\end{center}
\end{figure}
\end{center}

Of course, the first massive contribution appears at ${\mathcal
  O}(\alpha_s)$, so in order to assess whether indeed a threshold
 prescription does help we
 should compare the exact ${\mathcal
  O}(\alpha_s)$ massive (M) result, i.e. $F^{(n_l)}(x,Q^2)$
 Eq.~(\ref{eq:Fnlbar})
 to its massless (M0) approximation $F^{(n_l,\,0)}(x,Q^2)$, which is
 included in the massless (ZM) result, and to the putatively improved
 versions of the M0 which are obtained by supplementing it with a
 threshold prescription, such as the damping factor
 Eq.~(\ref{eq:thrfacts}) or $\chi$--scaling
Eq.~(\ref{eq:chiscaling}).
This comparison is performed in
Figs.~\ref{fig:thresholdlosx}-\ref{fig:thresholdlo} for the
$F_{2\,c}$ structure function.  We see that at
very small $x$ mass effects are small, but that for $x\gsim 0.01$
close to threshold the deviation between the M result and its M0
approximation becomes significant, the full result being suppressed by
mass effects in comparison to its massless approximation. The
threshold factor reproduces well this suppression especially at
smaller $x$ values, while
$\chi$--scaling provides a bit too much suppression and accordingly a
somewhat worse approximation. That $\chi$--scaling provides an  excess
of suppression for $F_2$ was already noticed in
Ref.~\cite{Nadolsky:2009ge}, where modified, improved $\chi$
variable were suggested to remedy this situation. It is interesting to
observe that the slightly different $\chi$--scaling prescription
Eq.~(\ref{eq:altchi}) (discussed in Ref.~\cite{Martin:2009iq})
leads instead to rather less suppression, and in
particular, at small $x$,  less than the damping factor.

\begin{center}
\begin{figure}
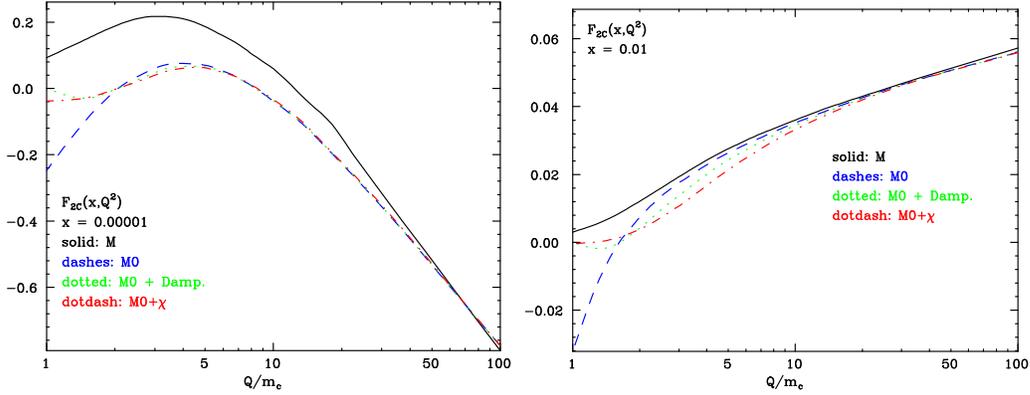

\begin{center}
\epsfig{width=0.45\textwidth,figure=F2-cthres-alpha2only-1000010.ps}
\epsfig{width=0.45\textwidth,figure=F2-cthres-alpha2only-1010000.ps}\\
\end{center}
\caption{\small \label{fig:thresholdnlosx}
Same as Fig.~\ref{fig:thresholdlosx} but for the ${\mathcal O}(\alpha_s^2)$ contribution to 
$F_{2\,c}(x,Q^2)$.}
\end{figure}
\end{center}

\begin{center}
\begin{figure}
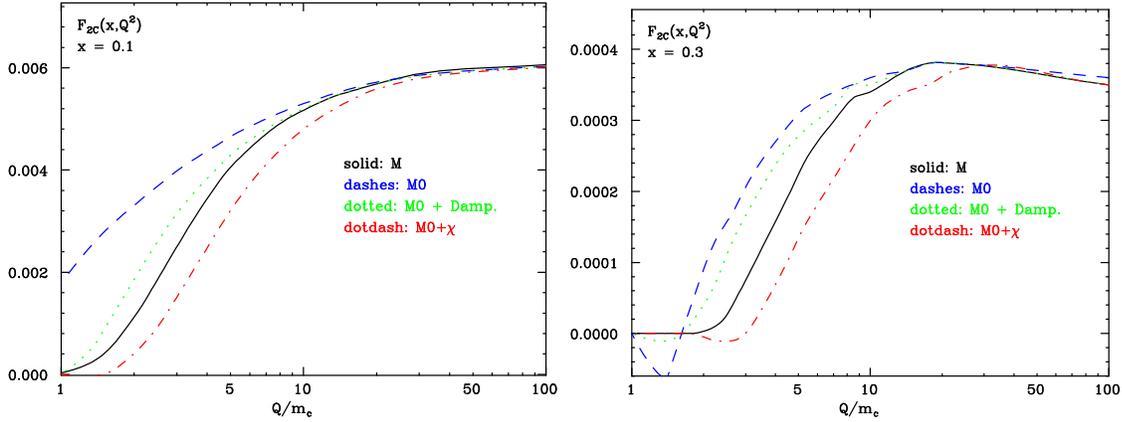

\epsfig{width=0.49\textwidth,figure=F2-cthres-alpha2only-1100000.ps}
\epsfig{width=0.49\textwidth,figure=F2-cthres-alpha2only-1300000.ps}\\
\caption{\small \label{fig:thresholdnlo}
Same as Fig.~\ref{fig:thresholdnlosx} but at large $x$.}
\end{figure}
\end{center}

Our results  so far, Figs.~\ref{fig:thresholdlosx}-\ref{fig:thresholdlo},
show that indeed a threshold factor may help, but they are  of academic
interest as soon as we implement the FONLL prescription at lowest
nontrivial order, because ${\mathcal
  O}(\alpha_s)$ terms are then treated exactly. We thus turn to the more
practically relevant case of the ${\mathcal
  O}(\alpha_s^2)$ terms: these are treated approximately if we adopt
the NLO FONLL method of scheme A. In
Figs.~\ref{fig:thresholdnlosx}-\ref{fig:thresholdnlo} we repeat the
comparison of the M, M0, and threshold--corrected M0 results, but now
for the ${\mathcal
  O}(\alpha_s^2)$ contributions to $F_{2\,c}$. Now the difference
between the exact M and approximate M0 result is significant even at
very small $x$. It is still true that, while results with threshold
prescriptions are closer to the exact one, $\chi$--scaling provides a
somewhat excessive suppression. However, now at large $x$ the
damping factor provides insufficient suppression, so the quality of
the approximation of either threshold prescription is similar, though
perhaps the damping factor is still slightly better on average.
%-
%-

If the FONLL method is treated including massive contributions up to ${\mathcal
  O}(\alpha_s^2)$, i.e. using schemes B or C the threshold
prescription only starts affecting ${\mathcal
  O}(\alpha_s^3)$ terms. We can then no longer assess its accuracy,
because exact ${\mathcal
  O}(\alpha_s^3)$ massive coefficient functions are unknown.

We can now study the impact of various threshold prescriptions on the
FONLL result. This is done in Figs.~\ref{fig:thr-F2}-\ref{fig:thr-FL}, 
where we compare
respectively the FONLL expression for the structure functions
$F_{2\,c}$ and $F_{L\,c}$, determined at different
perturbative orders, i.e. within schemes A-C, and either without any
threshold prescription, or with either of the two threshold
prescriptions that we discussed so far. The effect of the threshold
treatment is most visible just above threshold at $Q^2=4$~GeV$^2$, 
as expected based on
the discussion of Sec.~\ref{sec:compfmm} and
Figs.~\ref{fig:F2-0100}-\ref{fig:F2-0020}. At this scale,
in scheme A the effect of the threshold suppression
procedure is
visible for all values of $x$; in schemes B-C, where the threshold
treatment only affects ${\mathcal
  O}(\alpha_s^3)$ terms, the effect is generally small (except, in scheme C,
for very small or very large values of $x$). Therefore, we conclude
that the ambiguity related to subleading terms is substantially reduced
if massive terms are treated to ${\mathcal
  O}(\alpha_s^2)$, i.e. using schemes B or C. If massive terms are 
treated to ${\mathcal
  O}(\alpha_s)$, i.e. using scheme A, our results suggests
that the accuracy can be improved using a threshold
suppression, and in particular using a damping factor to treat
threshold effects. It is interesting to observe that, for $F_2$, right at
threshold the difference between scheme A and scheme B appears to be
non-negligible, i.e.  ${\mathcal
  O}(\alpha^2_s)$ have a significant impact at threshold, rather
larger than the typical size of the ambiguity related to threshold terms.

In the case of $F_L$ results at threshold in scheme A show numerical
instabilities due to imperfect cancellation between the ZM
and M0 contributions, which as shown in Fig.~\ref{fig:FL-0020} are
  much larger than the massive result. They are the consequence of the
  fact that the massless approximation is completely inadequate at
  threshold for this observable. As soon as the scale is raised
  above threshold this may lead to very large unphysical
  contributions even in the FONLL scheme unless a strong threshold suppression
is adopted, such as that provided by a damping factor. This is even
more the case in scheme C: due to the instability displayed in
Fig.~\ref{fig:FL-0020} results without threshold treatment, and even
those obtained using $\chi$ scaling are unphysical except at very
large $x$.

\subsection{Comparison of results at different orders}
\label{sec:compord}

\begin{center}\begin{figure}\begin{center}
\epsfig{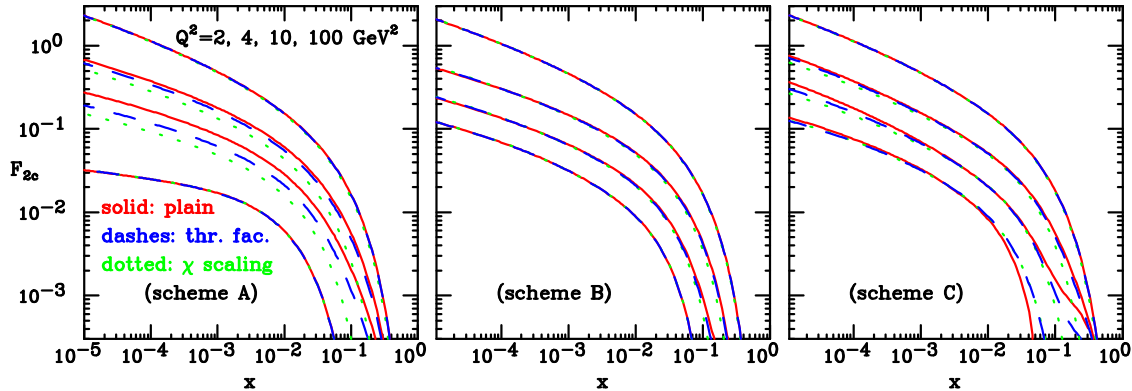}
\caption{\label{fig:thr-F2} The FONLL expression for $F_{2\,c}$ 
in the three schemes for
  perturbative ordering A--C of Sec.~\ref{sec:fonllsf}, 
and either with no threshold
  suppression terms, or with the damping factor or $\chi$ scaling
  suppression at threshold. The structure
functions are plotted as a function of $x$ for fixed
$Q^2=2,\>4,\>10,\>100$~GeV$^2$ (from bottom to top).}
\end{center}\end{figure}\end{center}

\begin{center}\begin{figure}\begin{center}
\epsfig{file=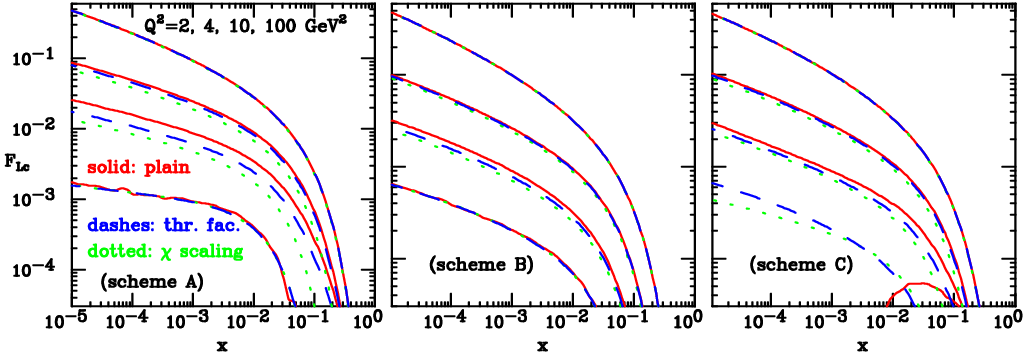,width=\textwidth}
\caption{\label{fig:thr-FL} Same as Fig.~\ref{fig:thr-F2} but for the
  structure function $F_{L\,c}$.}
\end{center}\end{figure}\end{center}

\begin{figure}
\begin{center}
\epsfig{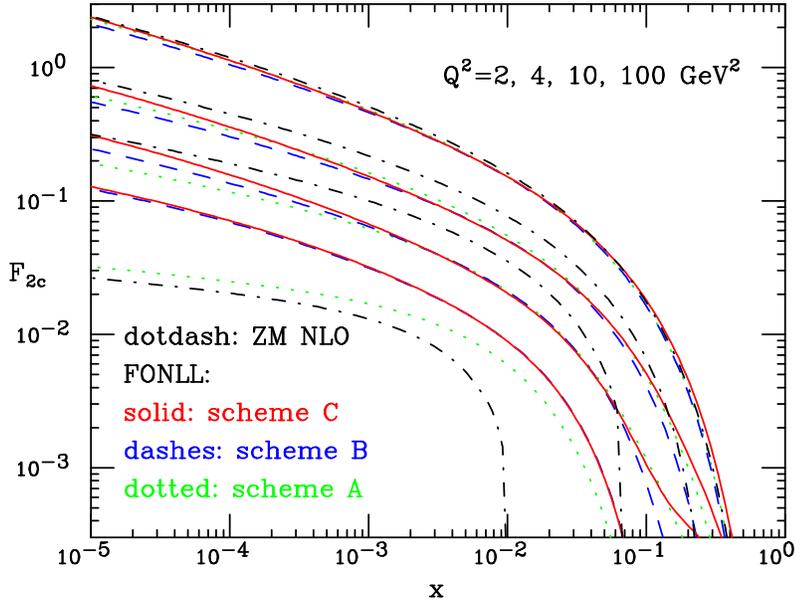}
\end{center}
\caption{\label{fig:thrfac-compABC-F2} The FONLL expression for the
  structure function $F_{2\,c}$ with the threshold treated using the
  damping factor Eq.~(\ref{eq:thrfacts}), in the three schemes for
  perturbative ordering A--C of Sec.~\ref{sec:fonllsf}. The structure
functions are plotted as a function of $x$ for fixed
$Q^2=2,\>4,\>10,\>100$~GeV$^2$ (from top to bottom). Results in
the NLO massless scheme result
(same as Figs.~\ref{fig:F2-0020}-\ref{fig:F2-0100})  
are also shown for comparison (dotdashed).}
\end{figure}

\begin{figure}
\begin{center}
\epsfig{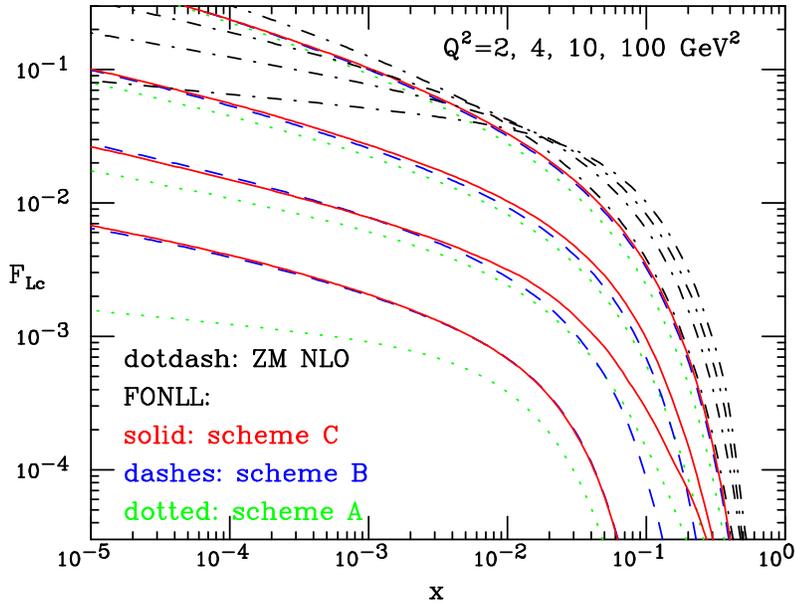}
\end{center}
\caption{\label{fig:thrfac-compABC-FL} Same as
  Fig.~\ref{fig:thrfac-compABC-FL}, but for the structure function $F_L$.}
\end{figure}

Finally, in
Figs.~\ref{fig:thrfac-compABC-F2}-\ref{fig:thrfac-compABC-FL}
we compare results for the structure functions $F_2$ and $F_L$ determined at
various perturbative orders, i.e.  in the three schemes A,B,C. All
results are now determined using our default option, namely, the FONLL
method with the threshold treated using a  damping factor: the curves
are thus the same as those shown in
Figs.~\ref{fig:thr-F2}-\ref{fig:thr-FL} for  the damping
factor prescription. For comparison, we also
show the simple massless--scheme result Eq.~(\ref{eq:Fnf}), determined
at NLO (${\mathcal O}(\alpha)$).

First, it appears that effects due to the heavy quark mass are not
only large in the threshold region, but in fact still
sizable at $Q^2\sim10$~GeV$^2$ for $F_2$, and for $F_L$ quite large
even 
at $Q^2\sim100$~GeV$^2$.
Also, it is clear that while the B-C scheme curves are very close to each
other, the curve in the A scheme differs substantially from them. This
means that scheme B is an improvement over scheme A, in that it
already includes most of the ${\mathcal O}(\alpha^2)$ effects which
are fully included in scheme C. 

Of course, because of asymptotic freedom, 
differences between results obtained using different perturbative
schemes become smaller at large $Q^2$. Note however that
here, for illustrative purposes, structure functions have been
determined  with a fixed set
of PDFs. In a realistic situation, the structure function would be
fixed by the data and the PDFs would be fitted. 

%%%%%%%%%%%%%%%%%%%

\section{Mass singularities in $F_{2\,h}$}
\label{sec:irs}

\begin{figure}
\begin{center}
\epsfig{figure=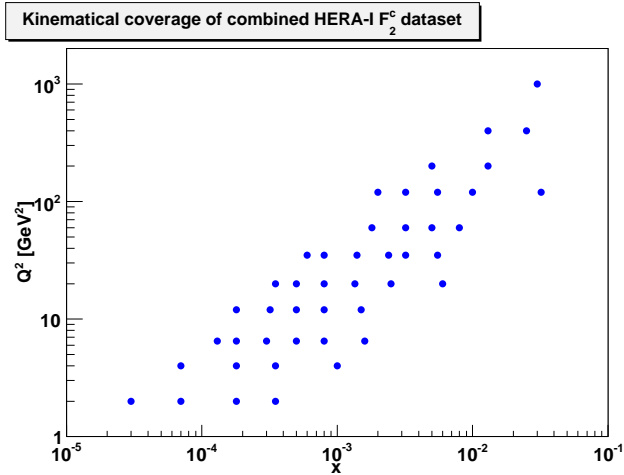,width=0.60\textwidth}
\caption{\small Kinematical coverage of $F_{2\,c}$ in the
combined HERA-I~\cite{H1comb} dataset.}
\label{fig:kin}
\end{center}
\end{figure}

The heavy contribution to structure functions is experimentally
accessible, and indeed the experimental determination of the charm and
beauty structure functions $F_{2\,c}$ and $F_{2\,b}$ has attracted
considerable interest recently, in particular at HERA: specifically, the
kinematic coverage of $F_{2\,c}$ data in the $(x,Q^2)$ plane for 
the combined HERA-I dataset~\cite{H1comb} is shown in Fig.~\ref{fig:kin}.
However, the experimental definition of  heavy structure functions
differs somewhat from the definition,    given in
Sec.~\ref{sec:fonllm}, where  $F_{h}$ was defined as
the contribution to the structure function 
$F$ obtained when only the
electric charge $e_h$ of the heavy quark is nonzero. Rather, the
experimentally measured heavy quark 
structure function, which we will denote by  
$\widetilde{F}_{h}$, 
is defined as the contribution to the structure function $F$ from   
all processes in which
there is a heavy quark in the final
state~\cite{Breitweg:1999ad,Adloff:2001zj,Aktas:2004az,Aktas:2005iw,Aktas:2006py,Chekanov:2008yd,Chekanov:2009kj,Aaron:2009ut}.

The difference between    $\widetilde{F}_{h}$ and $F_{h}$ is
potentially significant, because $\widetilde{F}_{h}$ is  
affected by mass singularities in the limit $m\to0$, due
to the fact that heavy--quark production 
contributions in which the virtual photon couples to
a light quark such as shown in Fig.~\ref{fig:feyn-nlo-ir} are included
in $\widetilde{F}_{h}$, but virtual corrections such as shown in
Fig.~\ref{fig:feyn-nlo-self} are not. In contrast, using the
alternative definition $F_{h}$ neither contribution is included, and
both contribute to $F_{l}$, leading to a cancellation of potential
mass singularities.

For finite values of the heavy quark mass  
these mass--singular contributions  to $\widetilde{F}_{h}$
are of course finite, but enhanced by double logs (powers of $L^2$):
the first diagram of
Fig.~\ref{fig:feyn-nlo-ir} leads to a contribution of 
order $\alpha_s^2\log^3 Q^2/m^2$~\cite{Buza:1995ie,Chuvakin:1999nx}.
The three logarithmic powers have the following origin: one arises from the collinear
singularity in the emission of the gluon from the light quark, one from
the collinear singularity of the gluon splitting into the $h\bar{h}$ pair, and one
is due to the gluon becoming soft. The latter log arises because the
contribution to the coefficient function from the diagram of
Fig.~\ref{fig:feyn-nlo-ir}  is singular at $z=1$: the convolution
integral with the PDF up to the kinematic limit $z=Q^2/(Q^2+m^2)<1$ then
leads to an extra log whatever the behaviour of the PDF.
At higher perturbative orders
$\widetilde{F}_{h}$ then receives double--logarithmic contributions
of the form $\alpha_s^{2+k} L^{3+2k}$. These contributions could in
principle be
controlled experimentally by tagging both the heavy quark and
antiquark 
and introducing
a cutoff on the invariant mass of the heavy quark--antiquar pair~\cite{Chuvakin:1999nx}.

In conventional parton fits, $F_{h}$ is usually computed and compared
to data, even though the data really refer to
$\widetilde{F}_{h}$. Furthermore, even if the theoretical expression
$\widetilde{F}_{h}$ were  implemented in a parton fit, 
one may still wonder whether
this quantity may be subject to large higher--order corrections,
because of the aforementioned double logs. It is thus interesting to assess the
size of the difference between $\widetilde{F}_{h}$ and $F_h$ both at
lowest nontrivial order and at higher orders. 

\subsection{Order $\alpha_s^2$}
\label{sec:ir_est}

\begin{figure}
\begin{center}
\epsfig{figure=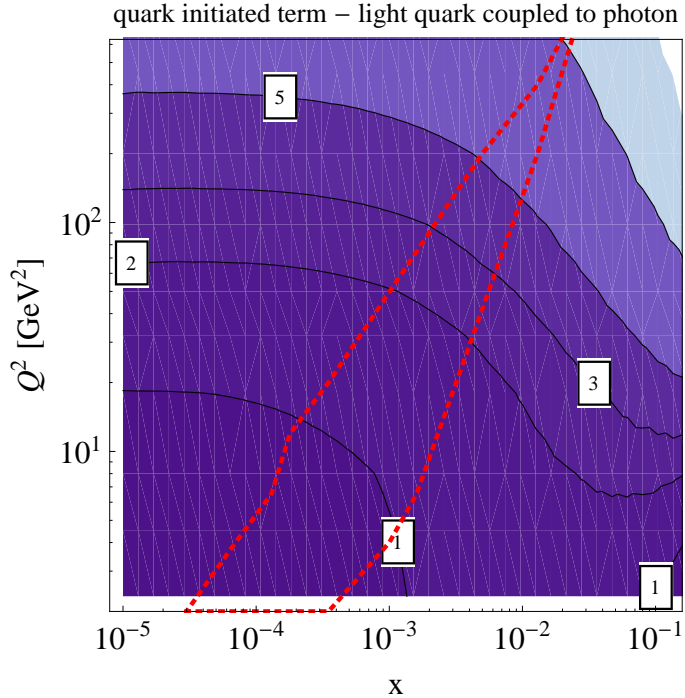,width=0.6\textwidth}
\caption{\small The
$\Delta F_{2\,c}^{(n_l)}$ percentage ratio Eq.~(\ref{eq:irc_el}), 
in the
$\lp x,Q^2\rp$ plane. The region bounded by the dashed lines is the
HERA kinematic range  of
Fig.~\ref{fig:kin}.}
\label{f2c-laenen}
\end{center}
\end{figure}

We have computed the ratio
\be
\label{eq:irc_el}
\Delta F_{2\,c}^{(n_l)}(x,Q^2) \equiv \frac{\widetilde{F}_{2\,c}^{(n_l)}(x,Q^2) - F_{2\,c}^{(n_l)}(x,Q^2) }{
F_{2\,c}^{(n_l)}(x,Q^2)}
\ee
for the more practically relevant  case of the charm
structure function $F_2$. We use
the ${\mathcal O} (\alpha_s^2)$ expressions for
$\widetilde{F}_{2\,c}^{(n_l)}(x,Q^2)$ and $F_{2\,c}^{(n_l)}(x,Q^2)$ of
Sec.~\ref{sec:alphasq}: the numerator of Eq.~(\ref{eq:irc_el}) then
is the contribution from the diagram of Fig.~\ref{fig:feyn-nlo-ir}.
Since we are mostly
interested in the behaviour near threshold, the massive scheme is
adopted throughout. The conventional Les Houches PDFs of
Sec.~\ref{sec:pheno} are used. Results are shown in
Fig.~\ref{f2c-laenen}.
For comparison, we also show
in Fig.~\ref{f2c-laenen2} the relative size
\be
\label{eq:irc_notg}
R_{\rm light}\equiv\frac{F_{2\,c}^{(n_l)}\Big|_{q_i,\bar{q}_i}(x,Q^2) }{F_{2\,c}^{(n_l)}(x,Q^2)} \ ,
\ee
of the  contribution from the light--quark initiated diagrams 
$F_{2\,c}^{(n_l)}\Big|_{q_i,\bar{q}_i}$ 
(shown in Fig.~\ref{fig:feyn-nlo-qinit})  
to the 
heavy structure function
$F_{2\,c}^{(n_l)}$. It appears 
 that the light--quark initiated contribution to
$F_{2\,c}^{(n_l)}$ of Fig.~\ref{f2c-laenen2}, 
though moderate, can amount to as
much as 10\% of $F_{2\,c}$ in the HERA region and it is thus 
rather larger than the
mass--singular contribution Fig.~\ref{f2c-laenen}, which remains at the percent
level in this region. Therefore, within this kinematics 
 the mismatch Eq.~(\ref{eq:irc_el}) 
between the two definitions of
the heavy quark structure function is negligible even when the measured
light--quark initiated countribution  (which includes both of the
contributions of Figs.~\ref{f2c-laenen}-\ref{f2c-laenen2}) is not small.

Even so, one may wonder whether these results still hold once
experimental cuts are accounted for. To this purpose, we have computed
\be
\label{eq:deltasigvis}
\Delta \sigma_{\rm vis,c} \equiv \frac{\widetilde{\sigma}_{\rm vis,c}-\sigma_{\rm vis,c} }{ \sigma_{\rm vis,c}} \ ,
\ee
where $\sigma_{\rm vis,c}$ is the so--called ``visibile'' 
reduced
cross section (after subtraction of the contribution from $F_L$),
i.e. the contribution to the DIS  charm production 
cross section with experimental cuts. We
have used
the {\tt HVQDIS} Monte Carlo program~\cite{Harris:1995tu}
and massive--scheme ZEUS-S PDFs~\cite{Chekanov:2002pv}; 
hadronization and charm decay corrections have been neglected,
though their inclusion would be straightforward. In order to define
the visible cross section, we have assumed $p_T^c\ge 1.5$ GeV
and $-1.6 \le \eta_c \le 2.3$, which correspond  to the acceptance of 
the recent
ZEUS muon analysis~\cite{Chekanov:2009kj} when hadronization and charm
decay corrections are neglected.

Our results are summarized in Table~\ref{tab:reshvqis2}, where we also
tabulate the percentage ratio Eq.~(\ref{eq:irc_el}) already shown in
Fig.~\ref{f2c-laenen},  but now also computed using {\tt HVQDIS} and
ZEUS-S PDFs. We have adopted
an $(x,Q^2)$ binning similar
to that of the upcoming combined HERA $F_{2\,c}$ dataset. 
We observe that, in qualitative agreement
with the results of Fig.~\ref{f2c-laenen}, the percentage discrepancy
is always $\le \mathcal{O}\lp 1\%\rp$, reaching 
$\ge \mathcal{O}\lp 2\%\rp$ only for the largest $Q^2$ bins,
where statistical uncertainties are anyway much larger.
Clearly,  with this choice of kinematics, results are only marginally
affected by
 experimental acceptances. We conclude that in the HERA kinematic
 range the discrepancy between  $\widetilde{F}_{h}$ and $F_h$ at
 ${\mathcal O}(\alpha_s^2)$ is at
 the level of the percent, even with experimental cuts. It is
 interesting to observe that at small $x$ and not too high $Q^2$ this
 contribution grows, and it should then be properly accounted for if
 $F_{2\,c}$ were measured at a future higher energy electron--proton
 collider~\cite{Dainton:2006wd}.

\begin{table}
\begin{center}
\footnotesize
 \begin{tabular}{|c|c|c|c|c|c|}
 \hline
 $x_{\rm min}$ & $x_{\rm max}$ & $Q^2_{\rm min}$ [GeV$^2$]& $Q^2_{\rm max}$
[GeV$^2$] & $\Delta F_{2\,c}^{\rm theo}$
 (\%) & $\Delta \sigma_{\rm vis,c}^{\rm theo} $ (\%) \\
 \hline
0.00002& 0.00005&     1.5&     3.0 & 0.2  & 0.3 \\
0.00005& 0.00013&     1.5&     3.0 & 0.2 & 0.2 \\
0.00013& 0.00026&     1.5&     3.0 & 0.2 & 0.6 \\
0.00026& 0.00044&     1.5&     3.0 & 0.3  & 0.4 \\
\hline
0.00005& 0.00013&     3.0&     5.3 & 0.4 & 0.2 \\
0.00013& 0.00026&     3.0&     5.3 & 0.3 & 0.1 \\
0.00026& 0.00068&     3.0&     5.3 & 0.3 & 0.2 \\
0.00068& 0.00125&     3.0&     5.3 & 0.3 & 0.1 \\
\hline
0.00010& 0.00015&     5.3&     9.3 & 0.4 & 0.5 \\
0.00015& 0.00024&     5.3&     9.3 & 0.4 & 0.6 \\
0.00024& 0.00040&     5.3&     9.3 & 0.2 & 0.5\\
0.00040& 0.00065&     5.3&     9.3 & 0.2 & 0.4\\
0.00065& 0.00120&     5.3&     9.3 & 0.3 & 0.2 \\
0.00120& 0.00200&     5.3&     9.3 & 0.4 & 0.2 \\
\hline
0.00014& 0.00025&     9.3&    16.0 & 0.3 & 0.9 \\
0.00025& 0.00041&     9.3&    16.0 & 0.3 & 0.9 \\
0.00041& 0.00065&     9.3&    16.0 & 0.2 & 0.2 \\
0.00065& 0.00115&     9.3&    16.0 & 0.4 & 0.4 \\
0.00115& 0.00187&     9.3&    16.0 & 0.3 & 0.3  \\
\hline
0.00026& 0.00043&    16.0&    27.5 & 0.3 & 0.7 \\
0.00043& 0.00065&    16.0&    27.5 & 0.4 & 0.4 \\
0.00065& 0.00108&    16.0&    27.5 & 0.4 & 0.4 \\
0.00108& 0.00193&    16.0&    27.5 & 0.4  & 0.3 \\
0.00193& 0.00425&    16.0&    27.5 & 0.4 & 0.3 \\
0.00425& 0.00750&    16.0&    27.5 & 0.4  & 0.3 \\
\hline
0.00045& 0.00070&    27.5&    47.5 & 0.4 & 0.5 \\
0.00070& 0.00110&    27.5&    47.5 & 0.3  & 0.2 \\
0.00110& 0.00190&    27.5&    47.5 & 0.4 & 0.3 \\
0.00190& 0.00280&    27.5&    47.5 & 0.4  & 0.3 \\
0.00280& 0.00435&    27.5&    47.5 & 0.4 & 0.3 \\
0.00435& 0.00687&    27.5&    47.5 & 0.5 & 0.3 \\
\hline
0.00135& 0.00250&    47.5&    90.0 & 0.4 & 0.4 \\
0.00250& 0.00410&    47.5&    90.0 & 0.5 & 0.5 \\
0.00410& 0.00650&    47.5&    90.0 & 0.6 & 0.5 \\
0.00650& 0.01000&    47.5&    90.0 & 0.6 & 0.5 \\
\hline
0.00150& 0.00260&    90.0&   160.0 & 0.5 & 0.5 \\
0.00260& 0.00435&    90.0&   160.0 & 0.6 & 0.6 \\
0.00435& 0.00775&    90.0&   160.0 & 0.6 & 0.6\\
0.00775& 0.02100&    90.0&   160.0 & 0.8 & 0.7 \\
0.02100& 0.04000&    90.0&   160.0 & 1.2 & 0.7 \\
\hline
0.00375& 0.00900&   160.0&   300.0 & 0.7 & 0.7 \\
0.00900& 0.01625&   160.0&   300.0 & 1.0 & 0.9 \\
\hline
0.00975& 0.01900&   300.0&   700.0 & 1.3 & 1.2 \\
0.01900& 0.03125&   300.0&   700.0 & 1.7 & 1.5 \\
\hline
0.02250& 0.03750&   700.0&  1500.0 & 2.4 & 2.4 \\
 \hline
 \end{tabular}
\end{center}
\caption{\small
 The percentage contribution of non-vanishing terms in the
$e_c=0$ limit to $F_{2\,c}$, Eq.~(\ref{eq:irc_el}),
and to $\sigma_{\rm vis,c}^{\rm theo}$, Eq.~(\ref{eq:deltasigvis}). 
The grid in $(x,Q^2)$ assumed
is similar to that of the combined HERA $F_{2\,c}$ dataset,
see Fig.~\ref{fig:kin}. \label{tab:reshvqis2}}
\end{table}
~

\subsection{All--order resummation}

When more gluon
splitting processes are inserted in the gluon propagator in Fig.~\ref{fig:feyn-nlo-ir}, before
the final splitting into the $h\bar{h}$ pair, they lead to
corrections enhanced by further
powers of $\alpha_s\log^2 Q^2/m^2$ (i.e. double log enhanced).
The effect of the complete resummation
of these double logarithms was studied in detail in Ref.~\cite{Mangano:1992qq}
(see also Ref.~\cite{Mueller:1985zp}), in the case of heavy flavour production
associated with the production of a gluon jet at a scale $Q^2$.
It amounts to an enhancement of the production cross section by a factor $n_g(Q^2,K^2)$,
where $K^2$ is the virtuality of the $h\bar{h}$ pair. The way $n_g$ is
determined and its precise definition are discussed
Ref.~\cite{Mangano:1992qq}; here, we can interpret it as the enhancement of
the light--quark initiated heavy quark contribution of
Fig.~\ref{fig:feyn-nlo-ir}, i.e. the numerator of
 Eq.~(\ref{eq:irc_el}),
whose relative 
 size is shown in Fig.~\ref{f2c-laenen}. 
We have $K^2\ge 4m^2$,
and $Q^2$ can be identified with the DIS $Q^2$ scale.

The reason why the scale $Q^2$  should be  chosen as the scale 
for the evaluation
of $n_g$, as opposed to $W^2$, deserves a comment, given that in the small $x$
limit these two scales are widely different. In fact, if we assume
a behaviour like $1/x^{1+\delta}$ for the light quark density in the small
$x$ limit, the
contribution of the first diagram of Fig.~\ref{fig:feyn-nlo-ir} has the form
\begin{equation}
\int_x^{z_{max}} \frac{d z}{z} \left(\frac{\alpha_s}{4\pi}\right)^2
L_q^{{\rm NS},(2)}(z,Q^2/m^2) q(x/z)\approx  q(x)
\left(\frac{\alpha_s}{4\pi}\right)^2 \int_0^{z_{\rm max}} d z z^\delta L_q^{{\rm NS},(2)}(z,Q^2/m^2)\;.
\end{equation}
Since $L^{{\rm NS},(2)}$ is a non-singlet coefficient function, its $z$ integral is finite at
the lower limit,
and the result grows like $\log^3 Q^2/m^2$ (see
Eq.~(D.8) and subsequent comment in Ref.~\cite{Buza:1995ie}, see also Ref.~\cite{Chuvakin:1999nx}).
This confirms
the scale choice of $Q^2$ rather than $W^2$. 

The function $n_g$ is plotted in Fig.~\ref{fig:mult} as a function of
$K^2$ for different scales $Q^2$: 
at a scale $K^2=4m^2=8\;{\rm GeV}^2$, we see that the enhancement is below 20\%{}
for $Q^2\le 100\;{\rm GeV}^2$, and below  90\%{} for $Q^2\le 1000\;{\rm GeV}^2$.
Hence, the effect is moderate in the HERA region, where the
contribution which is thus enhanced amounts to a few percent of the
structure function in the first place, as shown in
Fig.~\ref{f2c-laenen}. However, the effect becomes large as $Q^2$
increases and it could be a significant correction at a higher energy
electron--proton collider such as the LHeC~\cite{Dainton:2006wd}.
Also,
 in the very small
$x$ limit, single log enhanced contributions of the form $\alpha_s \log W^2/m^2$ will arise~\cite{Catani:1990eg},
and eventually prevail on the double logarithms discussed above. The effect of these small-$x$
logarithms is unlikely to be important the HERA energy regime, but it
may deserve further
studies.

\section{Conclusion and outlook}

We have presented a study of the theory and the phenomenology of the
inclusion of heavy quark corrections in deep--inelastic structure
functions, using the FONLL approach that had been previously proposed
in the context of heavy quark photo--- and hadroproduction. This approach is
suitable for the combination of fixed order heavy quark emission terms
with the all--order resummation of collinear logs which appear at
scales much larger than the heavy quark mass. A significant feature of
the method is that the perturbative order at which the fixed--order
and resummed results are obtained can be chosen independently of each
other in the most suitable way: in fact, we have explicitly considered
two different NLO implementations (denoted as scheme A and scheme B)
in which fixed order results of order $\alpha_s$ or 
$\alpha^2_s$ have been combined with NLO parton distributions. 

After discussing in detail the method and its implementation to
${\mathcal O}(\alpha_s^2)$, and verifying explicitly its consistency,
we have studied the impact of heavy quark corrections and their
ambiguity on the $F_2$ and $F_L$ structure functions. We have found
that charm mass effects have a significant impact, at the level
of 10\% on the charm structure function $F_{2\,c}$
(for fixed PDF) for scales as large as
$Q^2\approx 10$--$20$~GeV$^2$. The effect is
rather larger  in the threshold region, and also
for the $F_{L\,c}$ structure function, for which it is a sizable
correction even at $Q^2\approx 100$~GeV$^2$.
For scales $Q^2\approx 4$~GeV$^2$ there is an
ambiguity due to subleading terms which are not accurate as threshold,
which for $F_{2\,c}$ at ${\mathcal O}(\alpha_s)$ 
is almost as large as the whole heavy quark correction. 

We have
seen that introducing a suppression term near threshold for these subleading
contributions  (such as provided by a
damping factor or by so--called $\chi$--scaling) somewhat reduces this
ambiguity. Comparison to exact  ${\mathcal O}(\alpha_s^2)$  results
suggests that this threshold suppression improves the accuracy of the
calculation, though beyond these order there are no exact results to
compare to. The ambiguity is substantially reduced if ${\mathcal
  O}(\alpha_s^2)$  heavy quark production terms are used within the
NLO computation.

We have finally discussed issues related to mass singularities in the
experimental definition of the heavy quark structure functions, which
differs somewhat from the theoretically most natural definition: we
have seen that the impact of these corrections is small in the HERA
kinematic region, but could become relevant at higher energies.

Besides giving a simple, flexible and practically viable
implementation of heavy quark effects, our results  provide  a 
framework for the understanding of the
impact of heavy quark effects on the determination of parton
distributions and of the dependence of results on  details  of the
procedure due to its theoretical uncertainties.

\bigskip

{\bf Acknowledgements:} We thank Andrea Piccione for collaboration in
the early stages of this work.
We thank Jack Smith for correspondence, 
and providing computer programs. We are
especially grateful to Massimo Corradi for eliciting our interest in
issues related to the experimental definition of $F_{2\,c}$ and for
continuous assistance with the HVQDIS code. We thank Matteo Cacciari,
Joey Huston, Pavel Nadolsky,  Fred Olness and Robert Thorne
for several useful discussions, and Sergey Alekhin and Johannes
Bl\"umlein for clarifications about Ref.~\cite{Alekhin:2009ni}. 
This work was partly supported 
by the European network HEPTOOLS under contract MRTN-CT-2006-035505 and  by the Netherlands Foundation for Fundamental
Research of Matter (FOM) and the National Organization for Scientific Research
(NWO).

%%%%%%%%%% Start TeXmacs macros
\renewcommand{\tmop}[1]{\ensuremath{\operatorname{#1}}}
\renewcommand{\tmtexttt}[1]{{\ttfamily{#1}}}
%%%%%%%%%% End TeXmacs macros

\appendix
\section{Appendix:
Implementation of the FONLL scheme for $F_2(x,Q^2)$ up to $\mathcal{O}\lp \alpha_s^2\rp$.}
\label{sec:appendix}
In this appendix,
we collect the relevant explicit formulae for the practical computation of the
the DIS structure function $F_2(x,Q^2)$ within the FONLL approach
up to $\mathcal{O}\lp \alpha_s^2\rp$, which has been already discussed
more formally in Sec.~\ref{sec:alphasq}. 
Formulae for $F_L(x,Q^2)$ are simpler and can be
obtained in an analogous way, and thus are not
discussed here. 
Unlike the formulae found 
in Sec.~\ref{sec:implement}, given
for the purpose of illustration, here we focus on practical implementation
issues, i.e., we state which equations and computer codes we have used to arrive
to the numerical results for the FONLL structure
functions that have been presented in Sec.~\ref{sec:pheno}.

For simplicity, we denote in this Appendix
$F_2$ by $F$ and similarly for all the various coefficients,
and we write throughout
\begin{equation}
  \alpha_s\equiv \alpha_s^{(n_l + 1)} (Q^2) ,\quad\quad f_i(y,Q^2)=f_i^{(n_l+1)}(y,Q^2)\;.
\end{equation}
As has been discussed in Sec.~\ref{sec:fonllm}, in order
to obtain the full structure function 
$F$ we compute separately the ``light'' $F_{l}$ and 
``heavy'' $F_{h}$
contributions.
For $F_{h}$, up to order $\mathcal{O} (\alpha_s^2)$, the
relevant equations are
\begin{eqnarray} \label{eq:f2diffappfirst}
  F^{\tmop{FONLL}}_{h}(x,Q^2) & = & F^{(d)}_{h}(x,Q^2) + F^{(n_l)}_{h}(x,Q^2), \\
  F^{(d)}_{h}(x,Q^2) & = & F^{(n_l + 1)}_{h}(x,Q^2) - F^{(n_l, 0)}_{h}(x,Q^2), 
  \\
  F^{(n_l)}_{h} (x,Q^2)& = & x \sum_{i =g, q,\bar{q}} \int_{x_\chi}^1 \frac{d y}{y}
  C_{i,h}^{(n_l)} \left( \frac{x}{y}, \frac{Q^2}{m^2}, \alpha_s \right) f_i(y, Q^2),
\\ \label{eq:f2diffappmz}
  F^{(n_l, 0)}_{h}(x,Q^2) & = & x \sum_{i =g, q,\bar{q}} \int_x^1 \frac{d y}{y}
  C^{(n_l,0)}_{i,h} \left( \frac{x}{y}, L, \alpha_s \right)
  f_i(y, Q^2) \\ \label{eq:f2diffapplast}
  F^{(n_l+1)}_{h} (x,Q^2)& = & x \sum_{i =g, q,\bar{q},h,\bar{h}} \int_{x}^1 \frac{d y}{y}
  C_{i,h}^{(n_l+1)} \left( \frac{x}{y}, \alpha_s \right) f_i(y, Q^2),
\end{eqnarray}
where $x_\chi=x(1+4m^2/Q^2)$ and $C^{(n_l,0)}_{i,h}$ are the massless limit of the 
massive coefficient functions $C^{(n_l)}_{i,h}$. 
Note that in Eqs.~(\ref{eq:f2diffappfirst}-\ref{eq:f2diffapplast})
the PDFs and the strong coupling are always given in the $n_l+1$ flavour scheme.

\begin{figure}
\begin{center}
\epsfig{figure=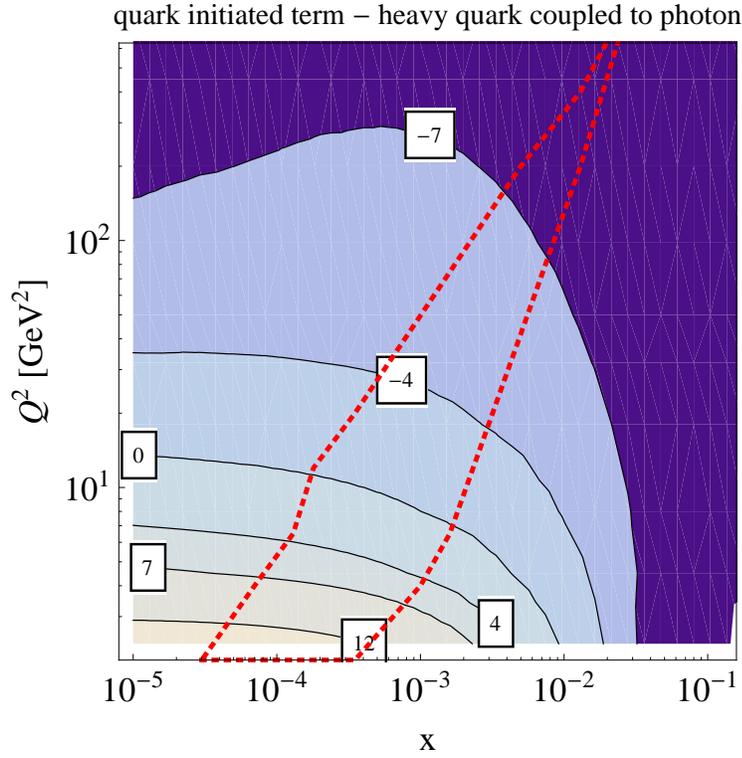,width=0.65\textwidth}
\caption{\small Same as Fig.~\ref{f2c-laenen}, but for the
percentage contribution  $R_{\rm light}$ of
light-quark initiated terms to $F_{2\,c}$
Eq.~(\ref{eq:irc_notg}).
}\label{f2c-laenen2}
\end{center}
\end{figure}

\begin{figure}
\begin{center}
\epsfig{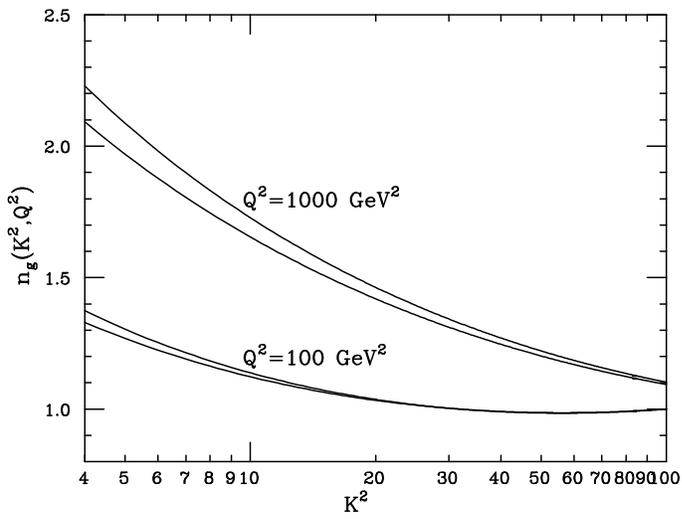}
\end{center}
\caption{\label{fig:mult}
The resummation factor $n_g(Q^2,K^2)$ as a function of $K^2$, for two values
of $Q^2$. Upper lines have $\Lambda=300$~MeV, lower lines have
$\Lambda=200$~MeV.}
\end{figure}
Up to  $\mathcal{O} (\alpha_s^2)$, the massive coefficient
functions $C_{i,h}^{(n_l)}$ are given by
\bea
  z C_{i,h}^{(n_l)} \left( z, \frac{Q^2}{m^2} \right) 
&=& \frac{e_h^2
  \alpha_s}{(2 \pi)^2}  \frac{Q^2}{m^2}  \Bigg[ c_{2,\,i}^{(0)} \left( z,
  \frac{Q^2}{m^2} \right) 
\nonumber \\&+& 4 \pi \alpha_s \left\{ c_{2,\,i}^{(1)} \left( z,
  \frac{Q^2}{m^2} \right) + \bar{c}_{2,\,i}^{(1)} \left( z, \frac{Q^2}{m^2}
  \right) \log \frac{Q^2}{m^2} \right\} \Bigg],
\eea
The massive coefficient functions
 $c_{2,\,g}^{(1)}$, $\bar{c}_{2,\,g}^{(1)}$, 
$c_{2,\,q}^{(1)}$,
 $\bar{c}_{2,\,q}^{(1)}$ 
were first computed in~\cite{Laenen:1992zk}, together with the 
analog expressions for $F_L$. 
They are given as phase-space integrals of
differential cross sections; the corresponding $c_T^{(1)}$,
$c_L^{(1)}$ coefficients (as defined in Ref.~\cite{Laenen:1992zk})
are plotted in Figs.~6--10 of that reference
(the quark coefficient has been subsequently corrected as Fig.~6 of 
Ref.~{\cite{Harris:1995tu}}). For the FONLL phenomenological studies 
of Sec.~\ref{sec:pheno} we have used
direct numerical integration of the expressions of
Ref.~{\cite{Laenen:1992zk}}, benchmarked against the plots in
Refs.~{\cite{Laenen:1992zk}} and~{\cite{Harris:1997zq}}; we have not
relied upon the interpolation provided in Ref.~\cite{Riemersma:1994hv}.
The massless--limit expressions
of the $C_{i,h}^{(n_l),1}$ coefficients, needed to compute $C^{(n_l,0)}_{i,h}$,
 were
taken from a Fortran implementation~\cite{smith} of  the results of
Ref.~{\cite{Buza:1995ie}}; they were tested at large values of $Q^2$ against our
massive expressions. The $\overline{\tmop{MS}}$ cross sections $F^{(n_l +
1)}_{h}$ at the NNLO order were computed using the {\tt QCDNUM}
program~\cite{qcdnum}.

For the ``light'' contribution $F_{l}$ instead, 
up to order $\mathcal{O} (\alpha_s^2)$, it is convenient to rearrange
the expression Eq.~(\ref{eq:FONLLnnlo}) of Sec.~\ref{sec:alphasq},
exploiting the fact that all contributions to $F^{(d)}$
Eq.~(\ref{eq:fdnnlo}) cancel except the heavy quark contribution, and
that the massive--scheme contribution with $n_l$ flavours
is most easily obtained by
writing it in terms of the corresponding contribution with $n_l+1$,
which can then be obtained from the \tmtexttt{QCDNUM} code. We thus wind up with
the following expression:
\begin{eqnarray}
  F^{\tmop{FONLL}}_{l}(x,Q^2) & = & F^{(d)}_{l} + F^{(n_l)}_{l}, \\
  F^{(d)}_{l}(x,Q^2) & = & x \sum_{i = h, \bar{h}} \int_x^1 \frac{d y}{y} C_{i,
  l}^{(n_l + 1)} \left( \frac{x}{y}, \alpha_s \right) f_i(y,
  Q^2),  \label{eq:F2dlfinal}\\
  F^{(n_l)}_{l}(x,Q^2) & = & F^{(n_l + 1)}_{l} - x \sum_{i = h, \bar{h}}
  \int_x^1 \frac{d y}{y} C^{(n_l + 1)}_{i} \left( \frac{x}{y}, \alpha_s
  \right) f_i(y, Q^2)  \label{eq:F2lfinal}\\
  & + & x \sum_{i \neq h, \bar{h}, g} \int_x^1 \frac{d y}{y} D_{i, l} \left(
  \frac{x}{y}, \frac{Q^2}{m^2}, \alpha_s \right) f_i(y, Q^2) \,
\nonumber
\eea
where we have defined
\bea
  D_{i, l} \left( z, \frac{Q^2}{m^2}, \alpha_s \right) & = & \left(
  \frac{\alpha_s}{4 \pi} \right)^2 e_i^2 \left[ L_{q}^{(2)} \left(
  z, \frac{Q^2}{m^2} \right) - \delta (1 - z) \int_0^1 d z L_{q}^{(2)}
  \left( z, \frac{Q^2}{m^2} \right) \right]  \label{eq:Dterm}\\
  & - & \left( \frac{\alpha_s}{2 \pi} \right)^2  \left[ e_i^2 K_{q q} (z, L)
  + \frac{2 T_R}{3} L C_i^{(n_l), 1} (z) \right] - \frac{\partial}{\partial
  n_l} C^{(n_l + 1)}_{i} \left( z, \alpha_s \right) \nonumber
\end{eqnarray}
where the function $L^{(2)}_{q}$ is given in formula (A.1) of Ref.~{\cite{Buza:1995ie}}, with the further assumption
\begin{equation}
L_q^{(2)} \left( z, \frac{Q^2}{m^2}, \alpha_s \right)=0\quad\quad \mbox{for}\quad
z\ge \frac{1}{1+\frac{4m^2}{Q^2}}\;.
\end{equation}

The function $K_{qq}$ is given by
\begin{eqnarray}
 K_{q q} (z,L) &=&K_{q q} (z) 
   +   \frac{L^2}{2}  \frac{2 T_R}{3}
  P^{((n_l),0}_{q q} (z) 
   -  L \Delta_{q q} (z),\\
K_{q q}(z) &=& C_F T_R \left[ \frac{1 + z^2}{1 - z} \left( \frac{1}{6} \log^2 z
   + \frac{5}{9} \log z + \frac{28}{27} \right) + (1 - z) \left( \frac{2}{3}
   \log z + \frac{13}{9} \right) \right]_+ , \\
\Delta_{q q} (z) &=& C_F T_R \left[ \frac{1 + z^2}{1 - z} \left(
  \frac{2}{3} \log z + \frac{10}{9} \right) + \frac{4}{3} (1 - z)
  \right]_+,
\end{eqnarray}
and the NLO quark coefficient function 
(see for example Ref.~\cite{Furmanski:1981cw}) is
\begin{equation}
C_i^{(n_l),1}=e_i^2 C_F\left[\frac{1+z^2}{1-z}\left(\log\frac{1-z}{z}-\frac{3}{4}
\right)+\frac{1}{4}(9+5z)\right]_+ \ .
\end{equation}
  The term proportional to
$\delta (1 - z)$  in Eq.~(\ref{eq:Dterm}) is the heavy flavour
virtual correction, and it is needed
to enforce the Adler sum rule, which thus determines its coefficient. 
We can easily see that
\begin{equation}
  D_{i, l} \left( \frac{x}{y}, \frac{Q^2}{m^2}, \alpha_s \right) = B_{i, l}
  \left( \frac{x}{y}, \frac{Q^2}{m^2}, \alpha_s \right) - C^{(n_l + 1)}_{i}
  \left( \frac{x}{y}, \alpha_s \right),
\end{equation}
so that, using also Eq. (\ref{eq:expcg}), we have
\begin{equation}
  F^{(n_l)}_{l} = x \sum_{i \neq h, \bar{h}} \int_x^1 \frac{d y}{y} B_{i,
  l} \left( \frac{x}{y}, \frac{Q^2}{m^2}, \alpha_s \right) f_i(y,
  Q^2) .
\end{equation}

In Eq.~(\ref{eq:F2lfinal}), the term $F^{(n_l + 1)}_{l}$ is evaluated
directly using the \tmtexttt{QCDNUM} program. All remaining terms must be
evaluated independently. Note that the second term on the r.h.s of 
Eq.~(\ref{eq:F2lfinal}) is identical to Eq.~(\ref{eq:F2dlfinal}): these
terms thus would cancel exactly, but the cancellation does not happen
if one uses a threshold prescription such as Eq.~(\ref{eq:threshold})
or Eq.~(\ref{eq:chiscaling}), because the threshold prescription only
affects 
$F^{(d)}$ but not
$F^{(n_l)}$. 
 The coefficients
of $n_l$ in $C^{(n_l + 1)}_{i}$ and $C_{h / \bar{h}, l}^{(n_l + 1)}$ have
been evaluated using the same equations as the \tmtexttt{QCDNUM} package,
given in Refs.~{\cite{vanNeerven:1999ca}} and~{\cite{vanNeerven:2000uj}},
since they have to compensate the \tmtexttt{QCDNUM} result in $F^{(n_l +
1)}_{l}$. Notice that,  according
to Eq.~(\ref{eq:mustvanish}) $D_{i, l}$ vanishes in the massless limit.

All the $\mathcal{O} (\alpha^2_s)$ results given so far, used with NNLO PDFs,
correspond to the scheme C of Sec.~\ref{sec:fonllsf}.
The $\mathcal{O} (\alpha_s)$ results (scheme A) were obtained
using the same formulae, truncated at order $\alpha_s$, and using NLO
evolution and massless coefficient functions in~\tmtexttt{QCDNUM}. 
The alternative NLO implementation, 
denoted by scheme B in Sec.~\ref{sec:fonllsf}, uses
 the NLO approximation in the massless result, and the full
 $\mathcal{O} (\alpha_s^2)$
result in the massive one 
is also obtained in a similar way.  The only
caveat is that in the subtracted term in Eq.~(\ref{eq:f2diffappmz}) we replace
the massless limit of the
 $C_{g,h}^{(n_l),1}$ (denoted by $C_{g,h}^{(n_l,0), 1}$)
with
\begin{equation}
  C_{g,h}^{(n_l,0), 1} \left( z, \frac{Q^2}{m^2} \right) \rightarrow C_{g,h}^{(n_l,0), 1} \left( z, \frac{Q^2}{m^2} \right) - C_{g,h}^{(n_l,0), 1} \left( z,
  1 \right),
\end{equation}
i.e. the non-logarithmic term is excluded from the subtraction,
according to the discussion in Sec.~\ref{sec:mismatch}.

\end{document}